%% file: manuscript_arxiv.tex
\documentclass{article}

\usepackage{lscape}

\usepackage{arxiv}

\usepackage{graphicx}
\usepackage{amsfonts} 
\usepackage{amsmath}

\usepackage[utf8]{inputenc} 

\usepackage[dvipsnames]{xcolor} 

\usepackage{tikz} 

\newcommand\bcmdtab{\noindent\bgroup\tabcolsep=0pt%
  \begin{tabular}{@{}p{10pc}@{}p{20pc}@{}}}

\newcommand\ecmdtab{\end{tabular}\egroup}
\newcommand{\vvset}[1]{{\mathcal{#1}}}
\newcommand{\vvector}[1]{\vv{\bm{#1}}}

\newcommand{\Acal}{\vvset{A}}
\newcommand{\Gcal}{\vvset{G}}
\newcommand{\Dcal}{\vvset{D}}
\newcommand{\Ccal}{\vvset{C}}
\newcommand{\Ecal}{\vvset{E}}
\newcommand{\Vcal}{\vvset{V}}
\newcommand{\Hcal}{\vvset{H}}
\newcommand{\Tcal}{\vvset{T}}

\newcommand{\Cfrak}{\mathfrak{C}}

\newcommand{\Imatr}{\mathbf{I}}
\newcommand{\Mmatr}{\mathbf{M}}

\newcommand{\eigenvector}[1]{\mathbf{e}^{(#1)}}
\newcommand{\kout}{k^{\mathrm{out}}}

\providecommand{\href}[2]{\texttt{#2}}
\providecommand{\url}[1]{\texttt{#1}}

\providecommand{\eqref}[1]{(\ref{#1})} 
\newcommand{\figref}[1]{Fig.\ \ref{#1}}
\newcommand{\Figref}[1]{Fig.\ \ref{#1}} 
\newcommand{\secref}[1]{section \ref{#1}}
\newcommand{\appref}[1]{Appendix \ref{#1}}
\newcommand{\tabref}[1]{Table \ref{#1}}

\newcommand{\vdef}[1]{\textsc{#1}}

\newcommand{\bea}{\begin{eqnarray}}
\newcommand{\eea}{\end{eqnarray}}
\newcommand{\beq}{\begin{equation}}
\newcommand{\eeq}{\end{equation}}

\newcommand*{\tran}{^{\mkern-1.5mu\mathsf{T}}}

\title{Cycle Analysis of Directed Acyclic Graphs}

\author{	{Vaiva VASILIAUSKAITE} \thanks{ vvasiliau@ethz.ch, funded by SoBigData++: European Integrated Infrastructure for Social Mining and Big Data Analytics (Grant agreement ID: 871042)}\\
	Computational Social Science, ETH Z\"urich, Z\"urich, Switzerland \\
	\And
	Tim S.\ EVANS\thanks{tim.evans@imperial.ac.uk} \\
         Centre for Complexity Science, Theoretical Physics Group,\\ Imperial College London, SW7 2AZ, U.K.\\
	\And
	Paul EXPERT\thanks{paul.expert08@imperial.ac.uk, funded by NIHR Imperial BRC (grant number NIHR-BRC-P68711)} \\Global Digital Health Unit, Imperial College London, SW7 2AZ, U.K.\\
World Research Hub Initiative, Tokyo Institute of Technology, Japan\\	
	$^{*}$ Corresponding author
}
\date{}

\begin{document}

\maketitle

\begin{abstract}
In this paper, we employ the decomposition of a directed network as an undirected graph plus its associated node metadata to characterise the cyclic structure found in directed networks by finding a Minimal Cycle Basis of the undirected graph and augment its components with direction information. We show that only four classes of directed cycles exist, and that they can be fully distinguished by the organisation and number of source-sink node pairs and their antichain structure. We are particularly interested in Directed Acyclic Graphs and introduce a set of metrics that characterise the Minimal Cycle Basis using the Directed Acyclic Graphs metadata information. In particular, we numerically show that Transitive Reduction stabilises the properties of Minimal Cycle Bases measured by the metrics we introduced while retaining key properties of the Directed Acyclic Graph. This makes the metrics consistent characterisation of Directed Acyclic Graphs and the systems they represent. We measure the characteristics of the Minimal Cycle Bases of four models of Transitively Reduced Directed Acyclic Graphs and show that the metrics introduced are able to distinguish the models and are sensitive to their generating mechanisms. 
\end{abstract}


\renewcommand{\thefootnote}{\arabic{footnote}}



\section{Introduction}\label{sec:intro}
Hierarchy is a landmark of complexity. In the words of H.\ Simon ``complexity frequently takes the form of hierarchy'' and it is ``one of the central structural schemes that the architect of complexity uses''~\cite{S91}. Hierarchy itself can take multiple forms, e.g.\ ``order'', ``level'', ``control'' or ``inclusion''~\cite{L06}. A unifying feature of all types of hierarchy is that they can be represented as a partially ordered set (poset) that represents the relationships between elements of the system. The term ``partial'' reflects the fact that not all pairs of elements need to be directly ordered. Directed Acyclic Graphs (DAGs) do not contain directed closed path and can thus be used to represent the relationships between the elements of a poset. A key distinguishing feature between a poset and a DAG representation of a poset is that while all transitive relationships intrinsically exist in a poset, they do not need to be present in a DAG representation of a poset. A good example is a citation network. The arrow of \vdef{time} constrains a paper to only cite older papers. The citation network thus represents a partially ordered set, where the order is given by the constraint imposed by time, yet not all transitive relations are present in the DAG as a paper only cites a fraction of all previously published papers.

DAGs are thus ``doubly-complex'' systems. The first level of complexity lies in the order relationship underlying the DAG, and the second comes from the ``missing information'' indirectly represented by the missing or unrealised edges of the poset underlying the DAG. In this paper, we characterise the mesoscopic structures present in DAGs that are carved in the underlying poset by the unobserved transitive edges.

Mesoscopic structures are rich descriptors to understand the organisation of complex systems, and the characterisation of connectivity patterns is one of the pillars of the study of complex networks. They are usually defined in terms of ``over-connectivity" and the notion of more densely connected subsets of nodes is at the centre of the definition of many mesoscopic organisation patterns. Such structures can take the form of core-periphery~\cite{KM18,BWRPMG13}, community~\cite{GN02,F10,ZMZSY18,JYSLQB18,YAT16,LLM10,LF09}, or clique~\cite{C18,Palla:2005cj,Evans_2010}. Cliques, cores and communities are important examples of emergent organisational patterns that can be interpreted as capturing beyond pairwise node organisation. The representation of higher-order interaction and their formalisation in terms of hypergraphs or simplicial complexes has recently seen a surge in interest
~\cite{Petri:2014hq,Battiston:2020kp}.

While over-connectivity can be used to define mesoscopic or higher order structures, it is not adapted to DAGs. In this case, it is more adequate to define structures by the \textit{absence} of connectivity. The notion of \vdef{antichains}---generalised ``anti-paths''--- which group together nodes that do not share direct connectivity, are relevant in DAGs, as they are composed of nodes at the same hierarchical level~\cite{VE20}. Another example are cycles: they can be described as a subset of nodes that have exactly two neighbours in their induced subgraph. A cycle can be seen as the boundary of a potential clique in a thresholded graph, or a hole in the fabric of a network. Cycles are also examples of structures that encode mesoscopic information as they form organised motifs. 

In this paper, we show that despite their names, DAGs can have cycles and that only DAGs that are directed trees are truly acyclic. Our starting point is the following, albeit informal, definition of a network: a network is \textit{a graph with something more}. This definition clearly highlights that graphs, being pure combinatorial objects, do not encode all the nuances and constraints of the complex systems that they represent. We formalise this definition by considering a DAG to be a combinations of two elements: 1) an undirected graph and 2) the metadata associated with its nodes and edges. 

The idea at the centre of our work is thus to use meta-data to give interpretability and contextualisation to the Minimal Cycle Basis associated with the undirected graph underlying a DAG. We then introduce metrics characterising the properties and organisation of generalised directed cycles forming said basis. In particular, we use Transitive Reduction on DAGs and show that the metrics we introduce differentiate between four different DAGs generating models; two deterministic ones: the lattice and Russian Dolls DAGs and two random: Erd\"os-R\`enyi and Price Model DAGs. Transitive Reduction on edges reveals the tree-ness of the graph underlying a DAG by removing non-essential cycles. We introduce a new compartment for ``acyclic'' network analysis in the rich topological data analysis toolbox.

This paper is organised as follows. In \secref{sec:methods} we first formally describe Directed Acyclic Graphs and Transitive Reduction. Then, we define and describe the generalised directed cycles and cycle bases and conclude by discussing the desirable properties Transitively Reduced DAGs have with respect to cycles bases. In \secref{sec:cycle_metrics} we introduce metrics for characterising cycle bases. In the following section, \secref{sec:algorithm}, we discuss the procedures used to compute the reduced minimum cycle basis. And finally, in \secref{sec:model_results}, we study the differences between Minimum Cycle Bases in several DAGs models, and show that our proposed metrics can differentiate and characterise network families.

\section{Methods}\label{sec:methods}
In this section, we review the basic properties of Directed Acyclic Graphs (DAGs), define four classes of generalised cycles in DAGs and introduce Minimal Cycle Bases (MCB), and discuss the effect of Transitively Reduction on the MCB.

\subsection{Graphs, Hierarchy, and Order}\label{sub_sec:DAG}

For many data sets, an undirected graph $\Gcal$ is an excellent way to capture important information. The set $\Vcal$ of $|\Vcal|=N$ objects of interest form the nodes, and the existence of a relationship between a pair of nodes is encoded by an undirected edge, gives us the edge set $\Ecal$ with $|\Ecal|=E$ edges. For instance, the nodes could be documents and we connect nodes by an edge if one document cites another giving us an undirected network representation of what is known as a citation network.

We are interested in the cycles in our networks, which are simply paths between vertices that run along the edges in a graph, as long as they end where they started without any other vertex being visited twice.  Formally, a path is a sequence of nodes in the network, $\{v_0,v_2, \ldots,v_\ell\}$, where every consecutive pair of nodes is an edge in the graph $(v_i, v_{i+1}) \in \Ecal$ \footnote{Note in some contexts, for example section 4.2.5 of \cite{WF94}, what we call a path is known as a `walk' while our self-avoiding paths are what others call a `path'.} \cite{N10}. A cycle is then a path that starts and ends at the same vertex ($v_0=v_\ell$, giving a \vdef{closed path}) but otherwise does not intersect itself,
 therefore a cycle can be treated as a subgraph $\Ccal$ in which all nodes have a degree of two.

However, in many cases we can go further as additional information in the meta-data gives a natural direction to the pairwise relationships.  In our citation network example, the edge could point from the document whose bibliography contains the second document. The set of nodes in the directed graph representation of the data is the same $\Vcal$. The edges in the directed case are also the same pairs of nodes as the edges in the undirected case but are now endowed with a direction, $\Ecal_{\textrm{dir}}$.  That is if $(v_i,v_j)$ is an edge from $v_i$ to $v_j$ in the directed case we know that $(v_i,v_j)=(v_j,v_i)$ is also an edge in the undirected representation. The paths in a directed graph must respect the direction of each edge so that in the formal definition $(v_i, v_{i+1})$ must be an edge, it is not sufficient for $(v_{i+1}, v_{i})$ to be an edge of a path in directed graph $\Dcal$. Closed directed paths are called directed cycles.

We are interested in a special case where there are no directed cycles in the directed network representation so that we have a Directed Acyclic Graph $\Dcal_{\textrm{DAG}}$.
This lack of directed cycles reflects an order implicit in our system, typically encoded in some additional information available in our data set. The simplest origin of such an order is a natural hierarchy. For instance, in some contexts it might be useful to direct the links in our citation network from high impact to lower impact papers (see journal ranking \cite{LE15} or code prerequisites \cite{VE20} for further examples). Another common form for this order information is a time stamp on the nodes.  For instance each publication has a publication date and we can only cite from newer to older documents. Thus in the directed graph representation of a citation network, the directed paths in the network will never form a directed cycle because that would require at least one document to cite a document published later in time\footnote{Of course, in practice data is never perfect. In our example, documents can be produced in different versions at different times so the directed citation network may produce a few cycles, less than one percent in some examples \cite{CE16}.  However, the order encoded in the meta data, here the time-stamp of the vertices, gives us several ways to produce a pure DAG representation.}. A couple of points worth noting.  The meta data may not always be able to order every pair of nodes. Equally, just because two nodes can be placed in a certain order by the meta data does not mean that there must be an edge between that pair of nodes. In our DAG citation network, two papers can have the same publication date and papers do not cite every other paper published earlier.

An undirected graph $\Gcal=(\Vcal,\Ecal)$ can thus be turned into a directed graph $\Dcal=(\Vcal,\Ecal_{\textrm{dir}})$ if its nodes are endowed with a meta-data such that any edge can be assigned a direction based on the order between its nodes and form a directed edge set $\Ecal_{\textrm{dir}}$. We define a function $F_{\textrm{dir}}$ that takes an undirected graph $\Gcal$ and all available pairwise order relations between nodes $\mathcal{O}$ and yields a directed graph $\Dcal$:
\begin{equation}
     F_{\textrm{dir}}(\Gcal,\mathcal{O}) = \Dcal.
\end{equation}
$F_{\textrm{dir}} $ maps the nodeset onto itself,
$F_{\textrm{dir}}(v\in\Vcal,\mathcal{O})\rightarrow v\in \Vcal_{\textrm{dir}}$ for all nodes in $\Vcal$, whereas for edges, $F_{\textrm{dir}} $ maps each undirected edge $(u,v)\in\Ecal$ onto one directed edge $\Ecal_{\textrm{dir}}$, based on how pairs of nodes are ordered in $\mathcal{O}$:
\begin{equation}
    F_{\textrm{dir}}((u,v), \mathcal{O}) = \begin{cases}
     (u,v) \in \Ecal_{\textrm{dir}} \text{ if } u\prec v \text{ in }\mathcal{O} \\
     (v,u) \in \Ecal_{\textrm{dir}} \text{ if } u\succ v \text{ in }\mathcal{O}. \\
      \end{cases}
\end{equation}
Furthermore, we define a function $F_{\textrm{undir}}$ that maps a directed graph $\Dcal$ to its underlying undirected $\Gcal$ by striping its edges from their direction, see \figref{fig:fig_1_panel}(a) for an illustration of both functions. As $F_{\textrm{dir}}$ is effectively acting at the edge level, it can take any subgraph as argument, and in particular cycles. In the next section, we will study in detail the images of cycles obtained by $F_{\textrm{dir}}$ and show that they can be classified into four general classes.

$\mathcal{O}$ only really needs to encode pairwise node order via some relation operator $\prec$ that only needs to possess reflexivity: $u \prec u$, and antisymmetry: if $u \prec v$ and $v \prec u$, then $u = v$. However, if it also possesses transitivity: if $u \prec v$ and $v \prec w$, then $u \prec w$, then $\mathcal{O}$ forms a partially ordered set ---\vdef{poset}. In this case $F_{\textrm{dir}}$ maps $\Gcal$ onto a $\Dcal$ that is a Directed Acyclic Graph, which in general is only a partial representation of the information contained in $\mathcal{O}$ since not all transitive edges are present in $\Gcal$, as we mentioned when discussing the citation network example above. A network representation of $\mathcal{O}$ itself would yield a DAG with all possible transitive edges present.

In this paper, we will focus our attention on a subclass of DAGs: Transitively Reduced DAGs, or reduced DAGs for short. Transitive reduction (TR), an operation that, when applied to a Directed Acyclic Graph (DAG), removes all edges representing transitive relationships and thus only keeps the edges that are essential to maintain connectivity. Specifically, if for an edge $(u,v)$ there is another, longer path which connects nodes $u$ and $v$, the edge $(u,v)$ is not essential and can be ``reduced'', that is such an edge is not present in the reduced DAG. This operation amounts to keeping the longest path(s) between every pair of nodes and thus the reduced DAG is a pruned version of a DAG that preserves nodes and the poset structure of $\mathcal{O}$~\cite{AGU72}, see \figref{fig:fig_1_panel} a). Moreover, the transitive reduction of a DAG yields a unique reduced DAG, contrary to general directed graphs. A reduced DAG is thus a well-defined sparsified version of the original DAG. TR has been applied to study citation networks, which form DAGs, to characterise the types of citations received by papers~\cite{CGLE15}.

\subsection{Generalised directed cycles}\label{subsec:gen_cycle}

Our definition of a graph $\Gcal$ and its directed counterpart $\Dcal$, means that for every cycle in the undirected graph, we can trace out the same sequence of vertices in the DAG, while retaining the sequence of directions. Since each cycle $\Ccal$ is a proper subgraph of $\Gcal$, we can apply $F_{\textrm{dir}}$ to obtain all possible directed cycle images. We will call these cycles \textit{generalised directed cycles}, a term we prefer to \textit{oriented cycles}~\cite{KM07,LR05} as it shows that they include directed cycles that are directed closed paths and also avoids confusion with the notion of orientation in algebraic topology. As every DAG $\Dcal$ has a unique underlying undirected graph $\Gcal$, this is the link we will exploit to give new insights into the organisation of data with a hierarchy by considering generalised directed cycles and the information they contain about the mesoscopic organisation of DAGs.

\begin{figure}[ht]
    \centering
    \includegraphics[width = \linewidth]{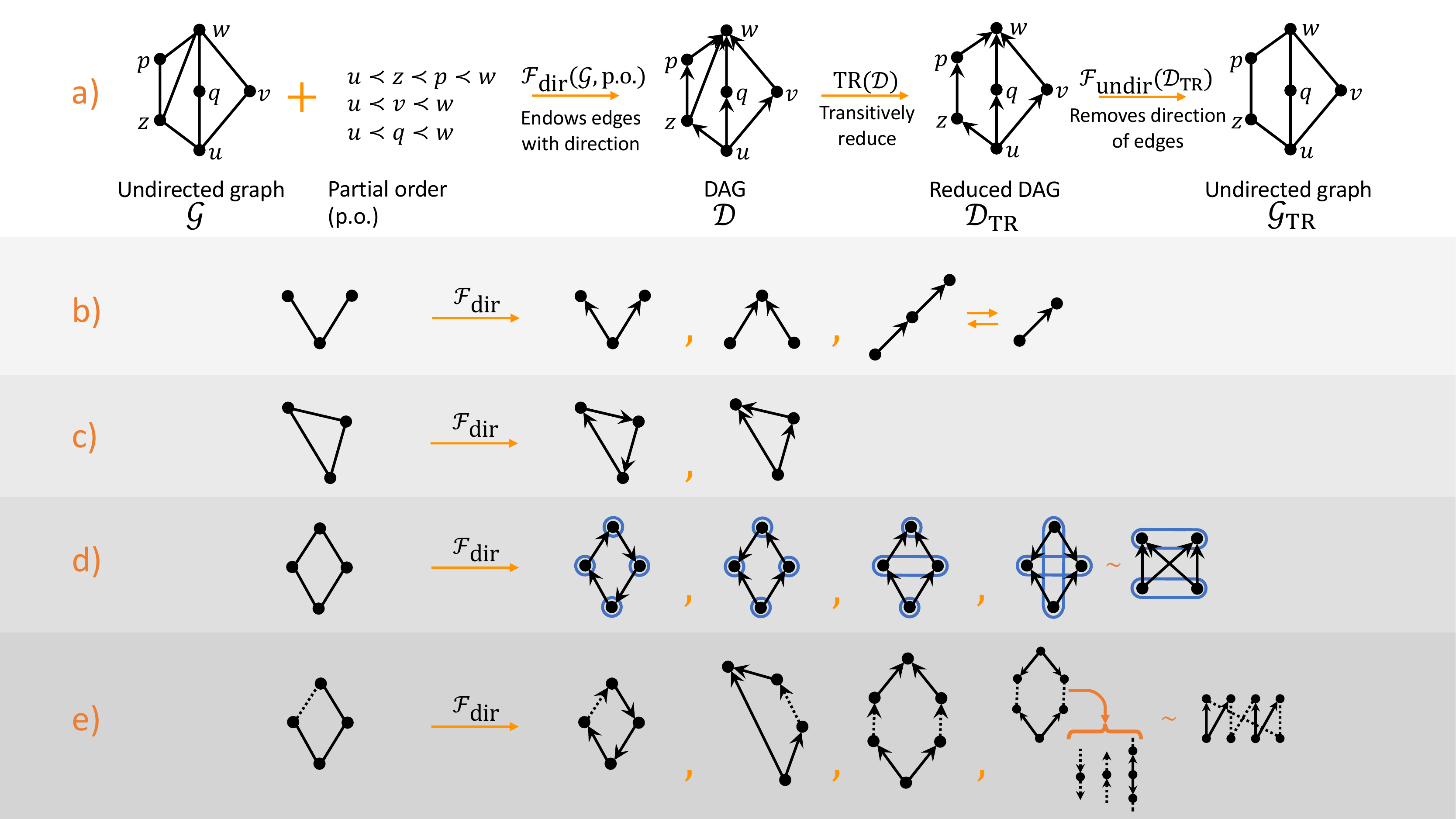}
    \caption{\textbf{a)} 
    Illustration of the central idea of the paper, a directed network is a combination of an underlying graph and metadata associated with directionality. The functions $F_{\mathrm{dir}}$ and $F_{\mathrm{undir}}$ map between the undirected and directed versions of a graph. Transitive reduction can be used to simplify DAG and their corresponding underlying graph. \textbf{b)} the 2 undirected wedge has three possible images: the source wedge, the sink wedge and the neutral wedge. The wedge contraction operation transforms a neutral wedge which has at most one non-neutral node as boundaries into an edge; the edge extension operation transforms an edge into a neutral wedge. \textbf{c)} An undirected triangle can only be mapped onto two classes of generalised directed cycles: feedback and shortcut. \textbf{d)} Size four undirected cycles can be mapped into four generalised cycles classes: feedback, shortcut, diamond and mixer. Antichains shown in blue. Shortcut, diamond and mixer generalised cycles can be present in DAGs, but only diamond and mixers can be present in Transitively Reduced DAGs. \textbf{e)} Undirected cycles of any size can be mapped into the same four generalised cycles classes as in \textbf{d}). Each class is fully characterised by its antichain structure in its contracted form, see \textbf{d}), and the relative positions and number of source and sink nodes pairs, see \tabref{t_4_classes}. Dashed line represent chains of nodes of arbitrary length. For feedback, shortcut and diamonds cycles, the dashed line can only be chains of neutral nodes pointing a set direction. For mixers, dashed lines can be made up of neutral nodes, but may also contain additional source and sink wedges attached to neutral nodes in coherent direction.}
    \label{fig:fig_1_panel}
\end{figure}

In this section, we will show that the directional organisation of generalised cycles of any size is well structured and can be classified into four classes, of which one is not possible in a DAG and two not possible in a Transitively Reduced DAG. Each class has a natural interpretation in terms of information processing. We will build our demonstration by inspecting directed motifs of size 3, then 4 and show that no other category appears for larger cycles.

First, let us note that nodes in a generalised directed cycle can be of three types, based on the way that edges, attached to the node, are oriented in the cycle subgraph: i) a \vdef{source node}: both edges point outwards, ii) a \vdef{sink node}: both edges point inwards, iii) a \vdef{neutral node}: one edge points inwards and the other outwards. These nodes naturally induce wedges (see \figref{fig:fig_1_panel}(b)): i) the \vdef{source wedge} $\left(\begin{minipage}{25pt}\includegraphics[width=25pt]{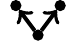}\end{minipage} \right)$, ii) the \vdef{sink wedge} $\left(\begin{minipage}{20pt}\includegraphics[width=20pt]{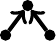}\end{minipage} \right)$, iii) the \vdef{neutral wedge} $\left(\begin{minipage}{20pt}\includegraphics[width=20pt]{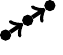}\end{minipage} \right)$. Closing a wedge by adding a third edge produces the smallest possible cycles. Let us close each wedge by adding an edge, giving in total 6 possibilities. It is easy to see by inspection that out of these 6 possibilities, there are in reality only two distinct types: directed cycles, comprising only neutral nodes, and shortcut triangles comprising one node of each type, see \figref{fig:fig_1_panel}(c). Moreover, if we define the \vdef{edge reversal} operation that flips the direction of an edge, we can turn one into the other.

We furthermore define the \vdef{edge-wedge extension} operation that transforms an edge into a neutral wedge by adding a node $\left(\begin{minipage}{30pt}\includegraphics[width=30pt]{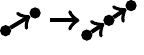}\end{minipage} \right)$. By adding a neutral node to the directed triangle, we obtain the directed/feedback cycle of size 4:
$\left(\begin{minipage}{20pt}\includegraphics[width=20pt]{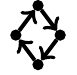}\end{minipage}\right)$. We can now apply the edge reversal operation. Reversing the direction of any one edge produces a shortcut cycle: $\left(\begin{minipage}{20pt}\includegraphics[width=20pt]{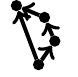}\end{minipage}\right)$. This cycle can also be obtained from the shortcut triangle by extending the arc that does not directly connect the source and sink nodes. Reversing the direction of an edge adjacent to the first one reversed yields a new type of cycle, the diamond: $\left(\begin{minipage}{20pt}\includegraphics[width=20pt]{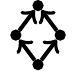}\end{minipage}\right)$. Finally, if we reverse the edge opposite to the first one instead of an adjacent one, we obtain another type of cycle, the mixer: $\left(\begin{minipage}{20pt}\includegraphics[width=20pt]{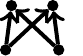}\end{minipage}\right)$. All other edge reversals lead back to a cycle isomorphic to one these. We note that the first two types are the same as the generalised directed triangle type, which do not exist in Transitively Reduced DAGs, and two are new and exist in Transitively Reduced DAGs. We will show below that no new class appear as the size of the cycle increases.

First let us characterise these four classes of generalised cycles, and then show that no large cycles can create a structure that falls outside this classification. We characterise each class with two properties, the first is the number of source-sink node pairs and the second reflects the hierarchical structure of the cycle. The way we constructed each class directly tells us the number of source-sink nodes pairs: 0 for the directed cycles, 2 for the mixer and 1 both for the shortcut and diamond cycles. While the shortcut and diamonds cycles both have one pair, they differ by the connection of the nodes in the pair: direct in the shortcut cycle, making it transitively reducible, and indirect in the diamond. This differentiation is clear when considering the second property: antichains. To formally define an antichain, we first note that reachability relationship in any graph gives rise to a preorder, where $u\prec v$ in the preorder if and only if there is a path from $u$ to $v$ in the graph. Conversely, every preorder is the reachability relationship of a graph. However, many different graphs may have the same reachability preorder as each other. For example, reachability in an undirected graph gives rise to an equivalence relation, as there exists a path between all nodes in a (strongly connected) graph. It is easy to see that many topologically different undirected graphs may have the same equivalence relation, based on the reachability criterion. In the same way, reachability of directed acyclic graphs gives rise to partially ordered sets. Where a relation $u\prec v$ (which indicates that a directed path from $u$ to $v$ exists) implies that $v\prec u$ cannot occur (indicating that a directed path from $v$ to $u$ does not exist).

An antichain $\Acal$ is as a mesoscopic structure that can be defined on an ordered set (and therefore on an ordered set that describes the reachability in a graph) and is a subset of elements of an order in which all elements are incomparable.

 Therefore, reflecting upon the order imposed by the reachability criterion, an antichain is a subset of nodes in a graph, such that none of the nodes are pairwise connected with edges or paths:
\beq
\Acal = \{u,v \in \Vcal|\forall u,v\in \Acal: d(u,v)=\infty ,d(v,u)=\infty  \},
\eeq
with $d(u,v)$ a length of the shortest path from $u$ to $v$.

We highlight two types of antichains: i) a maximal antichain is not a proper subset of any other antichain, ii) a unitary antichain is formed of a single node. The generalised cycles of length four can thus be characterised by their composition in terms of antichains, see \figref{fig:fig_1_panel}(d): directed and shortcut cycles are comprised of four unitary antichains, diamond cycles of two unitary and one non-unitary antichains and the mixer of two non-unitary antichains. The characterisation of the four classes is summarised in \tabref{t_4_classes}.

Before showing that no other class of generalised cycle possessing these properties exist, let us give an interpretation to each class in terms of information processing that justifies choice in properties. Directed cycles can be seen as feedback/reinforcement of information loop. Shortcut cycles are transitively reducible, reflecting, firstly, redundancy of the transitive edge in the context of message modification. Since no nodes exist on the transitive edge, no new information is created and there is no information mixing at the sink wedge of such cycle and therefore removal of this edge leads to no loss of topological information, nor no loss of information processing. Notably, transitive edges also reflect upon network's resilience, as they ensure that when longer paths are corrupted, some information is retained through transitive edges. Diamond cycles encode both resilience, as information set from one source has two paths to reach its destination, and diversity of information processing as the intermediary nodes might modify the information they receive. Finally mixers are providing information from two independent sources to each sink.

All that remains for us to do is to show that these two properties are enough to uniquely classify generalised directed cycles of any size. To do so, we will show that any cycle of any size i) can be generated by preserving the properties of the cycles of size 4 and ii) that any cycle of any length can be contracted into one of the four classes. From the edge direction manipulation in the directed cycles of size 4, we observe two key facts: first, reversing the direction of an edge bounded by two neutral nodes creates a pair of source/sink nodes and therefore reversing the direction of an edge bounded by a pair of source/sink nodes annihilate that a pair. Second, reversing the direction of an edge bounded by a source or sink node and a neutral node moves the source/sink node along the cycle. These imply that the edge-wedge extension operation preserves the number of source-sink pairs.

The edge-wedge extension operation can thus be used to grow these cycles types to an arbitrary size while conserving the number of source-sink pairs. We thus only need to show that no other source-sink node arrangements not covered in \tabref{t_4_classes} can appear in larger cycles. In a cycle of length four, it is not possible to have more than two source-sink pairs, but growing a mixer of size 4 twice with the edge extension operation can provide one edge with two neutral boundary nodes, which can be reversed to create a new source-sink nodes pair. In general, a cycle of length $L$ can have at most $L/2$ pairs of source-sink nodes. However, we show that any generalised cycle with more than two pairs of source-sink nodes are mixers as they can be contracted into two non-unitary antichains.

We define the \vdef{edge-wedge contraction} operation that reduces a neutral wedge into an edge $\left(\begin{minipage}{30pt}\includegraphics[width=30pt]{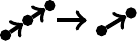}\end{minipage} \right)$ by removing the neutral node if at most one of the wedge boundary node is a source or a sink, ensuring no source nor sink nodes disappear in the contraction process, thus preserving the number of source-sink pairs as well as their fundamental relative organisation, i.e.\ a diamond cannot be turned into a shortcut. Moreover, it is clear that iterating the contraction process on any cycle with more than two pairs of source-sink nodes leads to a contracted form made of two non-unitary antichains, see \figref{fig:fig_1_panel}(e). We consider mixers of any (contracted) size to fall into the same class as they perform the same function as the size four mixer that is mixing into one sink node information coming from two independent ancestors. Thus any generalised cycle can be contracted into a form that falls into one of the four classes we defined.

The characterisation of the four generalised directed cycle classes we define and their interpretation are summarised in \tabref{t_4_classes}. Directed cycles do not exist in DAGs by definition, and shortcut cycles are killed by Transitive Reduction, leaving only diamonds and mixer cycles in Transitively Reduced DAGs, emphasising these two classes are key organising structures in DAGs.

We conclude this section by noting that the definition presented here generalise similar definitions introduced in the context of network motifs, i.e.\ cycles of a given length. The motifs introduced in~\cite{MSIKCA02} can be translated into our nomenclature: a ``bi-parallel'' motif is a diamond, a ``bi-fan'' is equivalent to the simplest contracted mixer of length 4, ``feed-forward'' loop is a triangle/shortcut, and a feed-back motif is a feed-back cycle. We emphasise that there is no notion of path-wedge contraction in motifs, as they are of fixed size: a cycle of a size larger that 4 would be a different motif, whereas we see cycles of all sizes as belonging to the same class, if, after path-wedge contraction, they share the same properties. Such motifs were also studied in DAGs in~\cite{C13} and~\cite{WH09}, where authors emphasised importance of strict triangles in acyclic graphs.

\begin{table}
\centering
\caption{The directed image of cycle of any size falls into one of four classes after the edge-wedge contraction has been applied: directed, shortcut, diamond, and mixer cycles. The classes are distinguished by the number of source and sink pairs, as well as their antichain characteristics that carry information about the hierarchy in the network.}\label{t_4_classes}
\begin{tabular}{c|ccc}
\hline\hline
Class & Number of source/sink pairs & Antichains & Interpretation\\
\hline
Directed \begin{minipage}{24pt}\includegraphics[width=24pt]{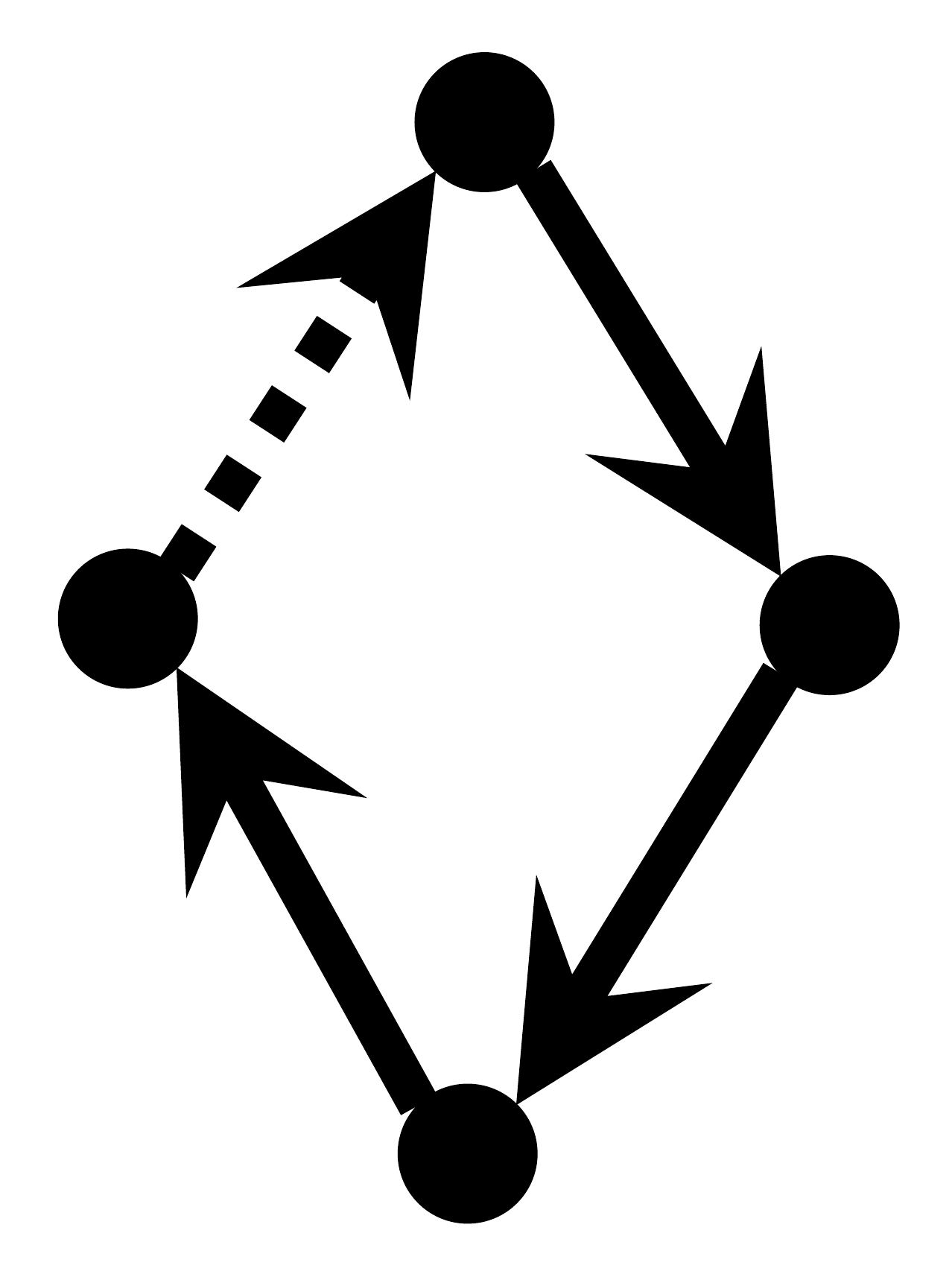}\end{minipage} & 0 & 4 unitary & Feedback loop\\
Shortcut \begin{minipage}{24pt}\includegraphics[width=24pt]{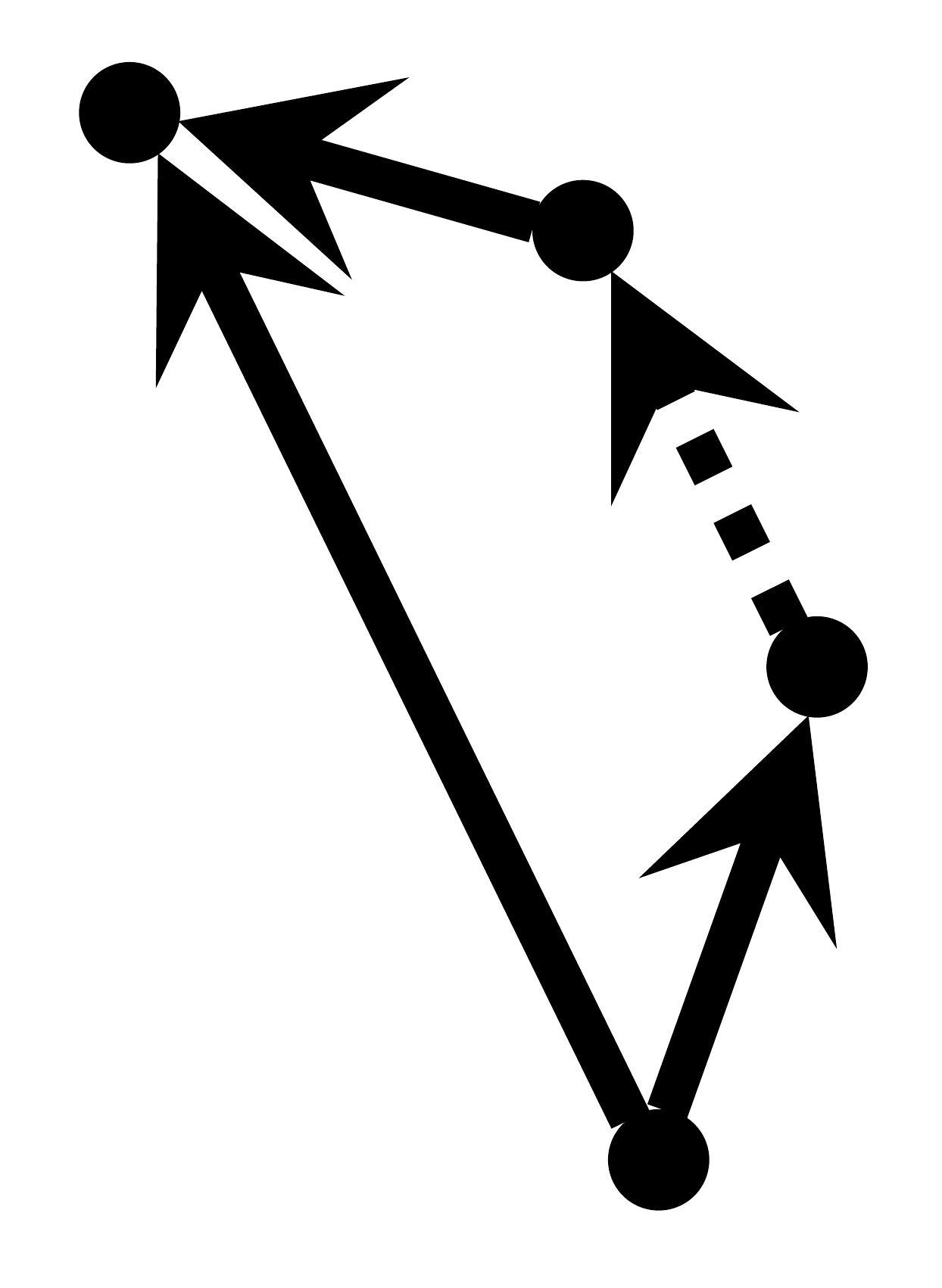}\end{minipage}& 1 (connected) & 4 unitary & Transitively reducible \\
Diamond \begin{minipage}{24pt}\includegraphics[width=24pt]{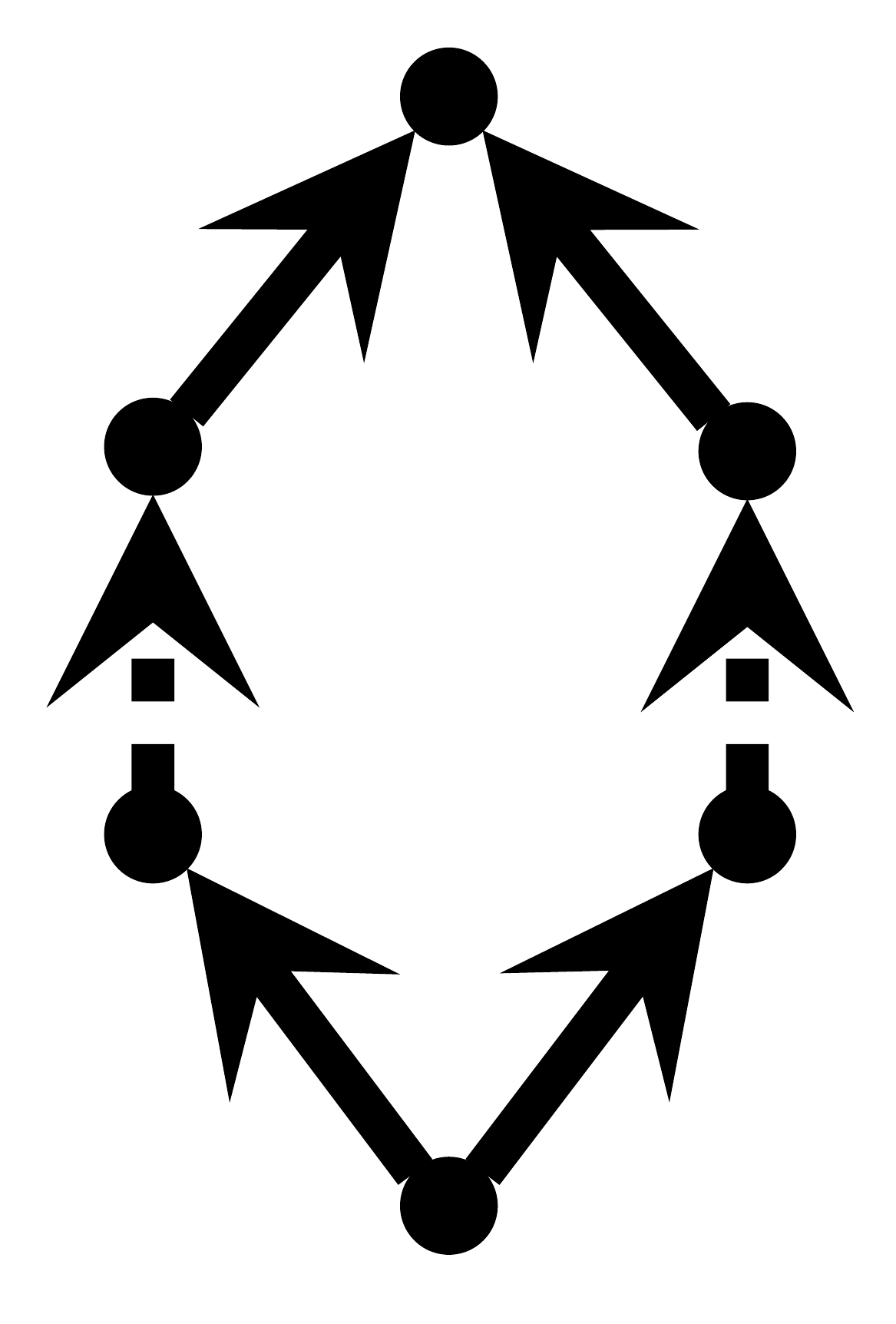}\end{minipage}& 1 (disconnected) & 2 unitary, 1 non-unitary & Resilience of information transfer\\
Mixer \begin{minipage}{24pt}\includegraphics[width=24pt]{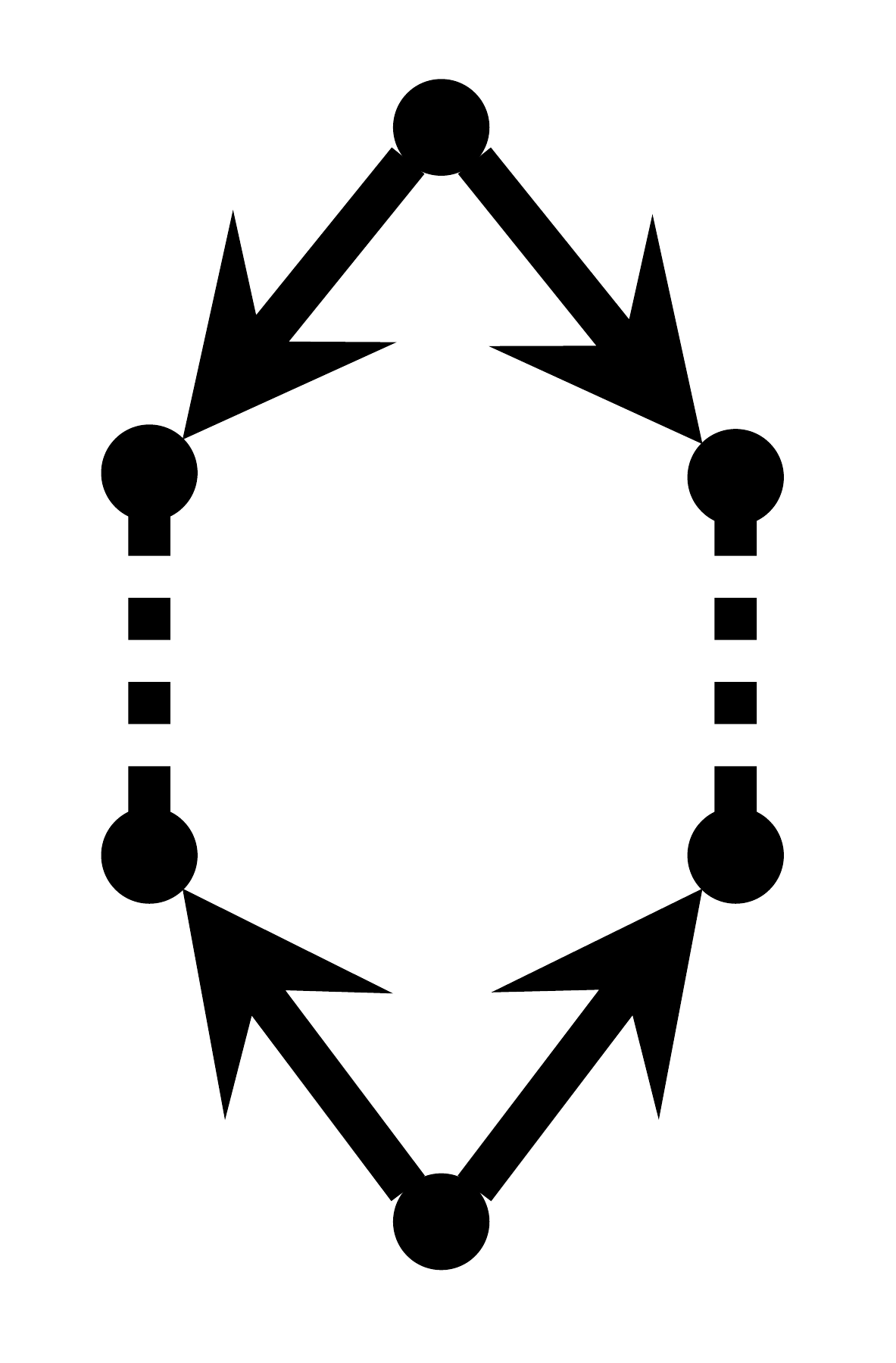}\end{minipage}& 2 or more &  2 non-unitary & Mixing of information
\end{tabular}
\end{table}

\subsection{Minimum Cycle Bases in undirected graphs}\label{sec:algorithm}
The cycle space of an undirected graph is the vector space of the set of its Eulerian cycles endowed with the symmetric difference, or equivalently with the element-wise addition over $\mathfrak{F}_2$ when representing cycles with an edge-cycle incidence matrix. The result of this operation is another subgraph that consists of edges that appear an odd number of times in the cycles taken as arguments, see \cite{D12} for more details. It is well-known that the number of independent cycles of an undirected graph is the rank of the cycle-edge incidence matrix (\eqref{eq:cycle_representation}), the circuit rank~\cite{H87}, and is equal to:
\begin{equation}\label{eq:number_cycles}
    d = E-N+n_c,
\end{equation}
where $n_c$ is the number of connected components in the graph.

Except in cases where no pairs of cycles overlap, i.e.\ the cycles are independent, the cycle basis is in general not unique. A strategy to reduce the number of possible cycle bases is to impose constraints on its elements. A simple constraint is to require minimality of the representation of the cycle basis. A \vdef{Minimum Cycle Basis} (MCB) is defined as a cycle basis in which the total length of the cycles in the cycle basis is minimal. While the minimality criterion reduces the possible choice of cycles in the MCB, there is no guarantee it is unique.

Many other types of cycle bases exist, an extensive study of the hierarchy of such bases is given in~\cite{KLMMRUZ09}. In particular it is possible to define Minimal Directed Cycle Bases, see \cite{HKM08}. We nevertheless decided to consider the MCB of undirected cycles underlying DAGs and then characterise them with directionality. Our decision was motivated on the one hand by our general approach to consider a DAGs as a combination of an undirected graph and meta-data encoding directionality, and on the other hand by computational considerations: Directed MCB algorithms typically run in $\mathcal{O}(E^3 N)$, against $\mathcal{O}(E^2N)$ for MCB, and codes to compute MCB are readily available in standard graph packages, e.g. \href{https://networkx.org/documentation/stable/reference/algorithms/generated/networkx.algorithms.cycles.minimum_cycle_basis.html}{networkx}, which is not the case for Directed MCB.

We now review the De Pina's algorithm to obtain a fundamental MCB. De Pina's algorithm finds \vdef{fundamental cycle bases}. Such cycle bases are computed from a minimum spanning tree of a graph~\cite{LR07}. To obtain a fundamental MCB, a spanning tree of a graph $\Tcal$ is considered along with the set of edges $\Ecal_B$ of $\Gcal$ that are not present in $\Tcal$. A fundamental cycle is defined as a cycle that consists of a path in $\Tcal$ whose endpoints are connected by one $e\in\Ecal_B$. A cycle basis $\Cfrak$ is fundamental if the following holds~\cite{S79}:
\beq
\Ccal_\alpha \not\subset \Ccal_\beta \quad \forall \quad \Ccal_\alpha,\Ccal_\beta\in\Cfrak
\eeq

\paragraph{De Pina's algorithm} (see \cite{KMMP07,P95}) begins by initialising a set of support vectors $S_{i} = \{e_i\}$, one for each  edge in $\Ecal_B$, i.e.\ there is a non-zero entry $e_\alpha$, otherwise the vector is zero. Two procedures are then iterated until a set of linearly independent cycles of size $d$ is obtained. First, a vector $C_\alpha$ representing a cycle $\Ccal_\alpha$ is computed. If $\langle C_\alpha,S_i\rangle  =1$, then $\Ccal_\alpha$ is added to the cycle basis.
To find $C_\alpha$, a new graph $\Gcal_\alpha$ is constructed. It contains two copies of each node, $v^+, v^-$ for all $v\in\Vcal$. The resulting network can be thought of as a multilayer network, where nodes are placed in three layers: nodes with positive sign ($v^+$), nodes with negative sign ($v^-$), and nodes that are neutral ($v$). Then for each $S_j$, $j>i$, $S_j \rightarrow S_j + S_i$ if $\langle C_\alpha,S_j\rangle=1$. This step ensures that the set of $\{S_j,..., S_{d}\}$ is orthogonal to $\{C_1,...,C_\alpha\}$, and the last $C_\alpha$ is orthogonal to the set $\{C_1,...C_{\alpha-1}\}$.

To add edges to $\Ecal_\alpha$, the edgeset of $\Ccal_\alpha$, for each edge $e_i = (u, v) \in \Ecal$ we check the following: if $S_i\neq1$, then add edges $(u^+ , v^+ )$ and $(u^-, v^-)$ to the edge set of $\Gcal_\alpha$. If $S_i=1$, then add edges $(u^+, v^- )$ and $(u^- , v^+)$ to the edge set of $\Gcal_\alpha$. Within each layer, we have edges $e_i$ which have $S_i=0$. Between the layers we have edges $e_j$ which have $S_j=1$. Any $v^+$ to $v^-$ path in $\Gcal_\alpha$ corresponds to a cycle in $\Gcal$ once positive/negative edges in $\Gcal_\alpha$ are matched to their neutral counterparts in $\Gcal$. If an edge $e \in \Gcal$ occurs multiple times we include it if the number of occurrences of $e$ modulo 2 is 1. In order to find the shortest cycle, we need to find the shortest path from $v^+$ to $v^-$ for all $v \in \Vcal$. The shortest cycle $C_\alpha$ which is orthogonal to the current basis is then appended to it. This concludes an iteration of De Pina's algorithm.

\subsection{Transitive Reduction of DAGs and Minimum Cycle Basis}\label{sec:TR_MCB_HB}
In this paper, we are particularly interested in the properties of the MCB of Transitively Reduced DAGs to characterise DAGs. MCB are not unique, but TR makes them defined well-enough that their characteristics are robust descriptors of reduced DAGs.

\begin{figure}[!ht]
    \centering
    \includegraphics[width=\linewidth]{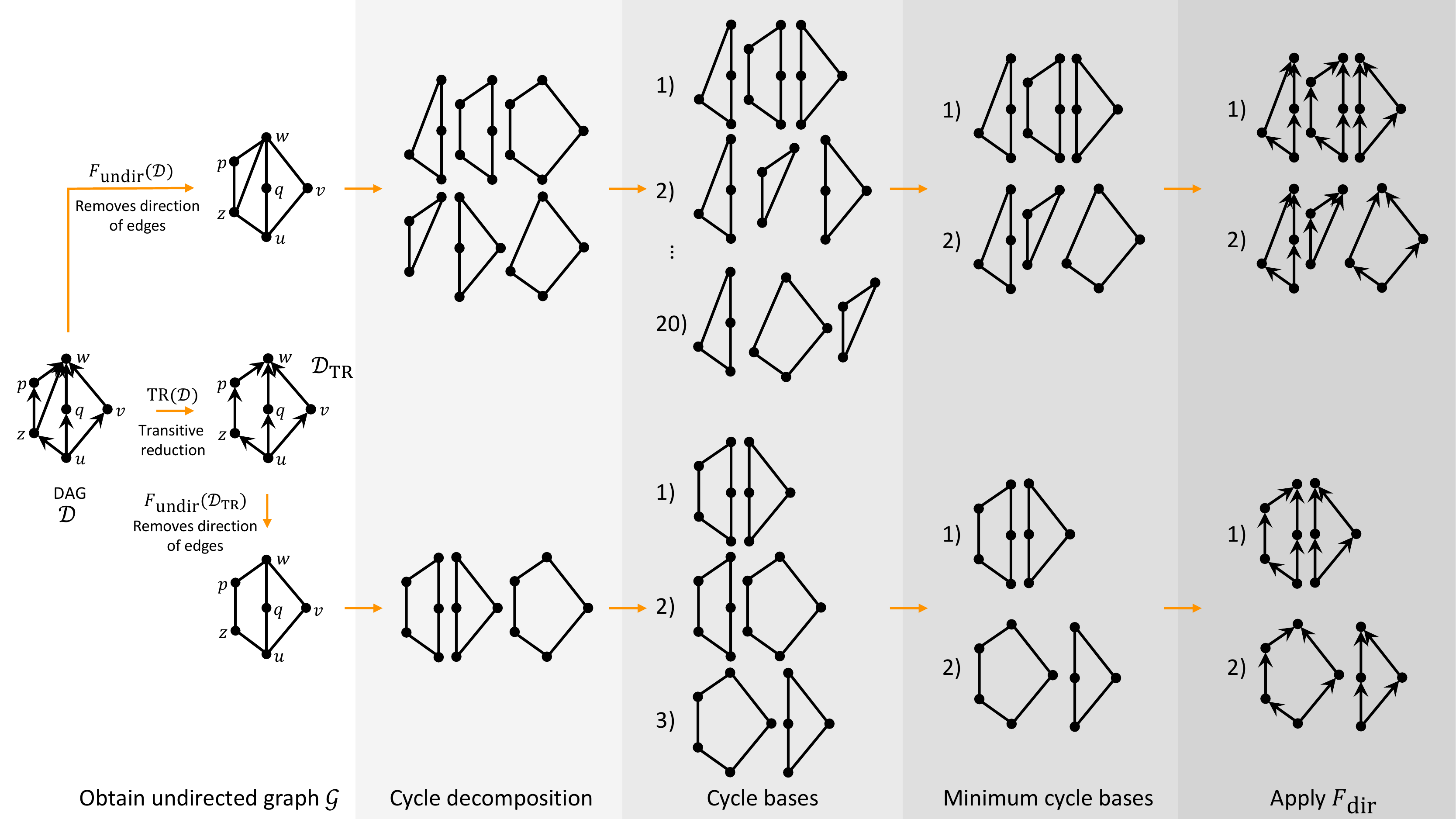}
    \caption{This figures compares the two routes to obtain the directed image of a Minimal Cycle Basis (far right column) of a DAG (far left column). In this paper, we follow the bottom route, that Transitively Reduce the DAGs, with the effect of limiting the directed images of cycles to mixers and diamonds. In large networks, it also has the effect of significantly reducing the dimension of the MCB and stabilising it, making its properties measured by the metrics introduced in section~\ref{sec:cycle_metrics} appropriate descriptors to characterise the cyclic structure of DAGs.}
    \label{fig:fig_2_panel}
\end{figure}

The procedure to obtain the directed images of the cycles of the MCB of a reduced DAG is illustrated in \figref{fig:fig_2_panel}. We apply $F_{\textrm{undir}}$ to the transitively Reduced DAG $\mathcal{D}_{\textrm{TR}}$ to obtain the underlying undirected graph $\Gcal$. We then compute an MCB of that graph and apply ${F}_{\textrm{dir}}$ to each cycle of the MCB to determine its properties.

Transitive Reduction limits the types of generalised cycles that can be present in the cycle basis to diamond and mixers, as all shortcut cycles are removed as they represent transitive closure. This brings a natural interpretation of TR in terms of information processing on a DAG: transitive reduction removes all structures that do not modify information nor are essential for information transfer in the network. TR therefore has a non-trivial effect on the dimension of cycle basis of the undirected graph underlying $\Dcal$ and reduces its dimension following \eqref{eq:number_cycles}. We also note that although the MCB of a Transitively Reduced DAG will in general be composed of both diamonds and mixers, it is possible to modify an MCB finding algorithm, Horton's algorithm~\cite{H87}, to obtain an MCB which is composed of only diamonds: a Minimal Diamond Basis. The details of this variation of Horton's algorithm is presented in \appref{app:mdb}.

\begin{figure}[!ht]
    \centering
    \includegraphics[width=0.3\linewidth]{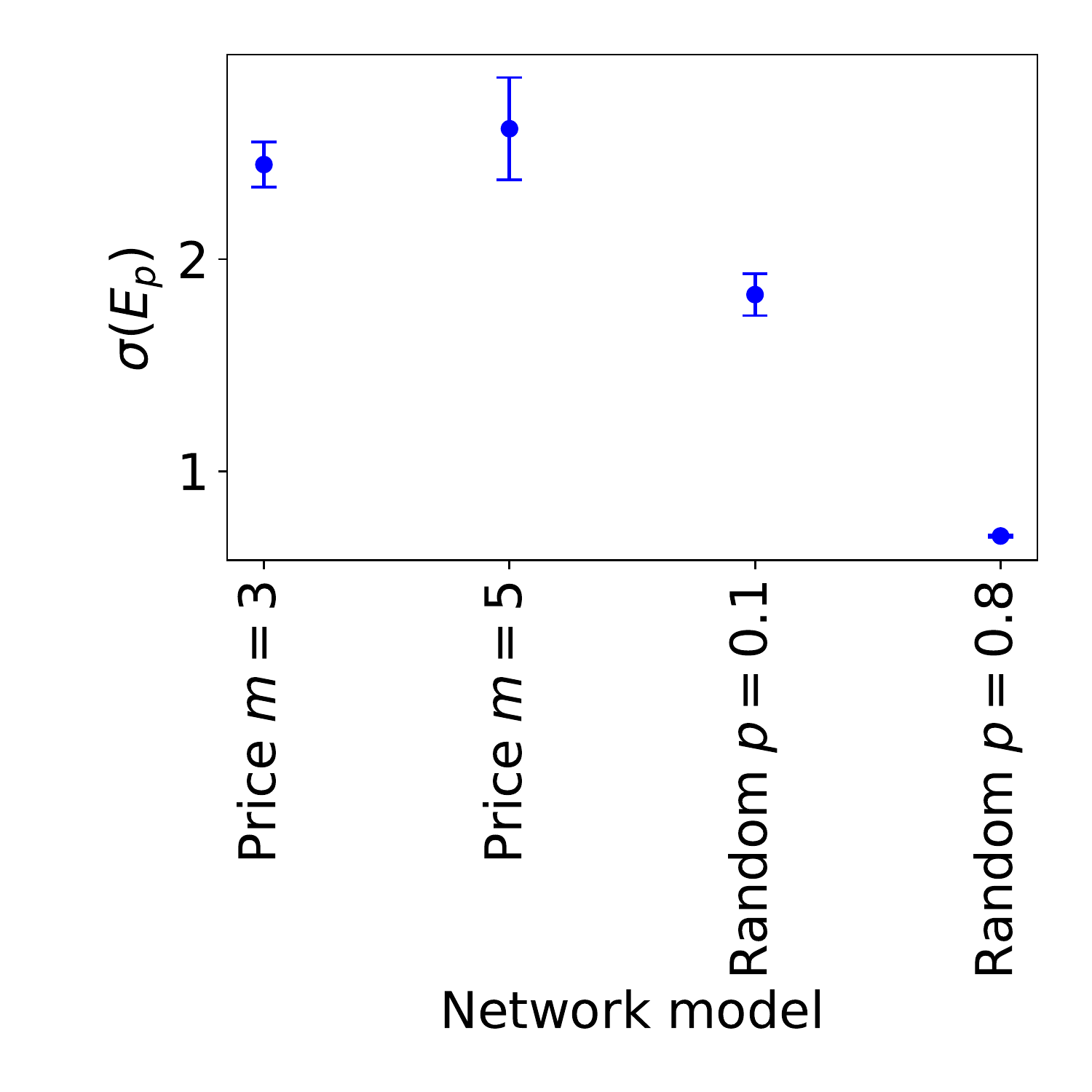}
    \includegraphics[width=0.3\linewidth]{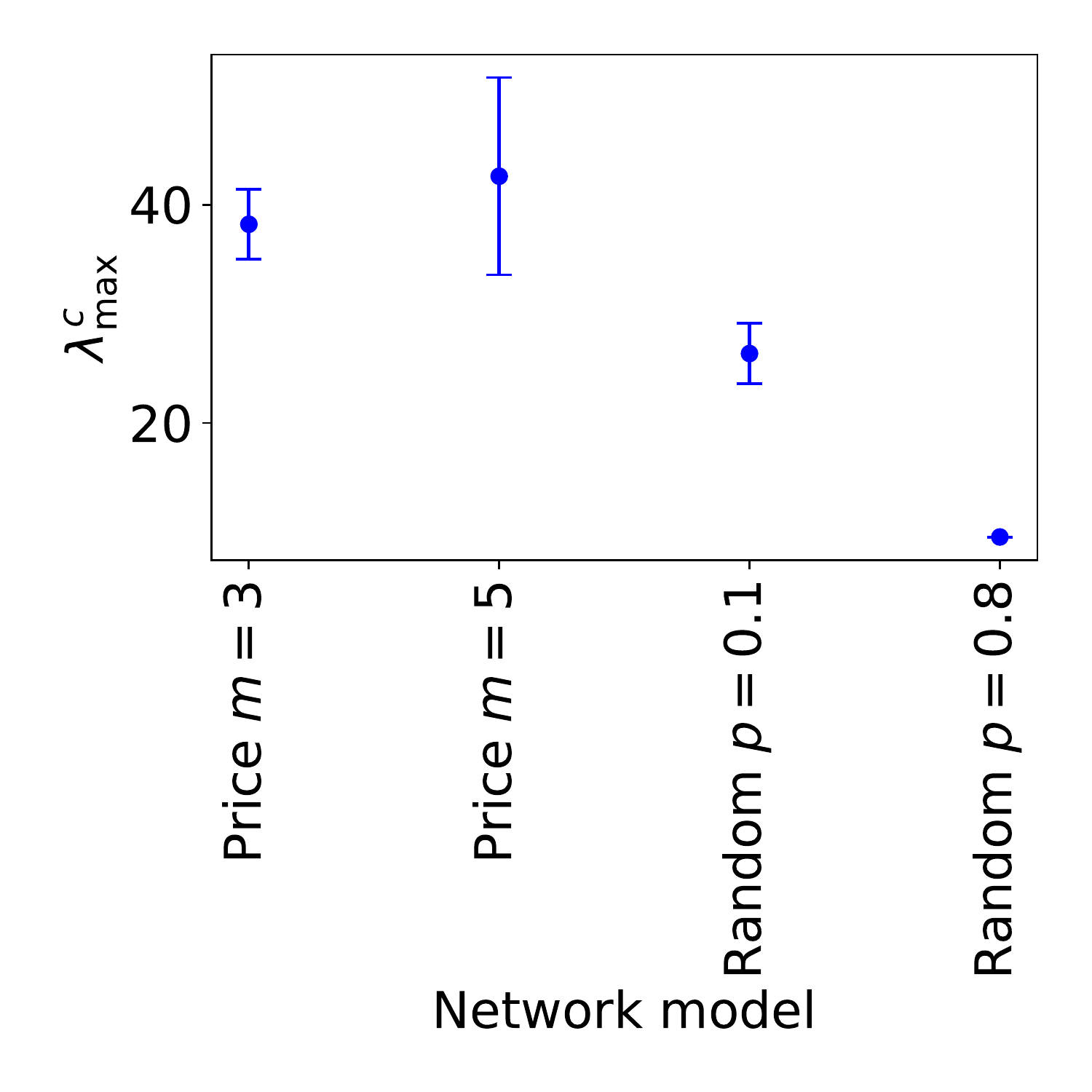}   \\
    \includegraphics[width=0.3\linewidth]{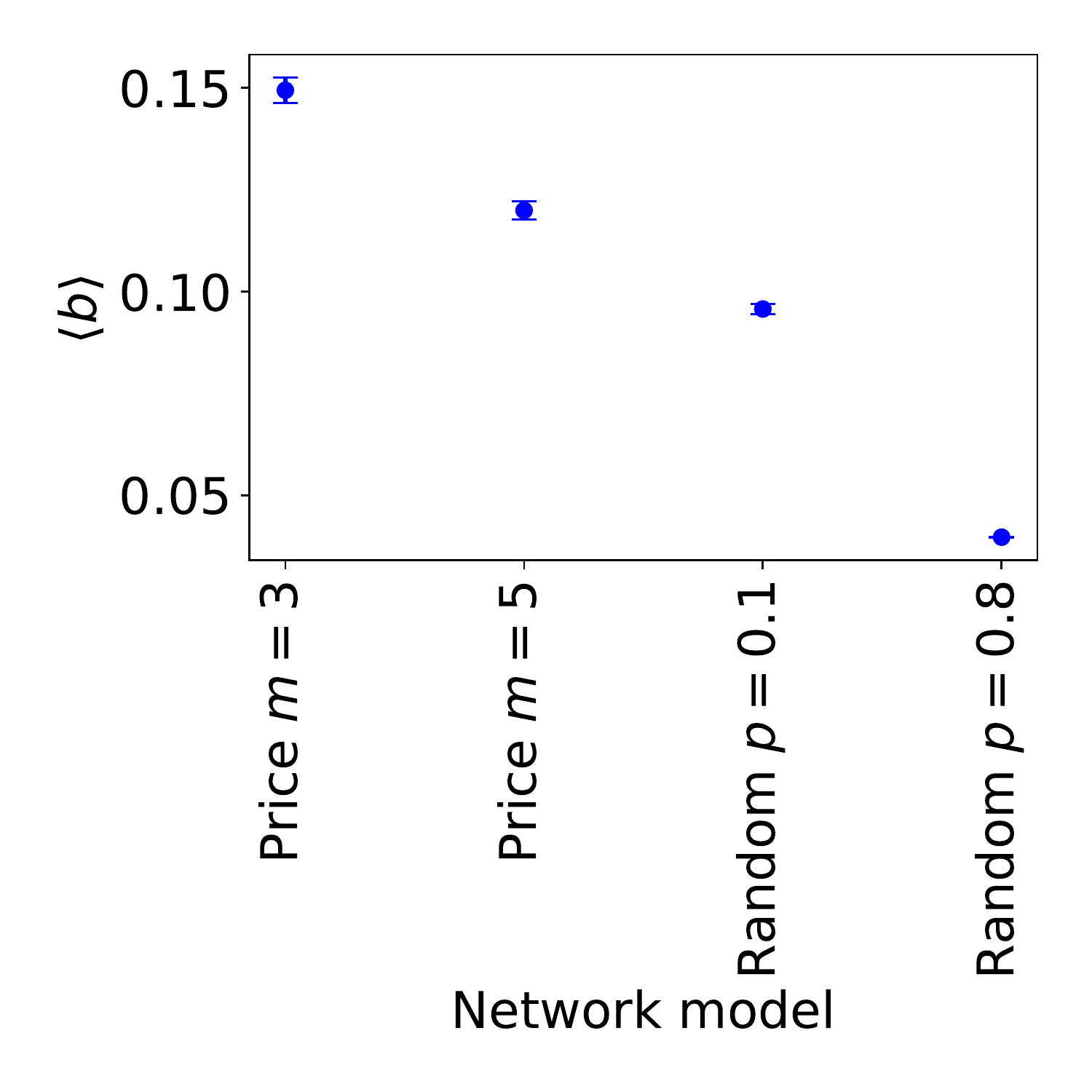}
    \includegraphics[width=0.3\linewidth]{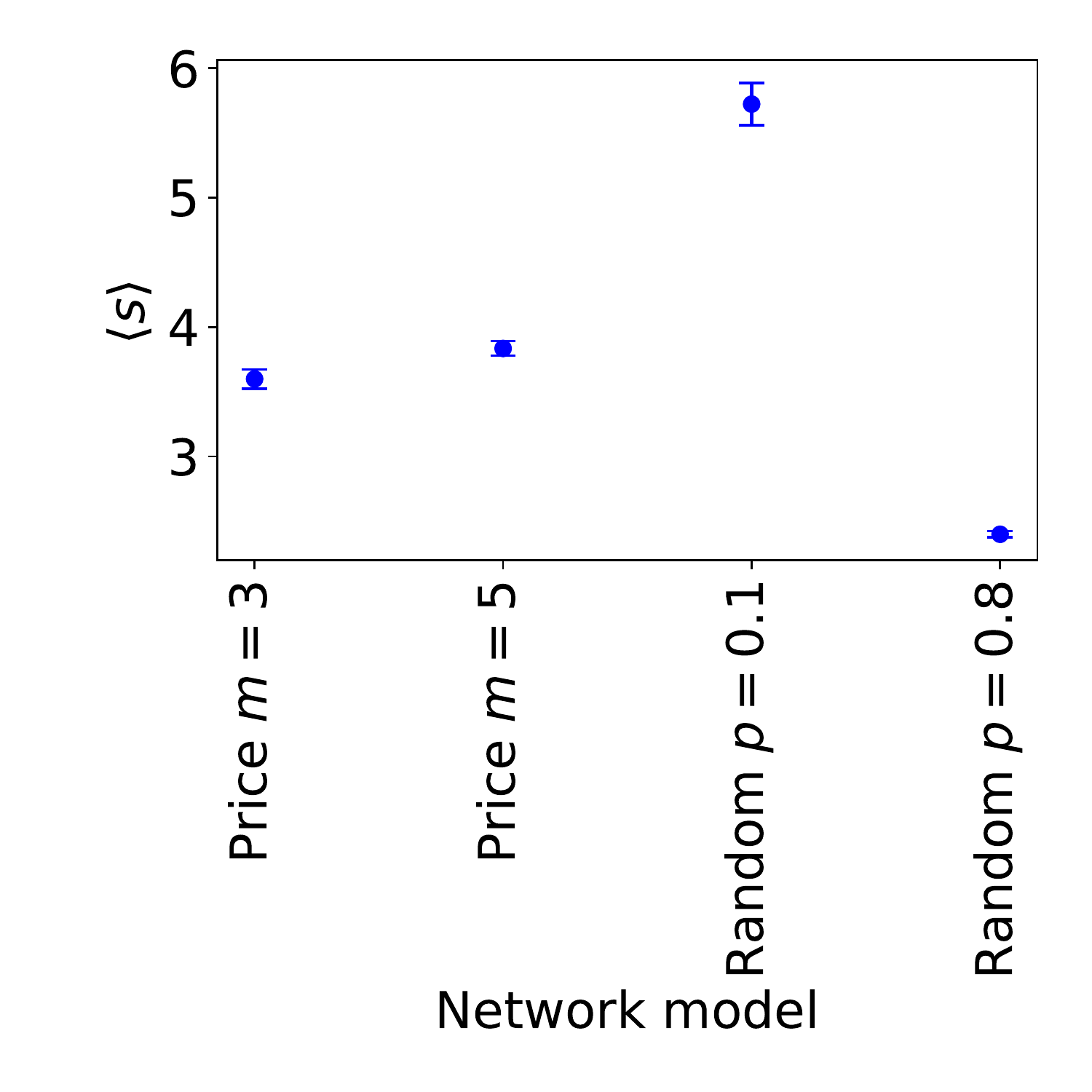}
    \caption{Statistics for the mean edge participation, the leading eigenvalue of $\mathbf{M}$, the average cycle balance and the average cycle stretch, obtained from running De Pina's algorithm $n=10$ times on four types of networks. The abscissa indicates a network model, the ordinate -- obtained values of several statistics of MCBs. The four different models considered are: random DAGs of \secref{sub_sec:random} with $p=0.1$ and $p=0.8$, and the Price model with $m=3$ as well as $m=5$. All networks have $N=300$ nodes.}
    \label{fig:mcb_variation}
\end{figure}

MCB algorithms are sequential and stochastic in nature, and as MCB are in general not unique, the MCB obtained will be run-dependent. Another effect of TR is the reduction the choices for the representative of cycles in the MCB as it greatly reduces the overlaps between cycles, and in some cases, the MCB is even unique, see \figref{fig:fig_2_panel} for the difference between the MCBs obtained when transitive reduction is employed before applying $F_{\textrm{undir}}$ and when it is not. To ensure that the MCBs found in reduced DAGs are stable and well-defined representatives of the underlying cycle space and therefore that their properties are true characterisations of the systems under study, we studied the statistics of the MCB statistics obtained for two Transitively Reduced DAGs model over $n$ runs. We considered four different types of networks of size $N=300$: two Erd\"os-R\'enyi DAGs with $p=0.1$ and $p=0.8$ and two Price models with $m=3$ and $m=5$, see \secref{sec:model_results} for precise definitions of these models. We used four different Cycle Bases statistics, see the next section for precise definitions: the mean edge participation, the value of the leading eigenvalue of $\mathbf{M}$, the average cycle balance, and the average cycle stretch. \Figref{fig:mcb_variation} shows the mean values and standard deviations of those statistics for each type of network considered. The results vary little for each type of DAG.

We now have a stable and consistent way to represent the cyclic structure underlying directed graphs in general and in particular Transitively Reduced DAGs. In the next section, we introduce a series of metrics that can be used to characterise the topological and geometrical properties MCBs.

\subsection{Cycle metrics}\label{sec:cycle_metrics}

Defining and finding cycles in reduced DAGs provides a representation of mesoscopic and higher-order structures. We can now explicitly use the meta data associated with DAGs to characterise cycle bases, and by extension characterise reduced DAGs themselves. We introduce simple, intuitive and yet insightful metrics to characterise cycle bases at three levels: cycle, cycle interactions and cycles embedding in the reduced DAG. Some of these measures can be computed for any cycle basis, while others are specific to DAGs. A summary of the metrics presented in this section can be found in \tabref{t_single_cycle_measures}.

\paragraph{Cycle-edge incidence matrix}
A fundamental object is the undirected edge-cycle incidence vector $c$ for a cycle $\Ccal$

\begin{equation}
    c_i = \begin{cases}
    1 \mbox{ if } e_i \in \Ccal, \\
    0 \mbox{ otherwise.}    \end{cases}\label{eq:cycle_representation}
\end{equation}

The information about all cycles can thus be represented by an $E\times d$ edge-cycle incidence matrix $\mathbf{C}$, with $E$ the number of edges and $d$ the number of cycles, see \eqref{eq:number_cycles}, that has the edge-cycle incidence vectors as its row. We can then define the matrix $\mathbf{M}$:
\bea
 \mathbf{M}
 =
 \mathbf{C}\, \mathbf{C}^{ \tran}
 \label{e:Mdef}
\eea
with entries $M_{\alpha\beta} = |\{e \in \Ecal_{\alpha}\}\cap \{e \in \Ecal_\beta\}|$, where $\Ecal_{\alpha}$ is the set of edges of the cycle $\Ccal_\alpha$. Its off-diagonal entries measure the overlap between two cycles and its diagonal elements the size of a cycle. This matrix can be interpreted in at least two ways: a cycle covariance matrix or a weighted cycle adjacency matrix with self loops. The spectral properties of the Laplacian matrix $\mathbf{L}^C=M-\mathrm{diag}(\mathbf{M})$ are also relevant for characterising cycle interactions and the organisation of cycles: the dimension of the nullspace of $\mathbf{L}^C$, $\mathrm{null}(\mathbf{L}^C)$ is equal to the number of cycle connected components. For a basis where there is no overlap between cycles, this is equal to the dimension of the basis.

\paragraph{Characteristics of isolated cycles} Let us introduce some basic cycle statistics. The only purely topological statistic we are going to introduce is the size $S_\alpha $ of a cycle $\Ccal_\alpha $, which is equal to the number of edges, equivalently of nodes, that it comprises:
\beq
 S_\alpha
 = M_{\alpha\alpha}
 = |\Ccal_\alpha|
 = \sum_{\beta} \mathbf{C}_{\alpha,\beta}
 = \sum_{\alpha} c_{\alpha}.
\eeq
In a Minimum Cycle Basis, the distribution of $S_i$, as well as its mean are minimal by definition.
All other metrics are going to be defined using the meta data associated with the DAGs.

The metadata, allows to study the paths (rather than walks) that constitute cycles. The length of a path $l$ is the number of edges between a source and a sink node. A cycle composed of paths that have equal length can be thought to be maximally ``balanced'', where \vdef{balance} reflects on the variation in the lengths of paths in the cycle. For example the diamond cycles in \figref{fig:fig_1_panel} are maximally balanced. The balance of the paths contained in a cycle $\Ccal_\alpha$, $b_\alpha$, can be defined using the coefficient of variation:
\beq
 b_\alpha
 =
 \frac{\sigma_\alpha(\ell)}{\langle \ell \rangle_\alpha},
 \label{e:bdef}
\eeq
where $\langle \ell \rangle_\alpha$ is the average length of the paths in $C_\alpha$ and $\sigma_\alpha(\ell)$ is their standard deviation.

In a DAG, each node $v$ also has a height $h(v)$, which is defined as the length of the longest path from any source node to $v$~\cite{VE20}. Heights of nodes can be used to localise cycles them with a ``vertical'' coordinate. We define the height of cycle $\Ccal_\alpha$ as the average height of its nodes:
\bea
 h_\alpha
 =
 \frac{1}{|\Ccal_\alpha|}\sum_{v \in \Ccal_\alpha}h(v).
 \label{e:hcycledef}
\eea
A cycle in a reduced DAG would always have $\sigma^2(h)>0$, as they are formed of at least three distinct maximal antichains.

To characterise the relative localisation of a cycle, it is interesting not only to consider its average position in the DAG, but also to capture its spread. We define the stretch of cycle $\Ccal_\alpha$ as the largest difference between the heights of any two nodes in $\Ccal_\alpha$:
\bea
  s_\alpha
  =
  \max(\{h_u|u\in\Ccal_\alpha\})- \min(\{h_u|u\in\Ccal_\alpha\}).
  \label{e:stretchdef}
\eea
If a cycle is treated as a subgraph, the stretch is then simply maximal height of the cycle. The stretch cannot be smaller than one for mixers and two for diamonds and shortcuts. The largest height in a cycle always belongs to one of the sink nodes, whereas the smallest height always to one of the sources.

\begin{table}
\centering
\caption{Measures which characterise an individual cycle (first block), cycle interaction with other cycles (second block), and the network as a whole (third bock) for Transitively Reduced DAGs.}\label{t_single_cycle_measures}
\setlength{\tabcolsep}{2pt}
\begin{tabular}{p{4cm}p{5.6cm}p{5.6cm}}
\hline\hline

    Measure & Notation & Explanation \\
\hline
\textbf{Cycle size} & $S_\alpha  = M_{\alpha\alpha} = |\Ccal_\alpha| = \sum_{\alpha}c_\alpha$ & Size of a cycle. \\
    \textbf{Cycle type} &type$=$\textit{diamond} if $n_{src}=n_{snk}=1$, \textit{mixer} otherwise.& Classifies cycles into three types, mixers, diamonds, and triangles.\\

     \textbf{Balance} &$\frac{\sigma_\alpha(\ell)}{\langle \ell \rangle_\alpha}$ & Variation in lengths of the paths $l$ that make up a cycle. \\

    \textbf{Height} & $(1/|\Ccal_\alpha|)\sum_{u \in \Ccal_\alpha}h_u. $& Average height of the nodes comprising a cycle.\\

    \textbf{Stretch} & $ \max(\{h_u|u\in\Ccal_\alpha\})\qquad\qquad\qquad$
    $\qquad\qquad\qquad - \min(\{h_u|u\in\Ccal_\alpha\})$
    & Maximal height difference of the nodes comprising a cycle.\\
     \hline
    \textbf{Eigenvalues of $\mathbf{M}$} &$\lambda_\alpha^C$ &``Effective'' cycle size: trade-off between the actual cycle size and the sizes of cycles it is adjacent to.\\
     \hline
      \textbf{Statistics of cycles} in MCB &$\langle\cdot\rangle,\sigma(\cdot),\cdot_{\textrm{max}}$, where $\cdot=S,b,s,h$& These statistics describe a characteristic cycle within a network; the maximum of characteristics defines an ``extremal'' cycle. \\
      \textbf{Number of cycle connected components}&$\mathrm{null}(\mathbf{L}^C)$ & Number of connected components of cycles.\\
        \textbf{Largest eigenvalue of $\mathbf{M}$} &$\lambda_{max}^C$& The largest effective cycle size. \\
      \textbf{Edge participation}  &$E_p=\frac{1}{E}\sum_{e\in\Ecal} |\{\Ccal_\alpha\in \Cfrak | e\in \Ecal_\alpha\}|$& Mean edge participation in cycles. If $E_p$ is small, a network is tree-like, whereas if it is large---it is lattice-like.\\
      \textbf{Variation in $E_p$}&$\sigma(E_p)$& Variation in edge participation. If there is large variation $\sigma(E_p)$, it indicates that there are denser and less compact cycle regions.

\end{tabular}
\end{table}

\paragraph{Cycle interactions} Cycles are adjacent if they share one or more edges. $M_{\alpha\beta}$ measures the overlap, i.e.\ the number of common edges, between two cycles $\Ccal_\alpha$ and $\Ccal_\beta$, and the diagonal entry $M_{\alpha\alpha}= S_\alpha$ the size of a cycle, as a cycle shares all edges with itself. The spectral properties and eigenvectors of $\mathbf{M}$ thus reveal the relative organisation of cycles in a DAG: the extent of cycle pairwise overlap/interaction. An in-depth interpretation and analysis of such covariance matrix is beyond the scope of this study, but we note that potentially interesting insights about their interconnectivity of cycles can be obtained from the structure of $\mathbf{M}$ and that results relating to the spectral properties of covariance matrices apply. In this study, we will look at one metric from the matrix spectra -- the largest eigenvalue of $\mathbf{M}$, $\lambda_{\textrm{max}}^C$. A large maximal eigenvalue of covariance matrix indicates strong variance direction in the data the matrix represents, meaning that a subspace spanned by few eigenvectors of the matrix can be used to describe the data~\cite{Hastie}.

Another pertinent indirect measure of cycle interactions is the average number of cycles per edge, $E_p$:
\beq
 E_p
 =
 \frac{1}{E}\sum_{e\in\Ecal} \left|\{\Ccal_\alpha\in \Cfrak | e\in \Ecal_\alpha\}\right|.
\eeq

This quantity is directly related to the ``treeness'' of a DAG: $E_p=0$ if and only if there are no cycles in a DAG, in which case the DAG is a tree. This equality echoes back to the title of this paper. Let us qualitatively explain the structure of a DAG for important parameters ranges: for $0<E_p<1$, there are edges in the network which do not participate in any cycle and the DAG is still locally a tree and cycles are clustered in branches, like grapes. When $E_p\geq  1$ branches might still exist, but $\sigma(E_p)$ must be considered as well to determine the statistical treeness of a DAG as, on average, each edge participates in one cycle. When $E_p\sim  2$ and $\sigma(E_p)\sim 0$, we find a regime equivalent to a lattice DAG that we will discuss in \secref{sec:model_results}.

\paragraph{Characterisation of a cycle basis}
The statistics of the metrics defined above for individual cycles naturally describe properties of the global network structure. For instance, the standard deviation of the height of cycles indicates how scattered or lumped together within a network cycles are, larger value indicating that cycles span all heights. On the other hand, a largely stretched-out cycle can have a height close to half of the largest height $h_{\textrm{max}}/2$, and yet this cycle is arguably different from a cycle whose all nodes have heights close to the average height of a cycle --- the latter cycle is much less stretched-out. Combinations of metrics can distinguish such differences between cycles organisation, as we show for two random DAG models in \secref{subsec:comparison}.

Many properties of a cycle basis are also encoded in $\mathbf{M}$. The leading eigenvalue of $\mathbf{M}$ contains information about the effective size of the largest cycle component, or in variance terms captures most of the interconnectedness of cycles. Intuition can be gained by observing that when two cycles share one or more edges their ``corresponding'' eigenvalues shift: one eigenvalue increases, while the other is reduced, the magnitude of the shift corresponding to extent of the overlap. Thus a large eigenvalue of $\mathbf{M}$ can indicate that it represent a small, highly interconnected cycle, or a large independent cycle: the localisation of the corresponding eigenvector deciding which case it is.

\section{Transitively Reduced DAG network models are characterised by different cycle statistics}\label{sec:model_results}

\begin{figure}[ht]
    \centering
    \includegraphics[width=0.2\linewidth]{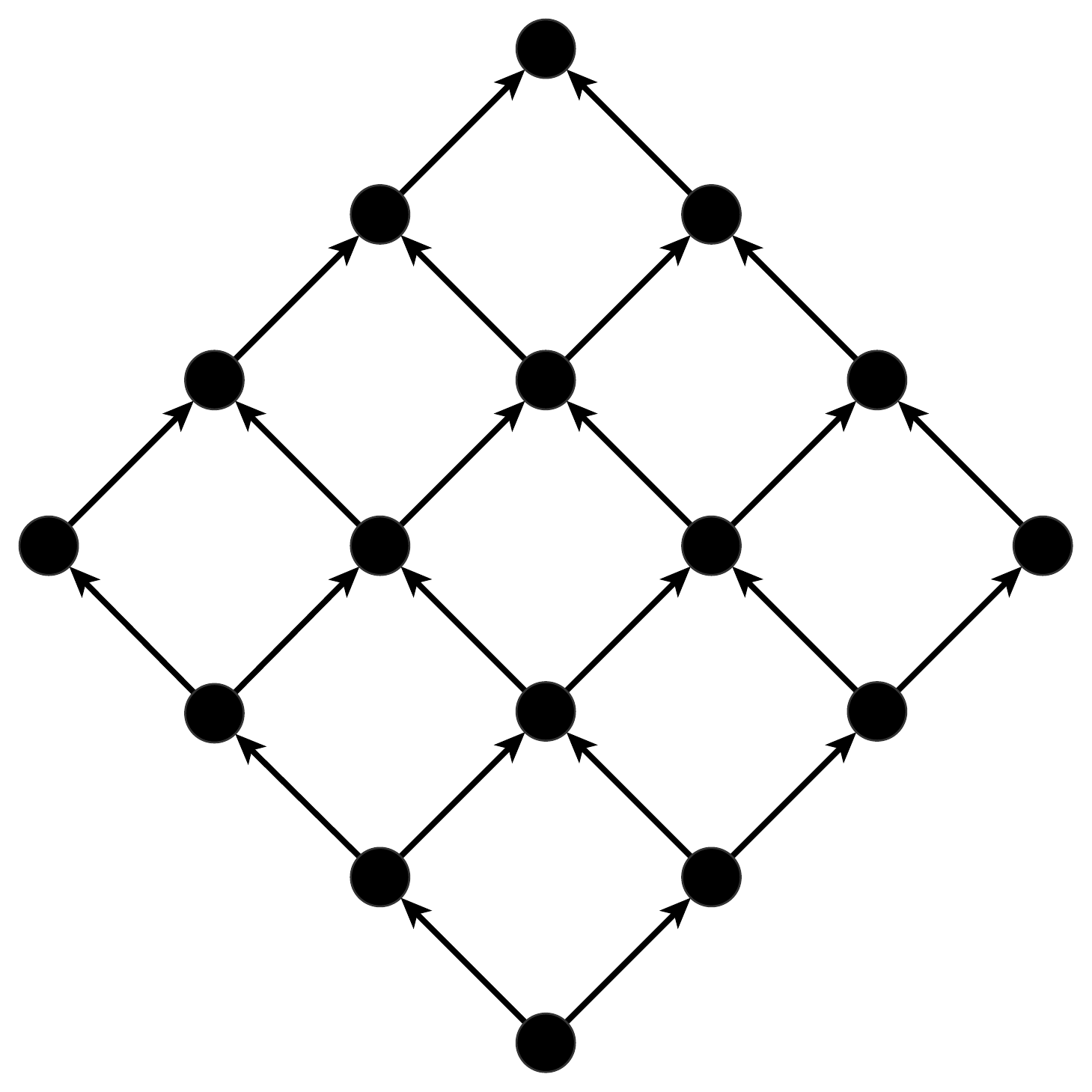}
    \hspace{0.1cm}
    \includegraphics[width=0.2\linewidth]{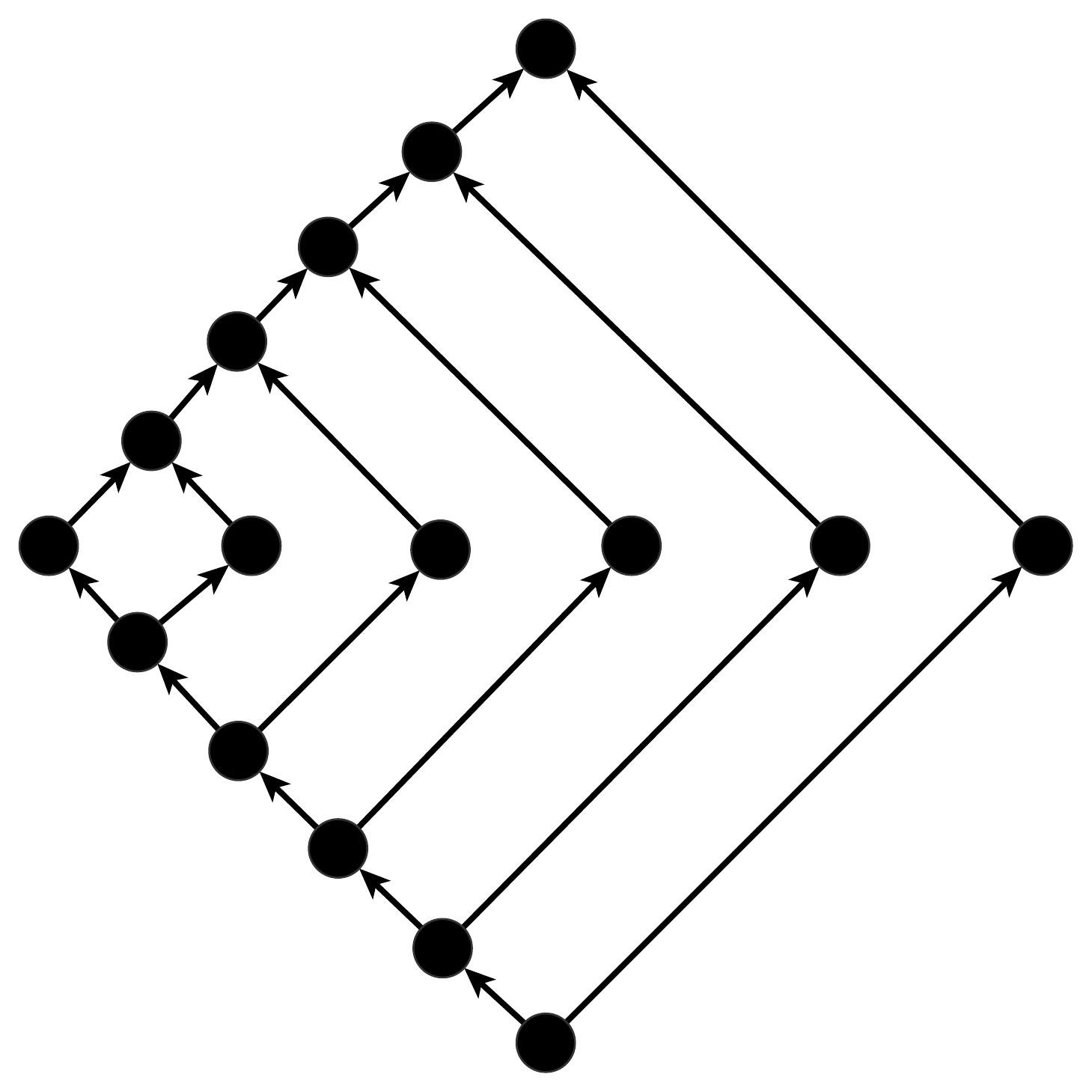}
    \hspace{0.1cm}\includegraphics[width=0.12\linewidth]{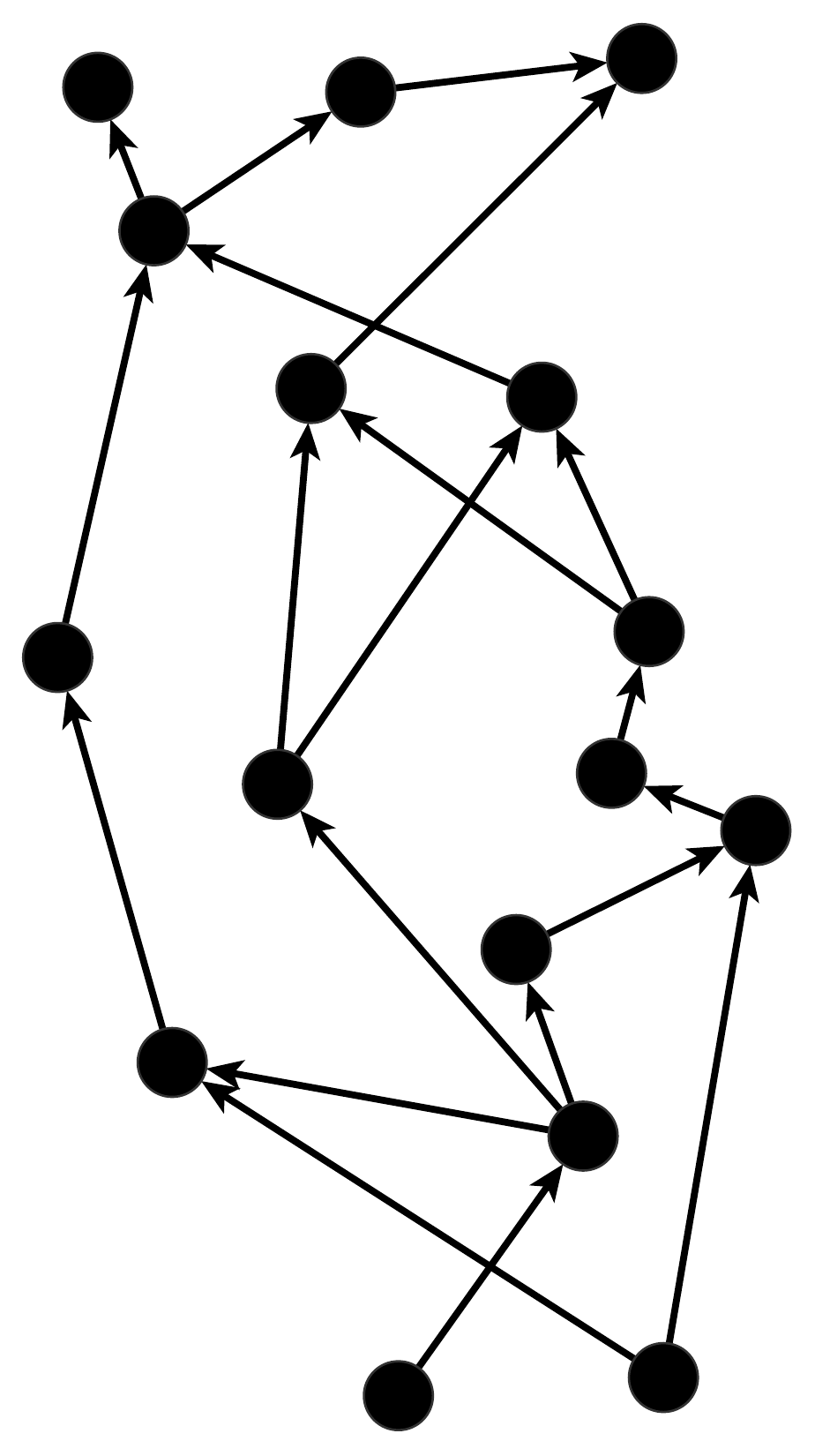}
    \hspace{0.1cm}\includegraphics[width=0.2\linewidth]{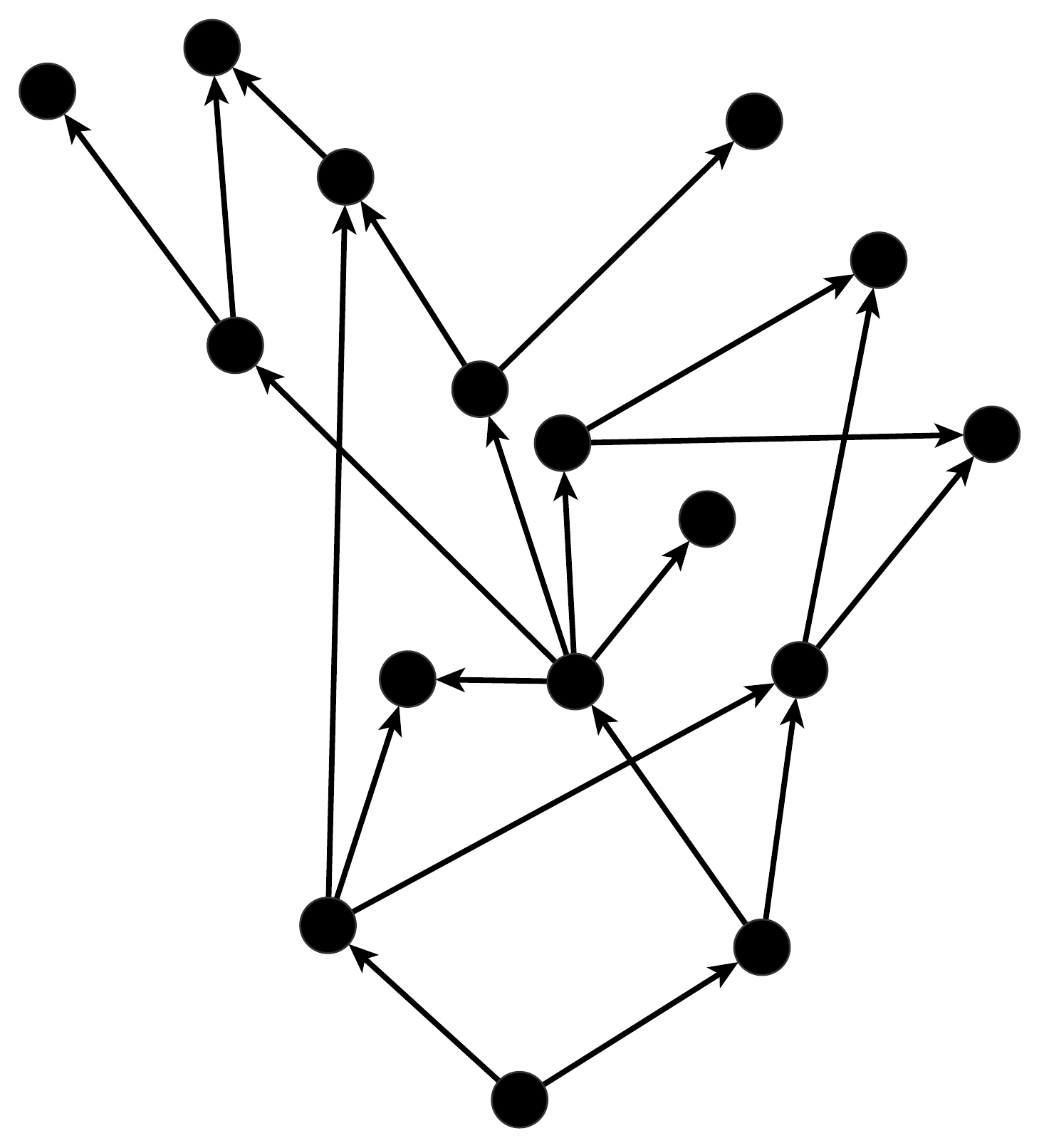}
    \caption{Illustrations of network models considered. From left to right: lattice, Russian doll, Erd\"os-R\`enyi DAG ($p=0.5$), Price model ($m=4, \delta =0.8$). All networks are Transitively Reduced, so the cycles seen are those considered in \secref{sec:model_results}. In all cases, the direction encoded in the DAGs is such that all arrows point up the page.}
    \label{fig:network_eg}
\end{figure}

We use the Transitively Reduced versions of four network models to show the usefulness and explanatory power of the metrics described in \secref{sec:cycle_metrics}. To determine the directed images of the MCB underlying the reduced DAG, we follow the procedure detailed in~\ref{sec:TR_MCB_HB} and illustrated in~\figref{fig:fig_2_panel}. Unless otherwise stated, we considered networks with $N=500$ nodes, and for stochastic network models, we generated $n=20$ realisations for each parameter value and computed one MCB for each realisation, see \secref{sec:algorithm} for a justification to use a single MCB as a representative of the ensemble of MCBs. An example of each type of network is given in \figref{fig:network_eg}. The code used to generate the results reported here can be found on \href{https://github.com/vv2246/tr-dag-cycles}{Github}.

\paragraph{Lattice model}\label{sub_sec:lattice}
The first network model we discuss is a --- finite --- lattice graph turned into a DAG by giving each node has two outgoing edges and two incoming edges, see Fig.~\figref{fig:network_eg} for an illustration. Specifically, to build our lattice DAG we assign a coordinate $(x_i,y_i)$ to each node $u_i$, where $x_i,y_i$ are non-negative integers less than some fixed size parameter $L$. We then have edges from $u_i$ to two nodes $(x_i+1,y_i)$ and $(x_i,y_i+1)$ provided these are allowed coordinates. This model is naturally Transitively Reduced.

The MCB is unique and its features are analytically tractable. Clearly, for our lattice DAG all cycles are diamonds of the same size $C_\alpha = \langle C \rangle =4$ so there is no variance, $\sigma(C)=0$, and so our balance statistic \eqref{e:bdef} of each cycle is also zero, $b_\alpha =0\ \forall \alpha$. The mean stretch of cycles \eqref{e:stretchdef} is always $2$. By symmetry, the mean cycle height is half the total height of the lattice DAG, $\langle h \rangle = \ell_{\textrm{max}}/2$. The variation in the height of the cycles, $\sigma(h)$ increases with a network size, since in a lattice all cycles are evenly distributed across a network. The mean edge participation $\lim_{N\rightarrow\infty}\langle E_p \rangle =2 $. It is also clear that $\textrm{null} (\mathbf{L}^C)=1$.

\begin{figure}
    \centering
    \includegraphics[width = 0.4\linewidth]{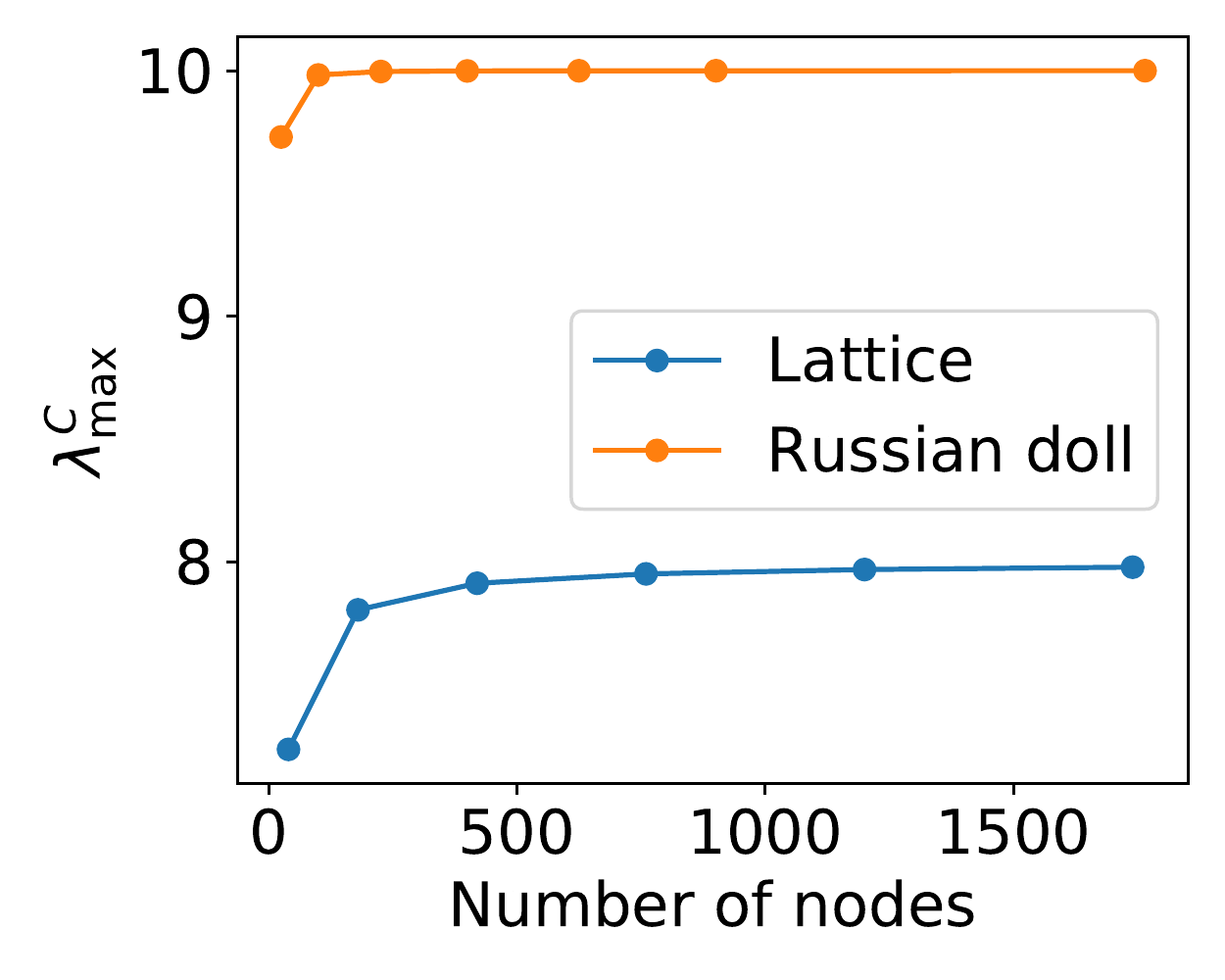}
    \caption{$\lambda_{\textrm{max}}^C$ as a function of a number of nodes in lattice DAGs (orange), and Russian doll DAGs (blue). Analytical results overlap with numerical results.}
    \label{f_lattice_eigvals}
\end{figure}

The eigenvalues of $\Mmatr$ for this lattice model can be solved exactly because of its structure, see \appref{app:Lspectral}. We can show that
\begin{eqnarray}
     \lambda_\alpha
     &=&
     4
     + 2\cos \left( \frac{ \pi m}{L} \right)
     + 2\cos \left( \frac{ \pi n}{L} \right)  \, ,
     \nonumber
     \\
     && \qquad
     \quad \alpha = (m-1)+(n-1)\,(L-1), \quad m,n \in \{ 1, \ldots, (L-1) \} \, .
\end{eqnarray}

Therefore the largest eigenvalue of $\mathbf{M}$ converges $\lim_{N\rightarrow\infty}\lambda_{\textrm{max}}^C=8$ for a large lattice. \Figref{f_lattice_eigvals} shows that $\lambda_{\textrm{max}}^C$ does approach the infinite lattice value $8$ rapidly.

\paragraph{Russian doll}\label{sub_sec:rd}
The Russian doll DAG $R_d$ has $N=(1+3d)$ nodes, $\{u_0,u_1,\ldots,u_{3d}\}$, and $4d$ edges. We start from $R_0$ which is the trivial DAG with one node and no edges. To grow the Russian Doll DAG $R_d$ we take the DAG $R_{d-1}$ and add the three nodes  $u_{3d-2}$, $u_{3d-1}$ and $u_{3d}$. Four new directed edges as also included: $(u_{3d-2},u_{3d-5})$, $(u_{3d-2},u_{3d-1})$, $(u_{3d-1},u_{3d})$, and $(u_{3d-3},u_{3d})$.
This ensures that the Russian doll DAGs $R_d$ are transitively reduced by construction. This construction shows that $u_{3d-2}$ is the source and $u_{3d}$ is the sink node for $R_d$. Since the new edge $(u_{3d-2},u_{3d-5})$ links the new source node in $R_d$ to the source node in the previous DAG $R_{d-1}$, we see this edge is part of the longest path in $R_d$. Likewise, the new edge from the sink node $u_{3d-3}$ of $R_{d=1}$ to the new sink node $u_{3d}$ of $R_d$ is also part of the longest path in $R_d$.  Thus the height of the Russian doll DAG grows by two for each step of this iterations giving us a network height of $2d$ for $R_d$.

Note also how this iterative  construction shows that we are adding one more cycle to those already in $R_{d-1}$. For a minimal cycle basis, we can define this to be the cycle $C_d$ formed by the four edges added when creating $R_{d}$ and two of the edges added a step earlier when creating $R_{d-1}$, namely $(u_{3d-5},u_{3d-4})$ and $(u_{3d-4},u_{3d-3})$. The $d$ cycles defined by this iterative process form the unique Minimal Cycle Basis. The cycles in the MCB of the Russian doll DAG only contains diamond cycles and the properties of the MCB are analytically tractable.

All the cycles in this MCB, except one, have size $6$. The `first' cycle, the one equivalent to $R_1$ containing $\{u_0,u_1,u_2,u_{3}\}$, has size 4. Thus $\lim_{N\rightarrow\infty}\langle S \rangle =6 $, $\lim_{N\rightarrow\infty}\sigma(S)=0$.
Each cycle is a diamond with balance \eqref{e:bdef} $b=1/2$ except for the first cycle, thus asymptotically, $\langle b \rangle =1/2$.
The cycles are arranged symmetrically so  the height of all cycles $h_\alpha$ \eqref{e:hcycledef} is identical and equal to half the height of the whole DAG, so $h_\alpha =d$.

However, the stretch \eqref{e:stretchdef} will increase with $N$. To see this, consider the extra cycle added when constructing $R_d$ from $R_{d-1}$ discussed above. As this contains the sink and source nodes for $R_d$ the stretch of this new cycle in the MCB is equal to $2d$, the height of $R_d$.  By induction we see that the $d$ cycles in this MCB have heights equal to the positive even numbers up to $2d$ so the mean value is simply $(1+d)$ and grows linearly as the size of the DAG.

Let us move from the statistical features of the Russian doll MCB to its spectral properties. First, it is clear that $\mathrm{null}(\mathbf{L}^C)=1$. The $\Mmatr$ cycles overlap matrix of \eqref{e:Mdef} is a simple tridiagonal matrix for which the eigenvectors and eigenvalues can be found.  In particular, the eigenvalues are given by
\begin{equation}
 \lambda_\alpha
 =
 6 + 4\cos \left( \frac{2 \pi \alpha}{(2d+1)} \right)
 \, ,
 \quad \alpha = 1,2 \ldots, d \,.
\end{equation}
as \figref{f_lattice_eigvals} shows, see also \secref{app:Rdspectral}. The largest eigenvalue has the limiting value of  $\lim_{d\to \infty} \lambda_1 = \lambda_{\textrm{max}}^C=10$. To intuitively understand the origin of this value, we first observe that all but one cycle are of size 6, giving the cycles size contribution. Moreover, the largest eigenvalue of $\mathbf{M}$ gets a contribution of $2$ from each overlapping cycles as they share 2 edges in the MCB. Since each cycle, besides the first and last, have two neighbours, we obtain $10=6+2+2$.

\paragraph{Erd\"os-R\'enyi DAG}\label{sub_sec:random}

The Erd\"os-R\'enyi (ER) network model~\cite{ER60} is probably the simplest random network model: each pair of nodes are connected with a probability $p$ to give a simple graph. We obtain an Erd\"os-R\'enyi DAG similarly to an Erd\"os-R\'enyi graph. We assign each node $u$ an unique integer $i_u$, and add a directed edge between each pair of nodes $(u,v)$ with probability $p$, with the direction of each edge going from the node with the smaller-valued integer to the larger-valued index, $i_u < i_v$. In what follows, we will call the transitive reduction of these random DAGs ``reduced Erd\"os-R\'enyi DAGs''.

\begin{figure}[h!]
\centering
    \begin{tabular}{p{0.45\textwidth}@{\hspace{0.05\textwidth}}p{0.45\textwidth}}
        \makebox[0.45\textwidth][c]{\textbf{Reduced Erd\"os-R\`enyi DAG}} & \makebox[0.45\textwidth][c]{\textbf{Reduced Price model}} \\
        \includegraphics[width=0.45\textwidth]{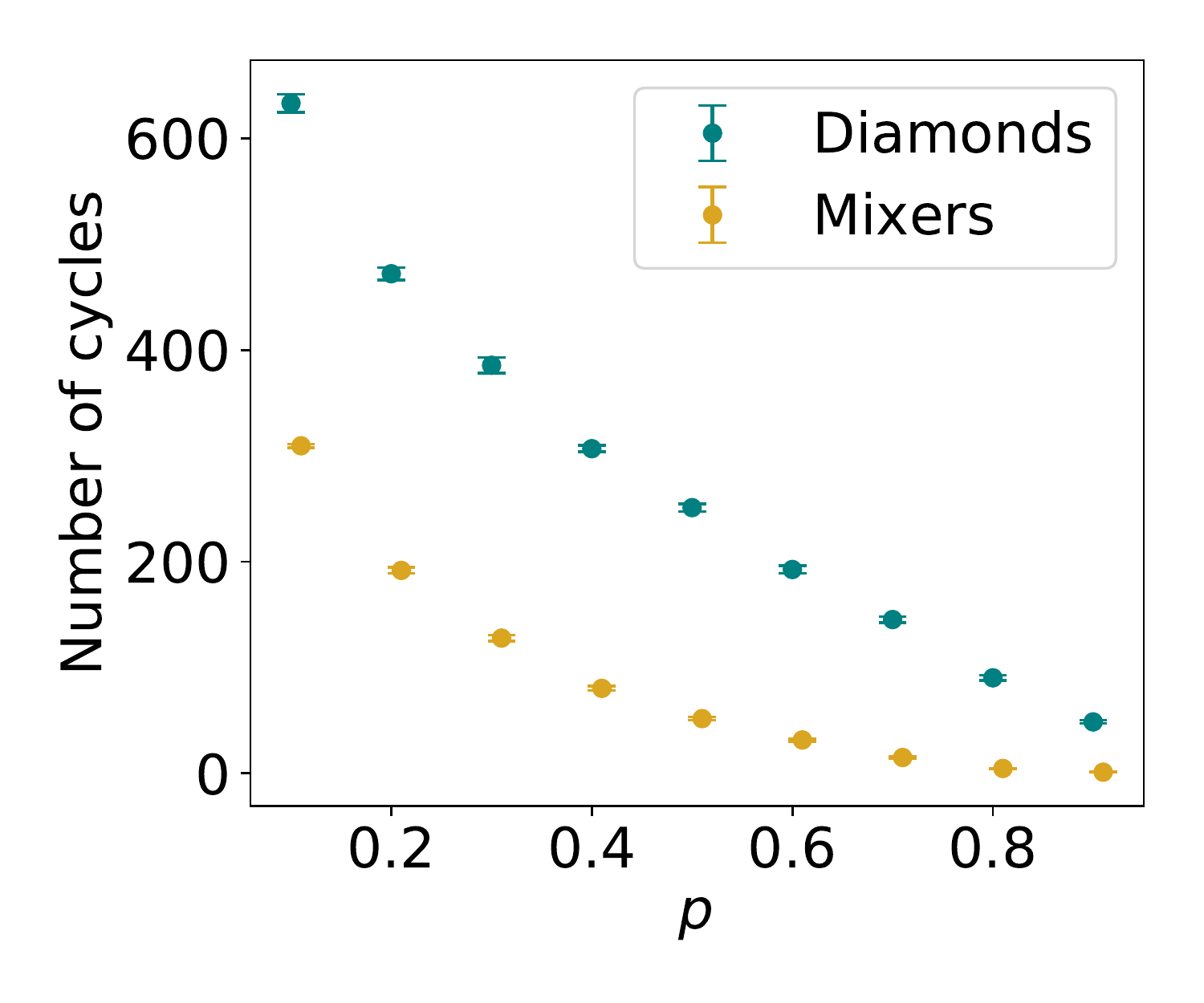} &
        \includegraphics[width=0.45\textwidth]{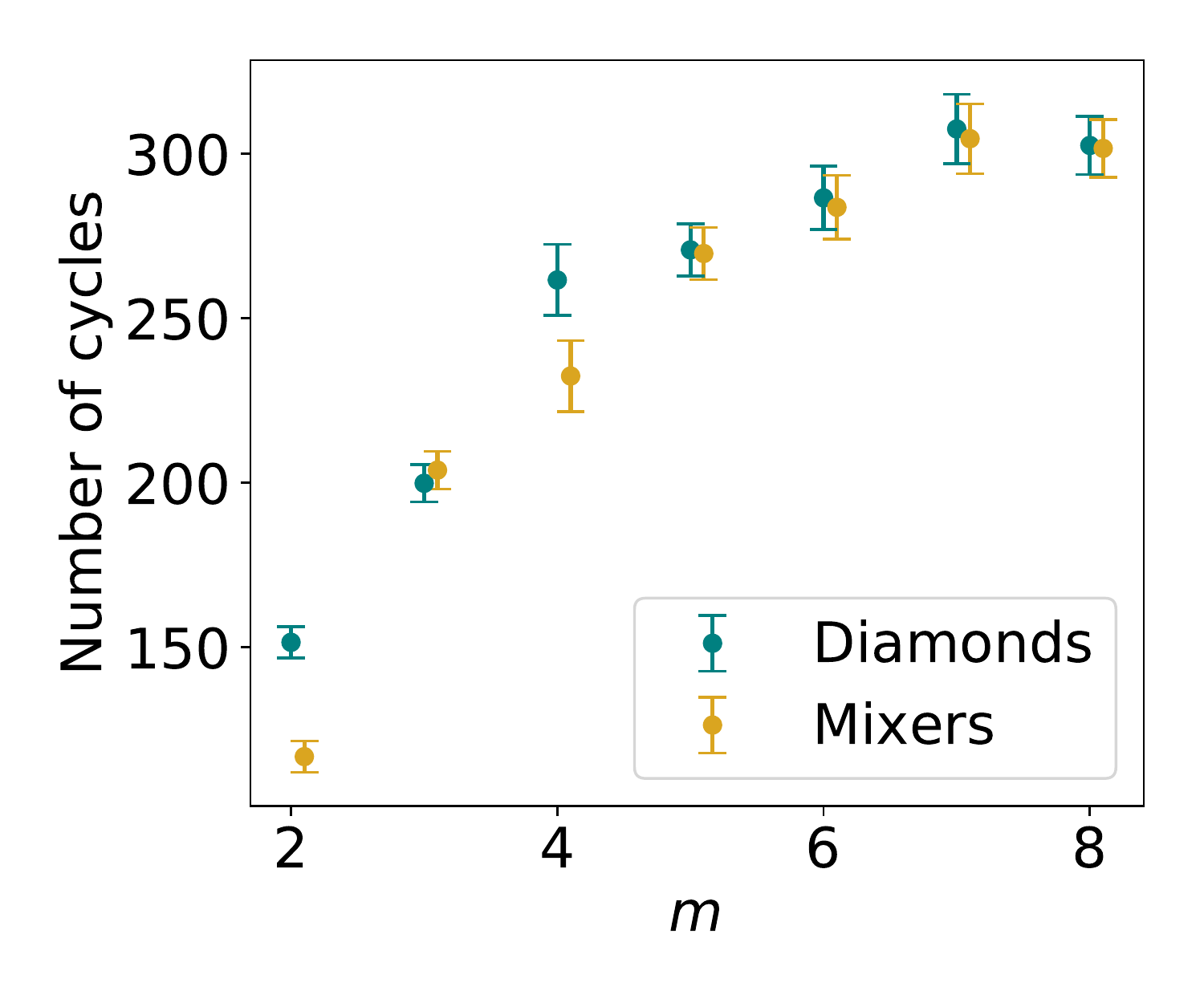}\\
        \includegraphics[width=0.45\textwidth]{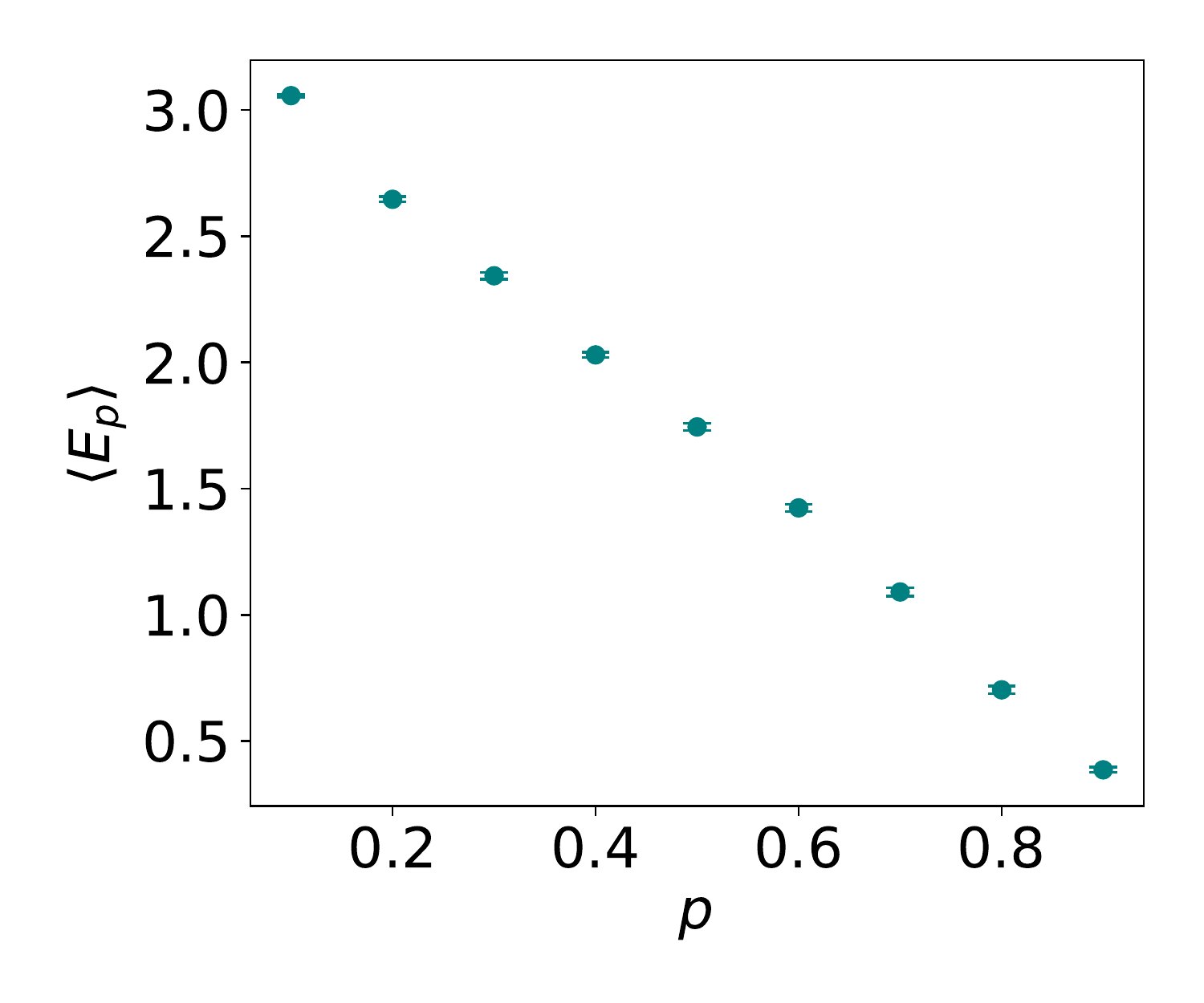} &
        \includegraphics[width=0.45\textwidth]{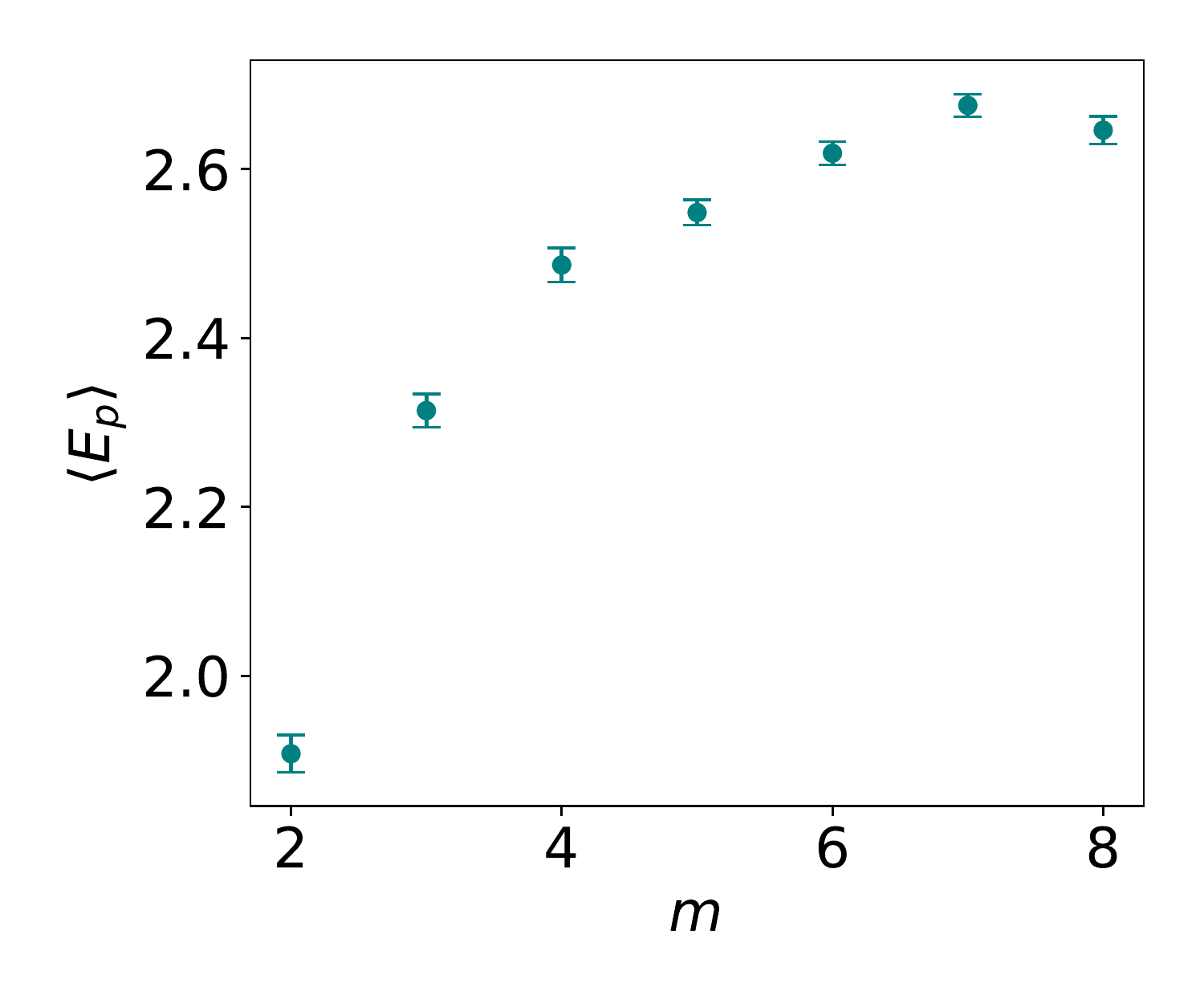}
\end{tabular}

    \caption{Number of cycles of mixers and diamonds in transitively reduced Erd\"os-R\`enyi DAGs and Transitively Reduced Price model (top) and the corresponding average edge participation $E_p$ (bottom). We considered networks with $N=500$ nodes, generated $n=20$ realisations for each parameter value and computed one MCB for each realisation, and $\delta=m(1+m)$ for the Price Model.}\label{f_ep_number_diamonds_mixers}
\end{figure}

The statistics of the minimal cycle basis of reduced Erd\"os-R\'enyi DAGs are richer than the previous simple models. Given the edge density parameter $p$ it is straightforward to estimate the expected number of cycles of a regular undirected random graph using \eqref{eq:number_cycles}. We note that in Erd\"os-R\'enyi DAGs, TR has the following suppression effect: the larger $p$ is, the more edges will be reduced by TR: the density of reduced Erd\"os-R\'enyi DAGs is monotonically decreasing with $p$, and consequently, there are more cycles at low $p$ than there are at high $p$, as is clear from \figref{f_ep_number_diamonds_mixers}. For instance, at $p=1$, we have a complete DAG which after transitive reduction is reduced to a simple path that follows the order prescribed by the $i_u$, with one pair of source and sink nodes, all other nodes being neutral. Therefore the reduced graph has the lowest possible density and no cycles.

\begin{figure}[h!]
\centering
\textbf{Reduced Erd\"os-R\'enyi DAG}\par\medskip
        \includegraphics[width=4cm]{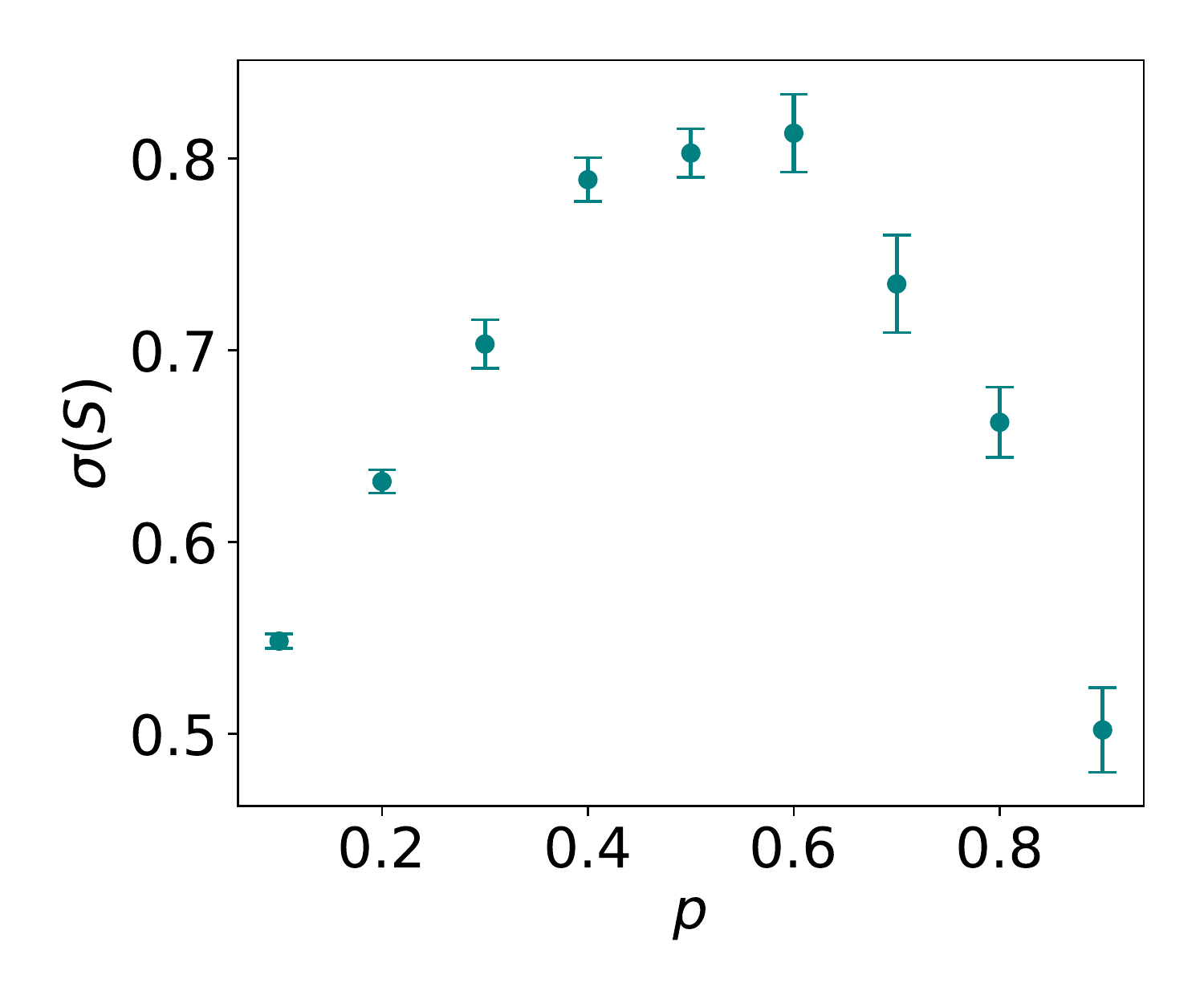}
        \includegraphics[width=4cm]{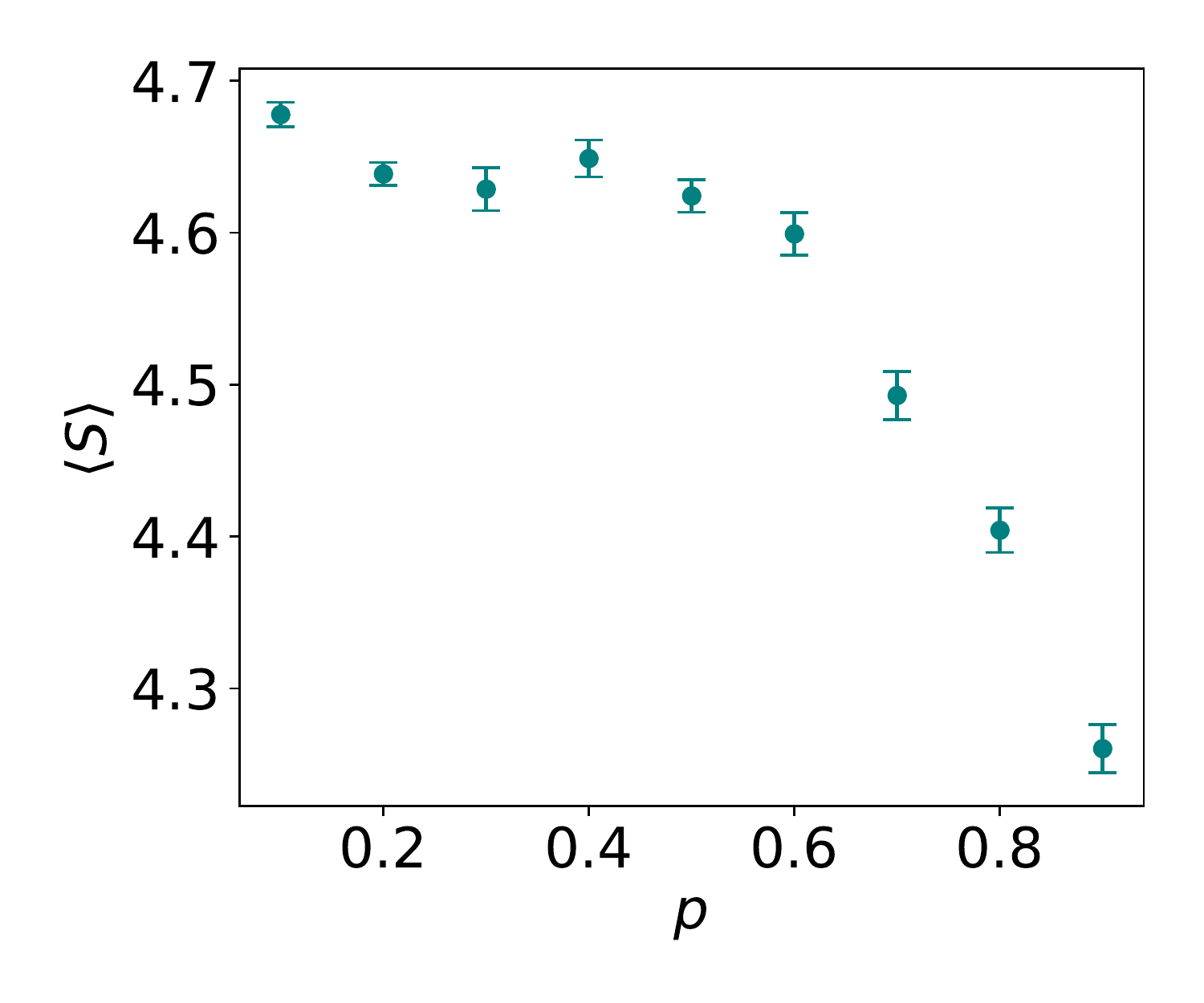}
        \includegraphics[width=4cm]{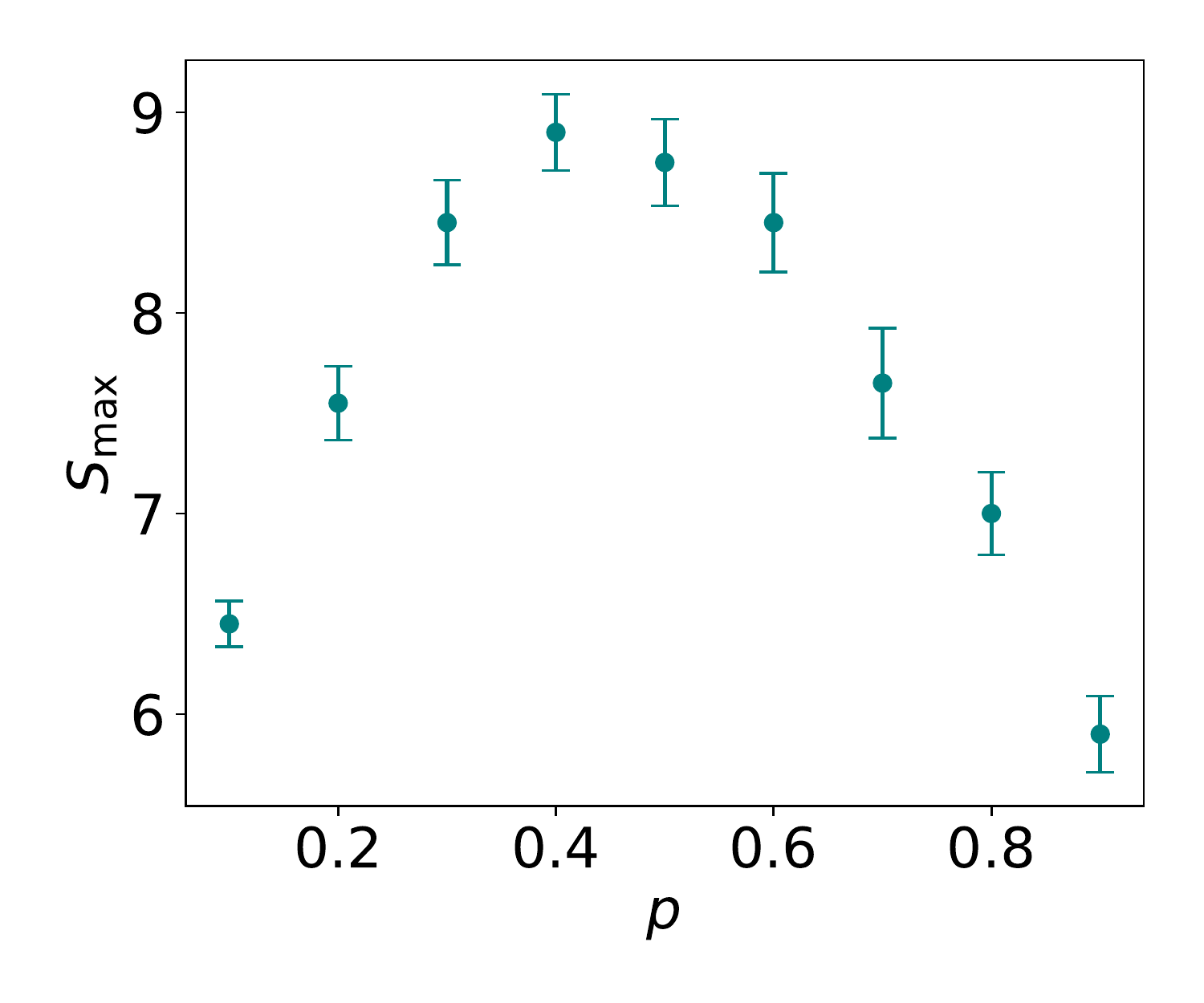}  \\
        \includegraphics[width=4cm]{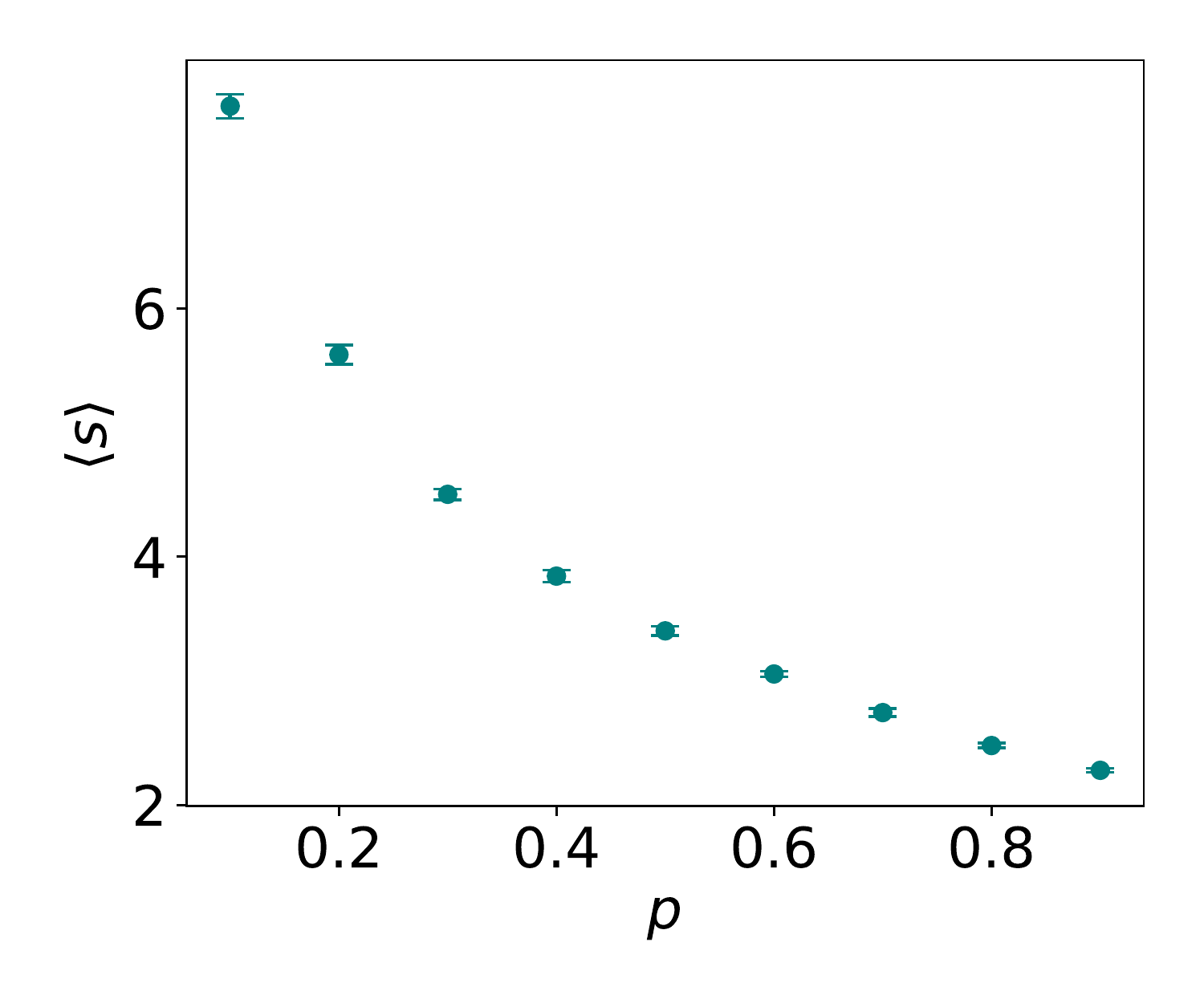}
        \includegraphics[width=4cm]{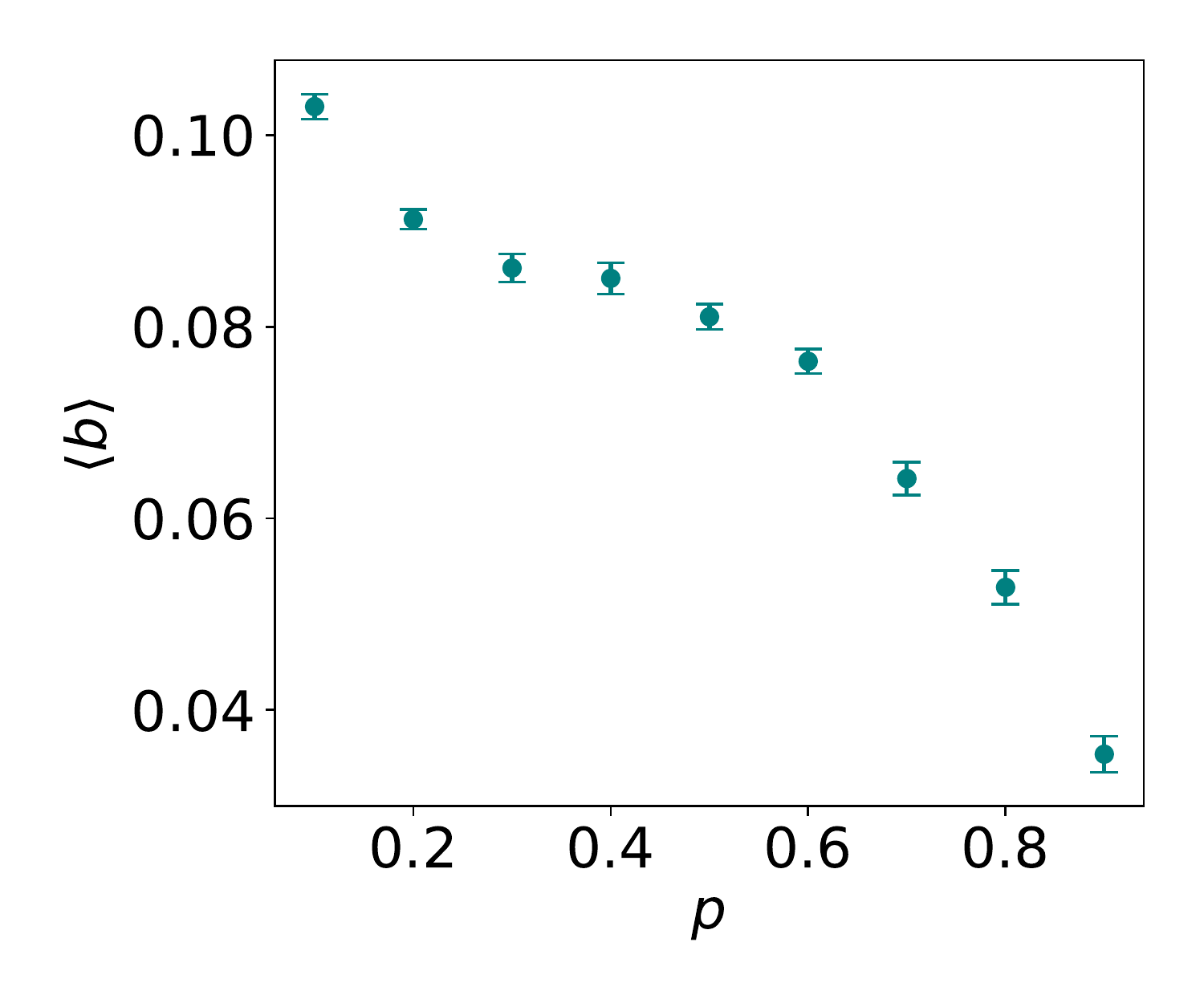}
        \includegraphics[width=4cm]{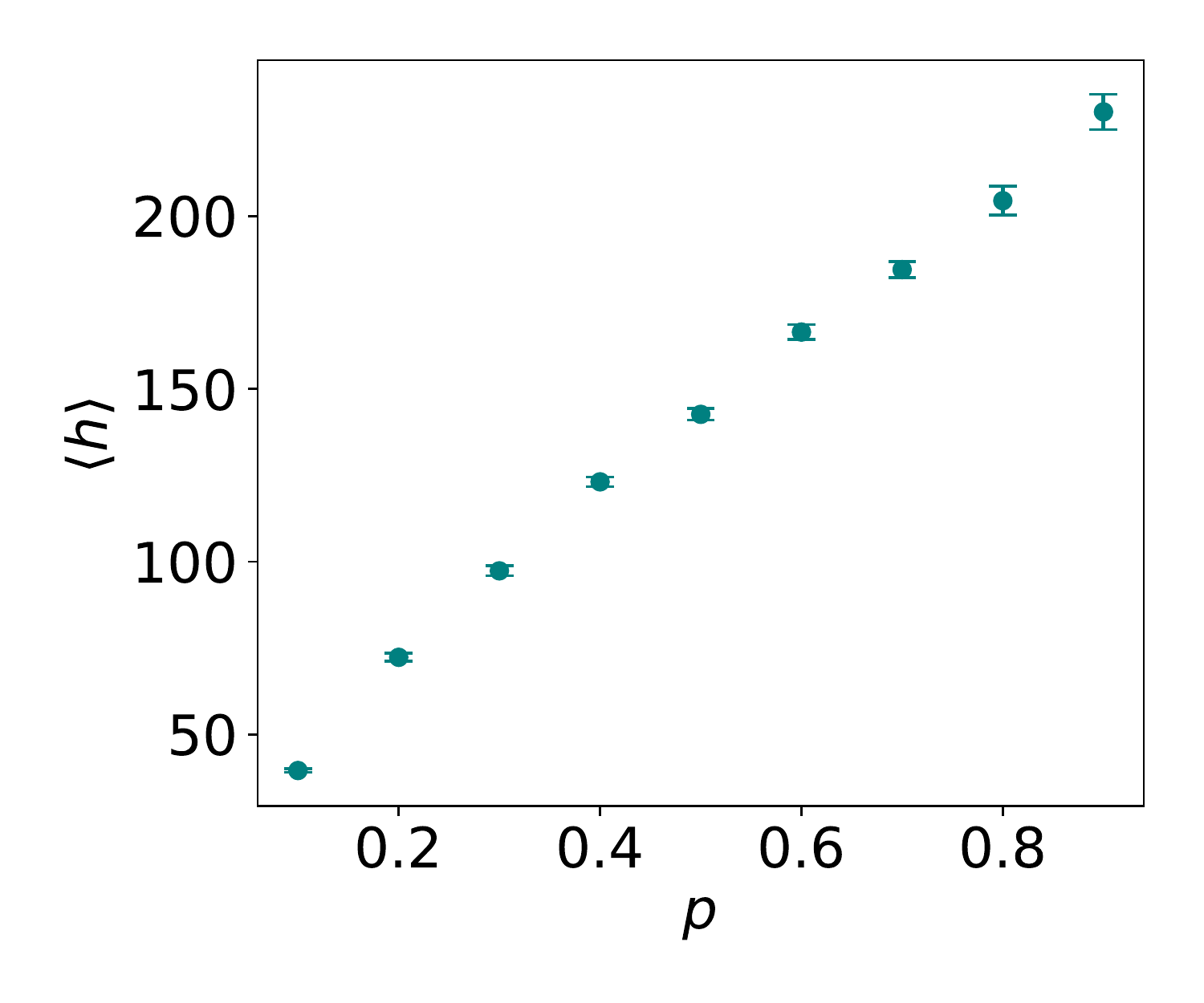}\\
        \textbf{Reduced Price model}\par\medskip
         \includegraphics[width=4cm]{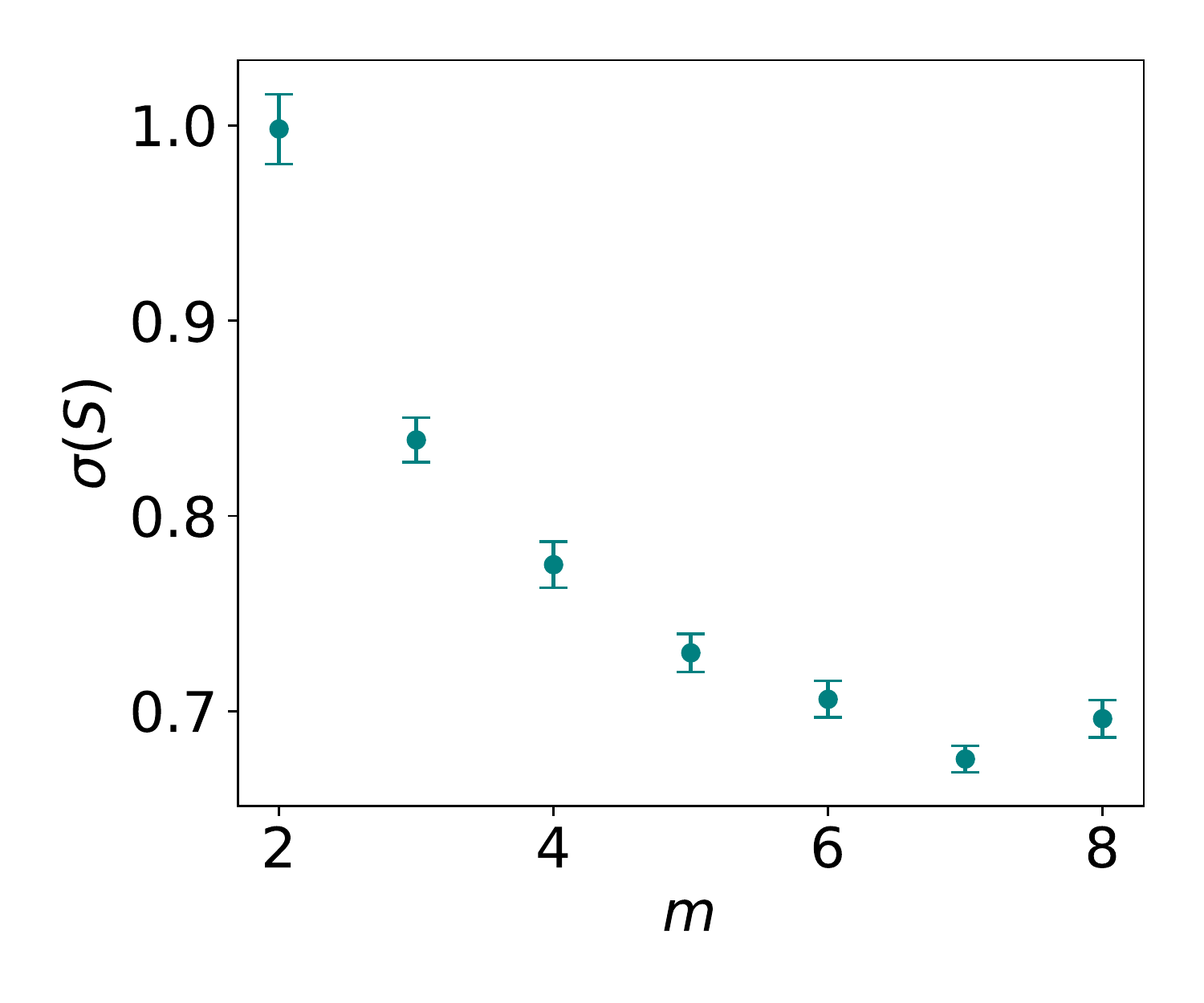}
        \includegraphics[width=4cm]{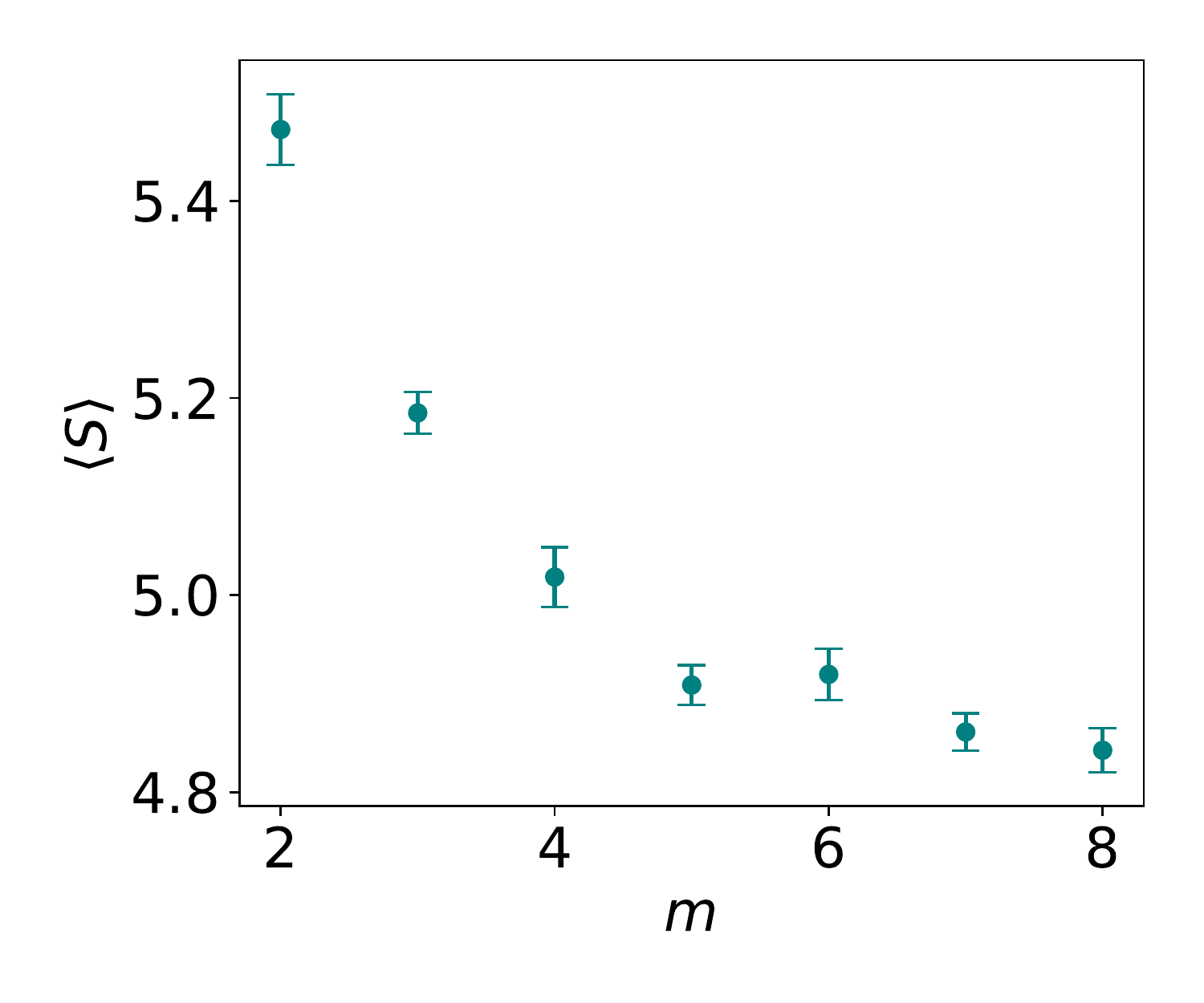}
        \includegraphics[width=4cm]{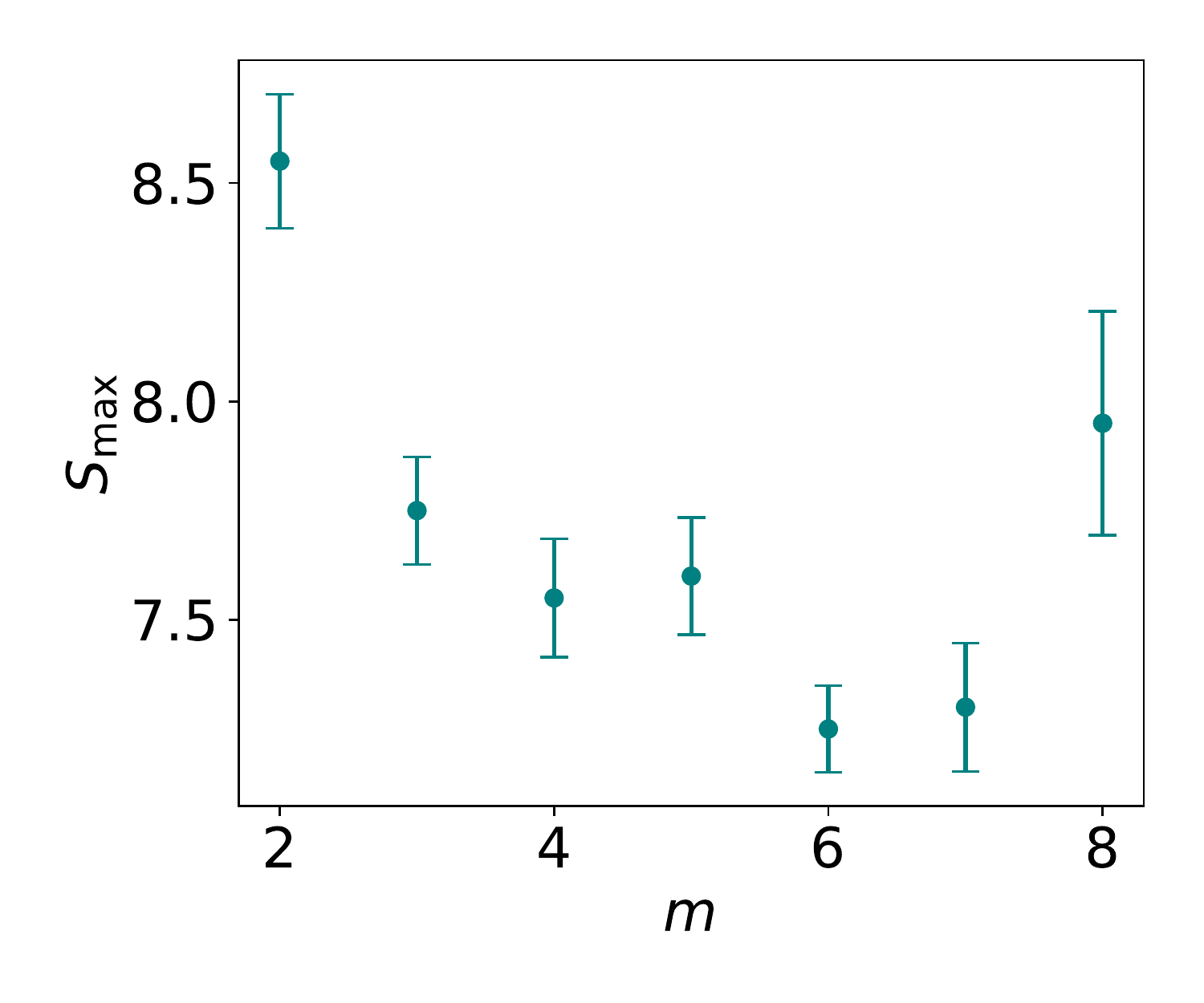}\\
        \includegraphics[width=4cm]{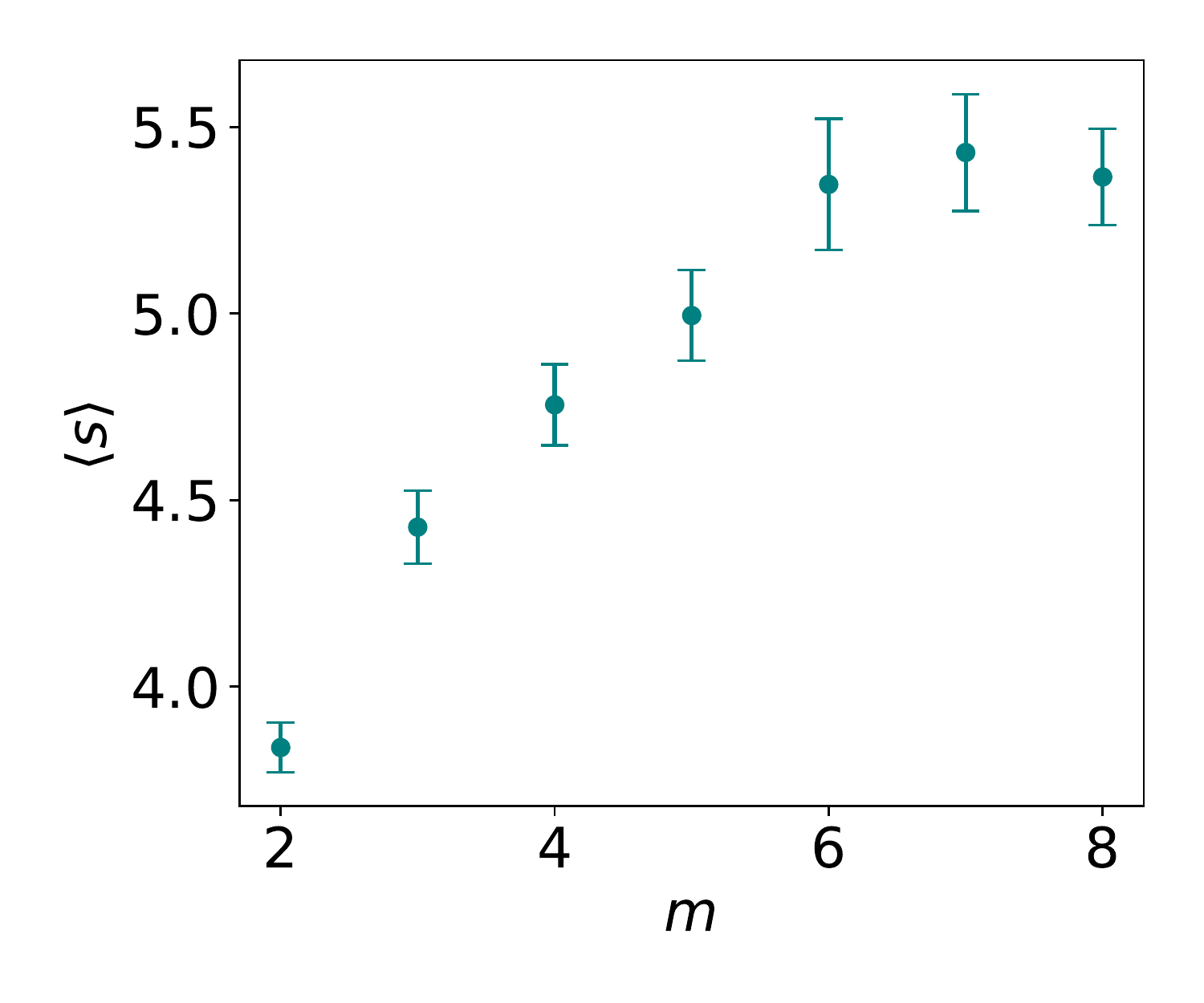}
        \includegraphics[width=4cm]{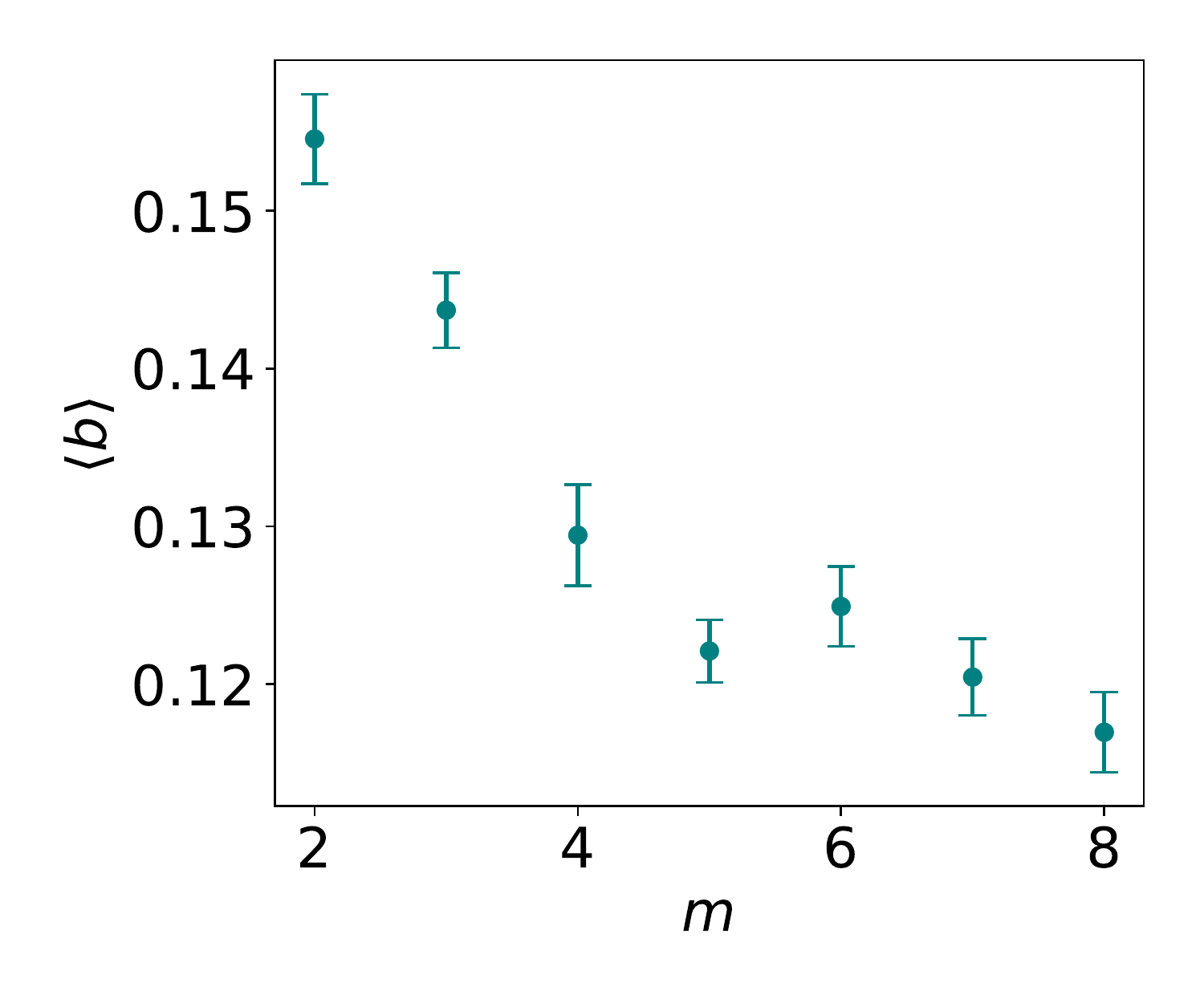}
        \includegraphics[width=4cm]{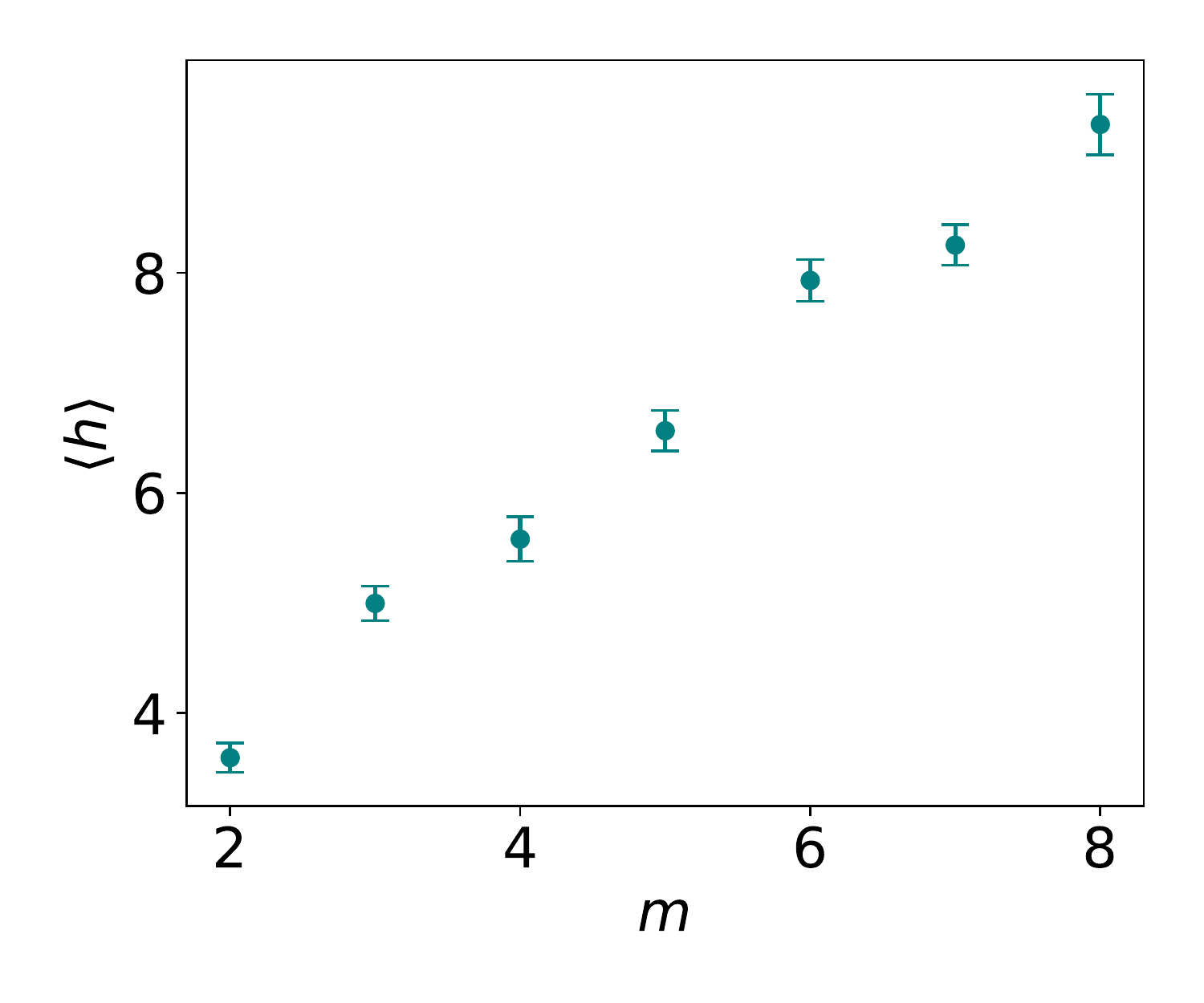}\\

    \caption{Collected cycle basis statistics - $\sigma(S)$, $\langle S\rangle$, $S_{max}$, $\langle s\rangle$, $\langle b\rangle$ and $\langle h\rangle$, see \secref{sec:cycle_metrics} for the definitions - for Transitively Reduced Erd\"os-R\`enyi DAGs (top panel) and Transitively Reduced Price model (bottom panel). We considered networks with $N=500$ nodes, and for stochastic network models, we generated $n=20$ realisations for each parameter value and computed one MCB for each realisation.}
    \label{f_random_price_dag_cycle_stats}
\end{figure}

We can further explain the behaviour we see in cycle statistics in reduced Erd\"os-R\'enyi DAGs, summarised in the top panel of \figref{f_random_price_dag_cycle_stats}, by considering cliques. Note that if $p$ is small, the network is sparse, in comparison to a large value of $p$. Thus the smaller the $p$, the smaller the number of cliques, as well as the average size of such cliques, as the probability of a clique of size $k$ is $p^k$. It follows that the dominant cliques at small values of $p$ are triangles, which are transitively reduced. 

\begin{figure}[h!]
\begin{center}
    \begin{tabular}{p{0.45\textwidth}@{\hspace{0.05\textwidth}}p{0.45\textwidth}}
        \makebox[0.45\textwidth][c]{\textbf{Reduced Erd\"os-R\`enyi DAG}} & \makebox[0.45\textwidth][c]{\textbf{Reduced Price model}} \\ 
        \includegraphics[width=0.45\textwidth]{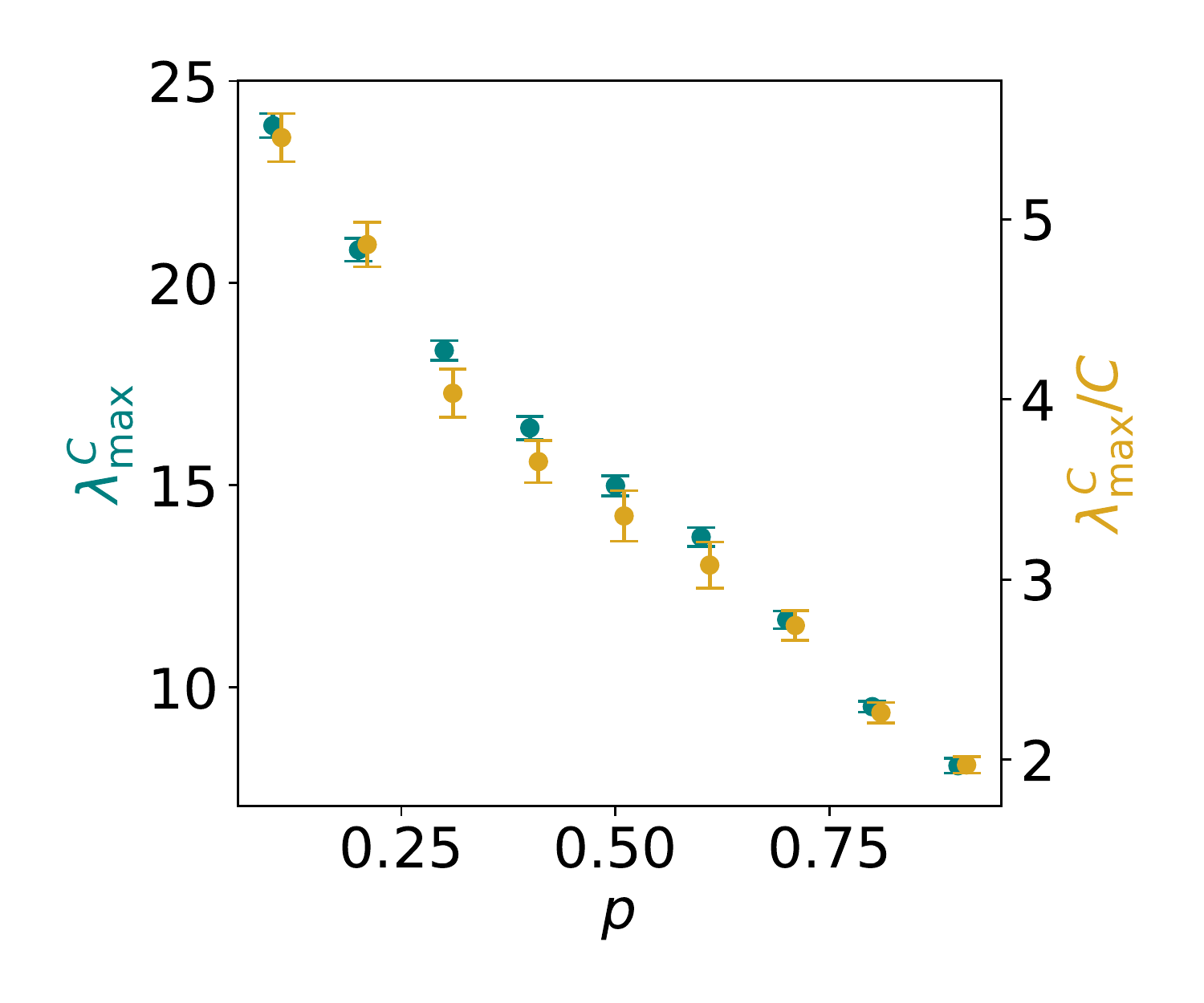}  &
        \includegraphics[width=0.45\textwidth]{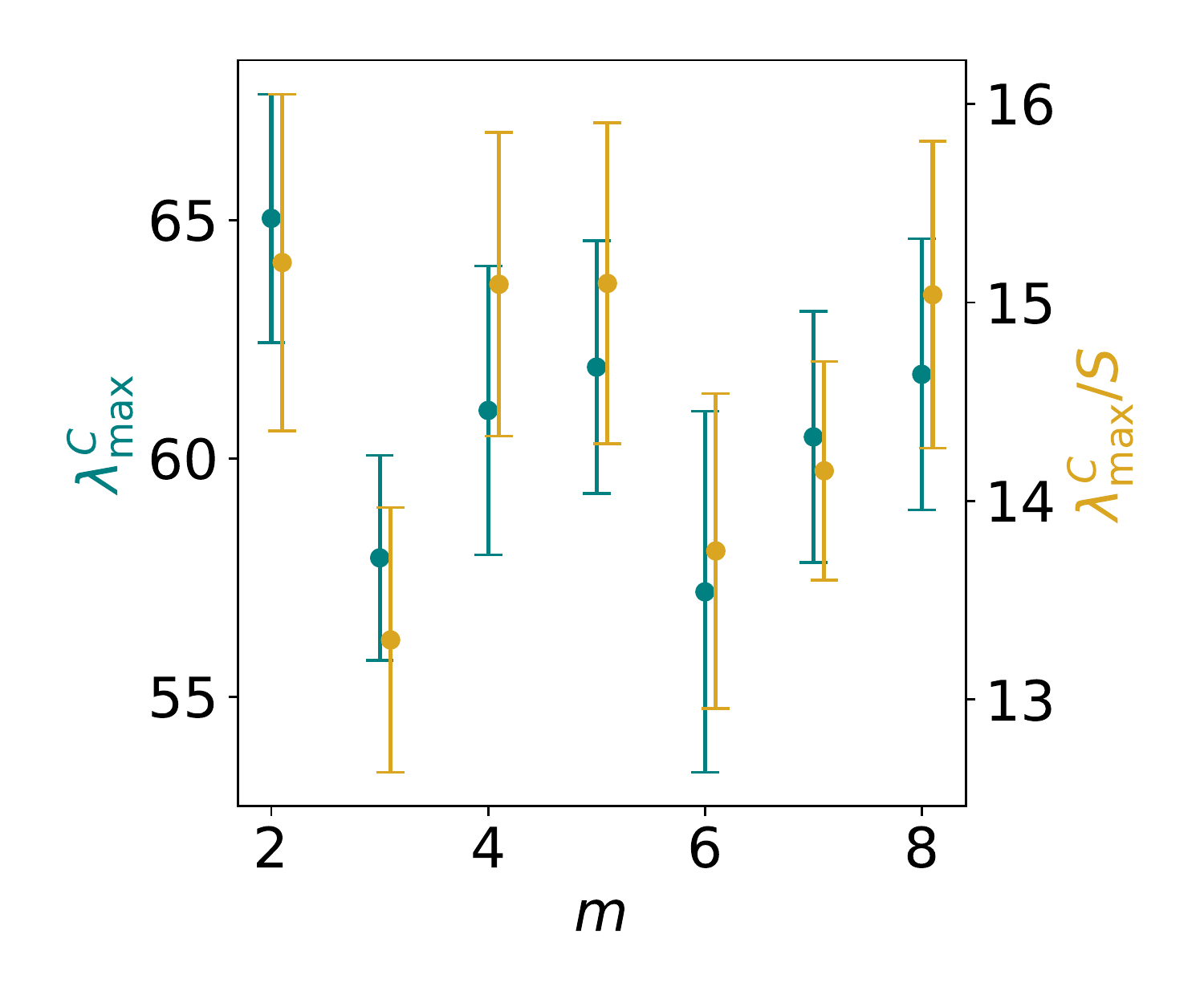}\\
        \includegraphics[width=0.45\textwidth]{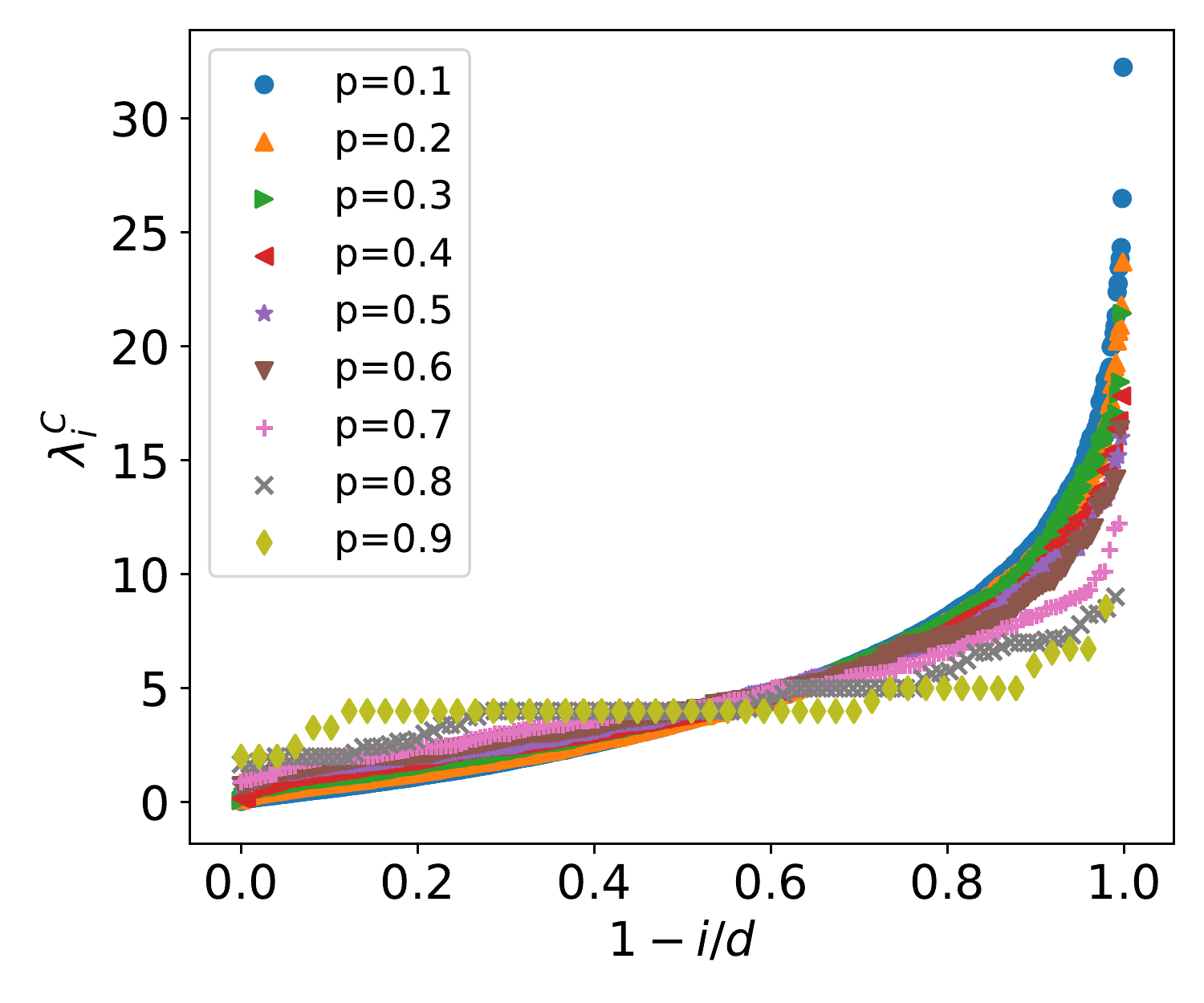} &
        \includegraphics[width=0.45\textwidth]{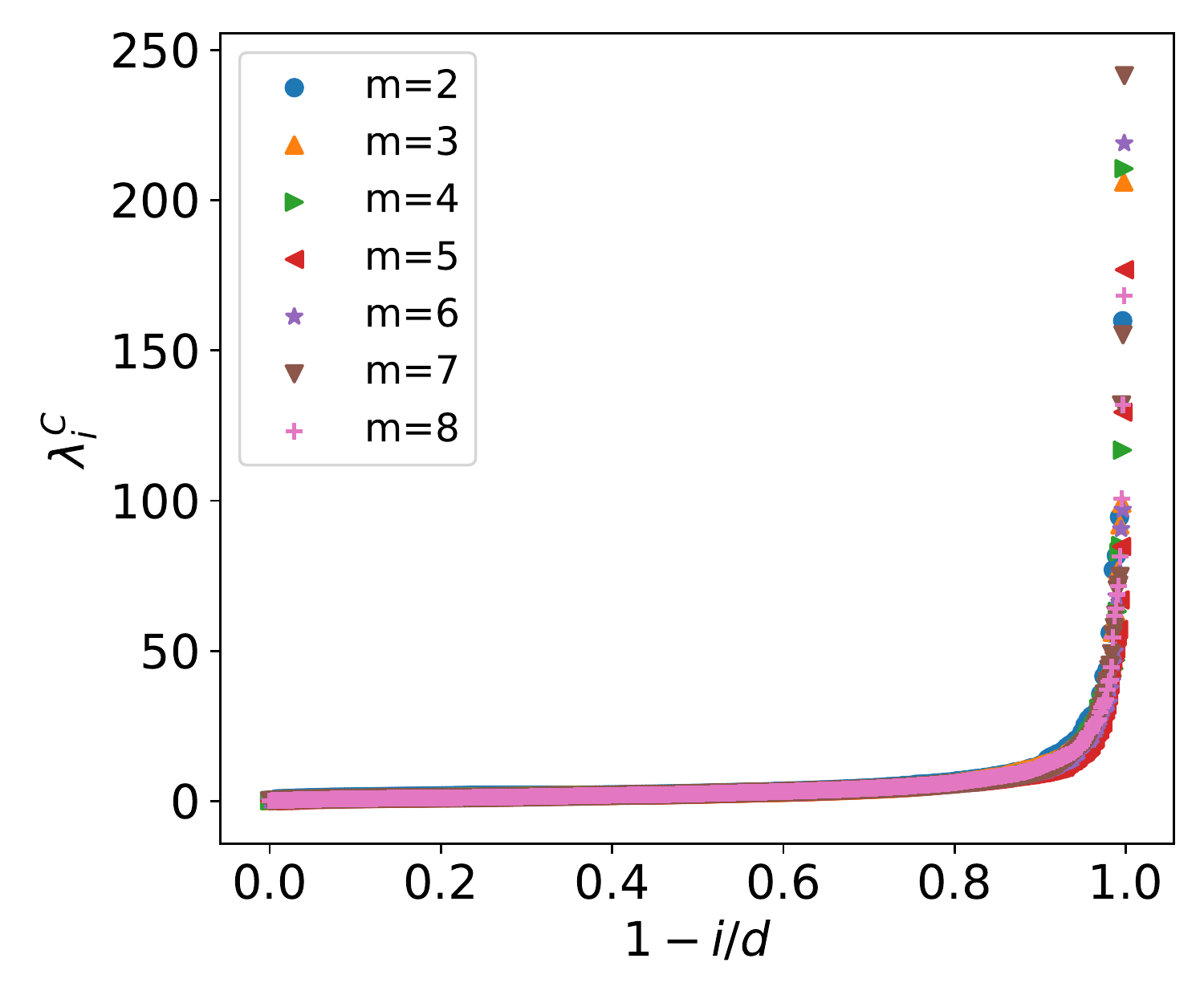} \\
        \includegraphics[width=0.45\textwidth]{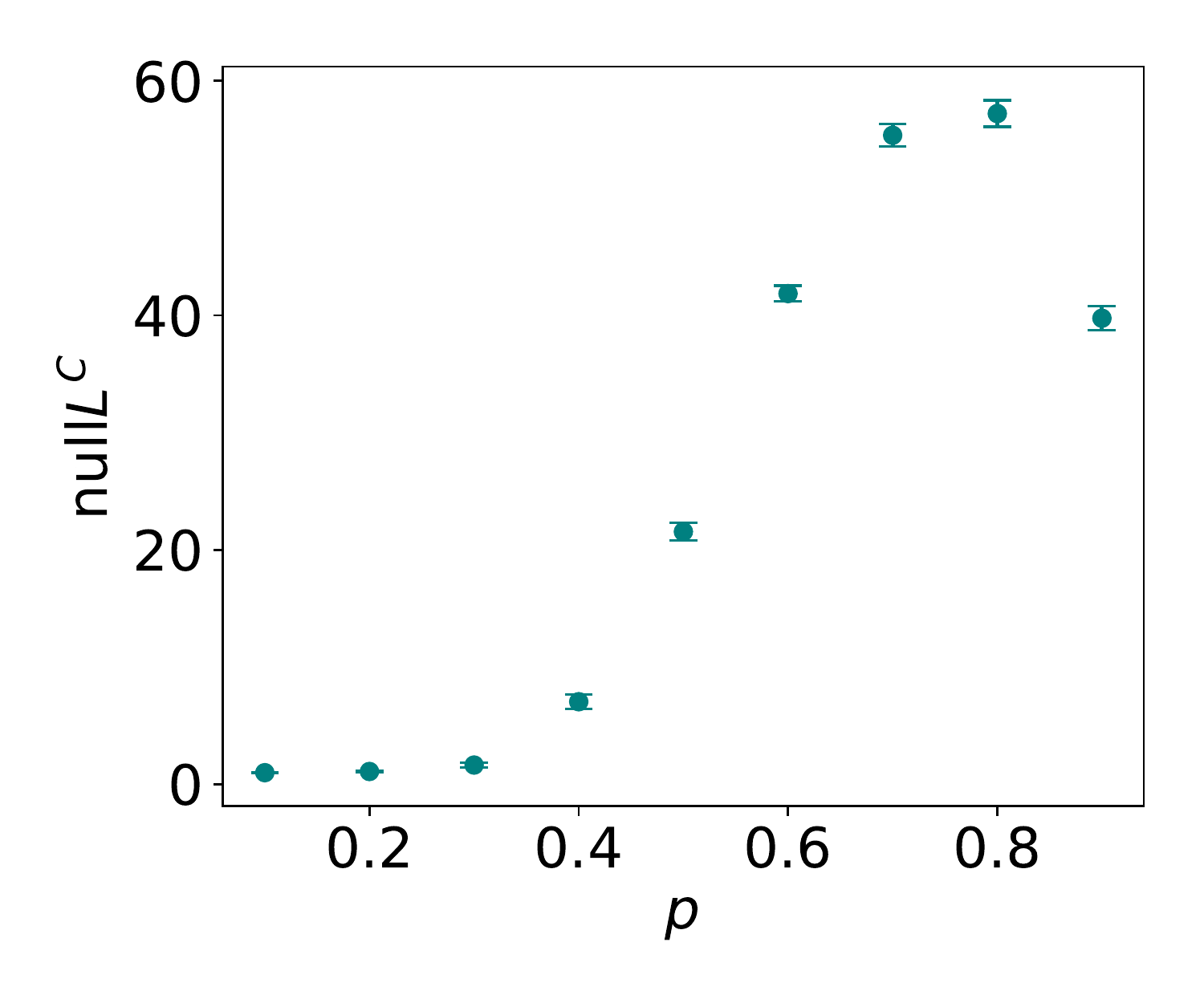} &
        \includegraphics[width=0.45\textwidth]{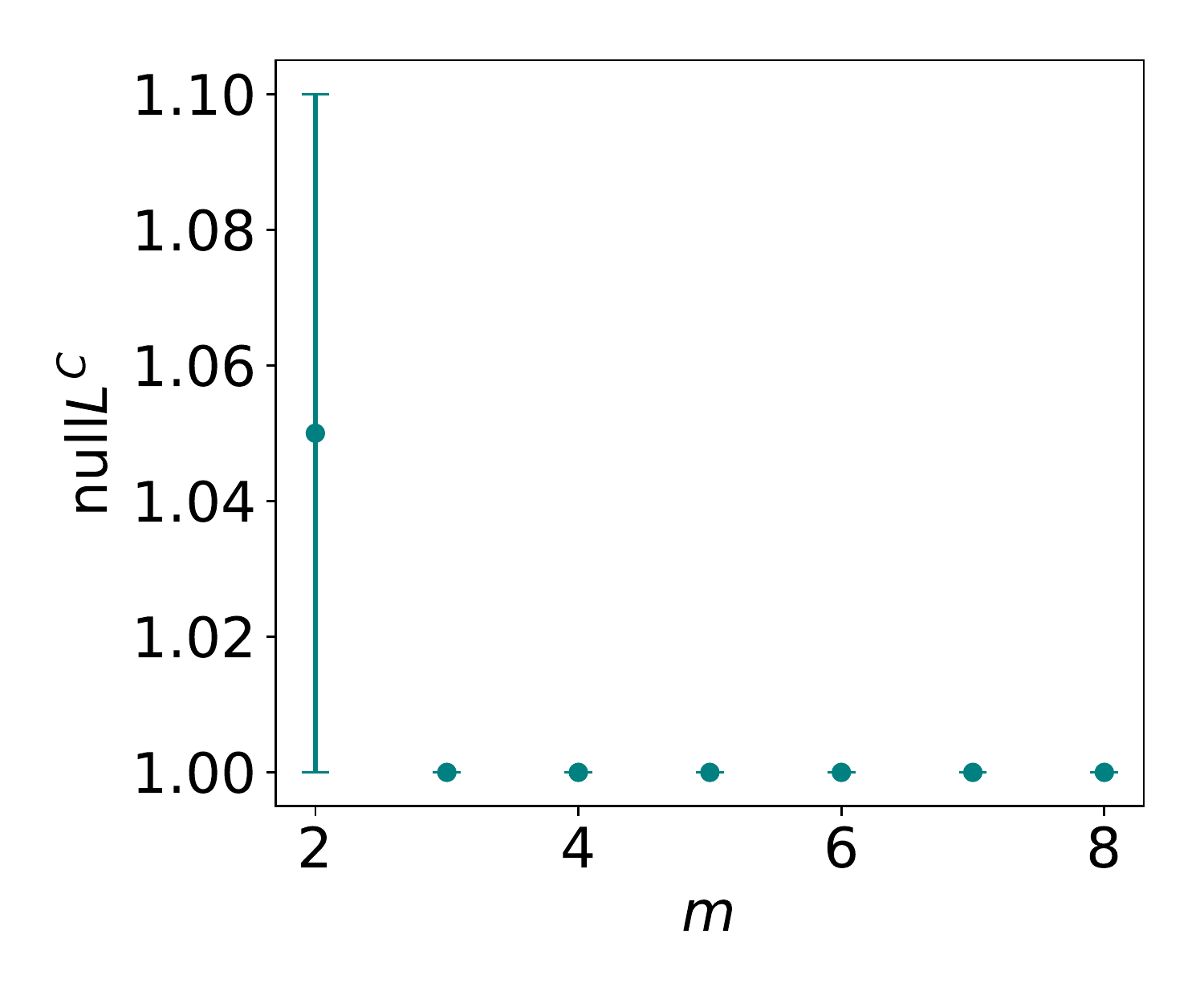} 
    \end{tabular}
    \caption{Spectral properties: dominant and normalised dominant eigenvalue, full spectrum for $\mathbf{M}^C$ and size of $\mathrm{null}\textbf{L}^{C}$ - of matrices derived from the Minimal Cycle Bases of the reduced Erd\"os-R\`enyi DAG model (left) and reduced Price model (right). We considered networks with $N=500$ nodes, generated $n=20$ realisations for each parameter value and computed one MCB for each realisation, and $\delta=m(1+m)$ for the Price Model.
    }\label{f_spectral_random_price}
\end{center}
\end{figure}

In the large $p$ regime, we see that the number of cliques, and their size, become so large that cycles are segregated in the reduced Erd\"os-R\'enyi DAG, as indicated by the behaviour of $\textrm{null} (\mathbf{L}^C)$ in \figref{f_spectral_random_price}. Since a clique of any size is always transitively reduced to a path element, it does not contain any cycles, and thus does not contribute to the circuit rank of the graph. If a network initially contains many large cliques, we are left with little space where cycles can ``form''. We expect cycle sizes and the number of edges shared between cycles to reduce as $p$ increases in the reduced DAG.

The largest eigenvalue $\lambda^C_{\textrm{max}}$ follows that same pattern as $E_p$: large for small $p$ and decrease linearly with $p$, see \figref{f_spectral_random_price} (top left). On the contrary, the average cycle size $\langle S_i \rangle$ decreases with $p$, see \figref{f_random_price_dag_cycle_stats} (top centre). Taken together, this indicates that the high value of $\lambda^C_{\textrm{max}}$ is driven by the interconnectedness of the cycles. This is supported when the full eigenspectrum is considered, see \figref{f_spectral_random_price}. For large $p$ the eigenvalues follow a ``flat'' distribution, i.e.\ there is no isolated large eigenvalue, whereas for small $p$ the distribution is steeper, with one significantly dominant eigenvalue. This can be understood if we consider the cycle overlap matrix $\Mmatr$ as the adjacency matrix of some effective cycle graph.  For small $p$, the matrix is dense with high value entries. If used to describe a broadcast process inherent in the definition of eigenvalue centrality, the process will spread fast, one highly dominant eigenvalue, and will reach equilibrium quickly due to the large gap between dominant and other eigenvalues. For large $p$, while the original network is dense, after transitive reduction we are left with a relatively small and empty cycle matrix $\Mmatr$ with mostly low value entries. The Perron-Frobenius theorem tells us that the bound on the largest eigenvalue goes down with increasing $p$.

\paragraph{Price Model}\label{sub_sec:price}
The Price model is one of the oldest models of citation networks~\cite{P65} and naturally defines a DAG with a fat-tailed distribution for the number of papers with a given citation count\footnote{The undirected version of this model is the Barab\'{a}si-Albert model, see~\cite{N10} for a discussion.}.

The Price model is a growing network model and nodes are labelled with a discrete time $t$ which represents the step at which a node was introduced to a network. The network growth starts from a seed network $\Dcal(t)$. The graph $\Dcal(t+1)$ is obtained by first adding one new vertex, labelled with $(t+1)$. This new vertex is connected to $m$ vertices $\{v\}$ chosen with probability $\Pi(t,v)$ from the set of vertices in $\Dcal(t)$. Our convention is that the edge runs from node $v$ to node $(t+1)$.

In this model, $\Pi(t,v)$ is composed of two terms. With probability $\delta$,  the node $(t+1)$ is connected to a vertex $v$ chosen with a probability proportional to the number of edges $\kout(t,v)$ leaving $v$ at the time $t$. Price called this \vdef{cumulative advantage} and, after normalisation, we have that the probability of choosing $v$ is $\kout(t,v)/E(t)$. The second process (random attachment) happens with probability $(1-\delta)$ and in this case we choose the source vertex $v$ uniformly at random from the set of vertices in $\Dcal(t)$, i.e.\ with probability $1/N(t)$. So the probability of connecting the vertex $(t+1)$ to an  existing vertex $v$ is $\Pi(t,v)$ where
\beq
  \Pi(t,v) =
  \delta \frac{\kout(t,v)}{E(t)} + (1-\delta) \frac{1}{N(t)}
  \mbox{ if } t \geq v \geq 1  \, .
  \label{PricePidef}
\eeq
For simplicity, we always chose $\delta$ such that $\delta=m/(1+m)$, the choice originally made by Price.
The result of this choice is that for larger values of $m$ the amount of preferential attachment increases, and the amount of random attachment decreases. For an in-depth discussion of Price model see~\cite{ECV20}. Let us now consider the minimal cycle bases of the DAGs obtained from this Price model. Our first remark is about the effect of transitive reduction. The number of edges after TR, $E_{\textrm{TR}}$, increases with $m$ in the Price model, as \figref{f_ep_number_diamonds_mixers} (bottom right) shows. Since edge participation increases with $m$, together with the decrease of the average cycle sizes, we can conclude that networks tend to have an increasing number of cycles with increasing $m$. It is important to note that even for small $m$, $E_p>1$ indicates that, on average, each edge participates in more that one cycle. Although on average each edge participates in at least one cycle, there are edges which are ``more active'' than others, as the average value of $E_p$ shows on the bottom right figure of \figref{f_ep_number_diamonds_mixers}. Thus transitively reduced Price model creates networks that are ``more lattice-like'' as the density parameter $m$ increases.

\begin{figure}
    \centering
    \includegraphics[width = 0.5\linewidth]{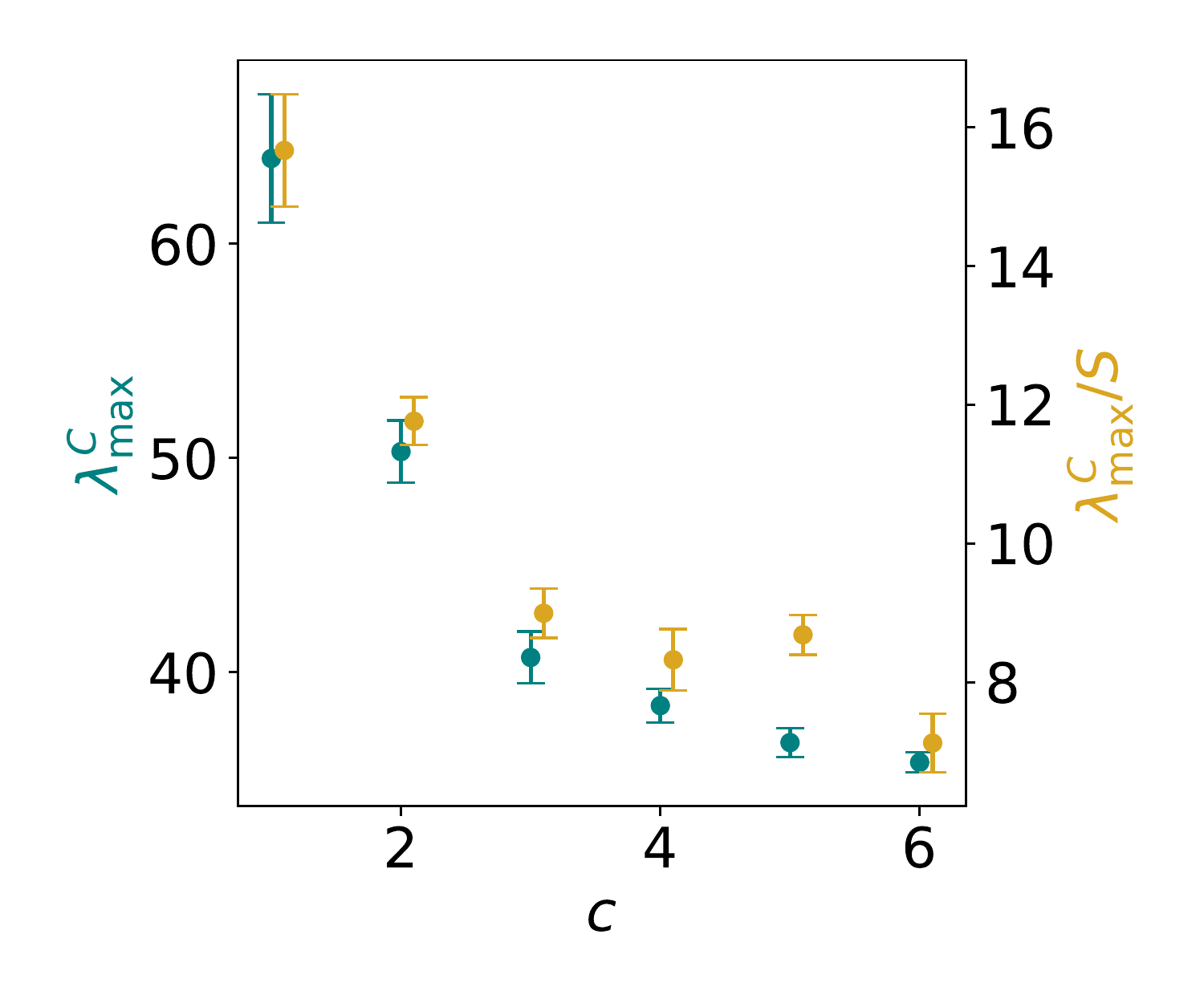}
    \caption{The largest eigenvalue of the Price model DAGs generated with fixed $m=7$ and varied $c= m(1-\delta)$. Here large $c$ values relate to networks in which many edges are attached due to random attachment as opposed to preferential attachment.}
    \label{fig:price_varyc_eigvals}
\end{figure}

Interestingly, the eigenvalues of $\mathbf{M}$ for the reduced Price model DAGs are independent of $m$, contrary to the eigenvalues of the reduced Erd\"os R\`enyi DAGs which depend on $p$. The largest eigenvalue $\lambda^C_{\textrm{max}}$ also does not fluctuate significantly with $m$. One plausible explanation of this phenomenon is that a large value of $\lambda^C_{\textrm{max}}$ can be a result of either strong interconnectedness of a cycle with other cycles, or its large size. In \figref{f_ep_number_diamonds_mixers} we saw that $E_p$ increases with $m$, whereas $\langle C\rangle$ decreases, indicating that cycles tend to be more interconnected, but smaller with increasing $m$. Thus the quasi-stationarity of $\lambda^C_{\textrm{max}}$ indicates that a balance between edge participation $E_p$ and average cycle size $\langle S_i\rangle$ must exist. In \figref{fig:price_varyc_eigvals} we consider $\lambda_{\textrm{max}}^C$ of Price DAGs with fixed $m$ and varied parameter $c$ which relates to the amount of random attachment of edges. Here we see that as randomness is increased in Price DAGs, the largest eigenvalue $\lambda_{\textrm{max}}^C$ decreases, indicating that the quasi-stationarity between the size and cycle interconnectedness is broken. Since $\langle S\rangle$ varies by a small amount as $c$ is varied, see \figref{fig:varyc_price}, and $\langle h\rangle$ increases, it means that random attachment allows for cycles to occur in more varied areas of a network thereby suppressing $\lambda_{\textrm{max}}^C$.

Finally, we see that in the reduced Price model DAGs, contrary to reduced Erd\"os R\`enyi DAGs, diamonds are as prevalent as mixers, see \figref{f_ep_number_diamonds_mixers}.

\paragraph{Comparing DAG models}\label{subsec:comparison}
We now turn to the question of the utility of the metrics to differentiate the MCBs for the DAG models and therfore the model themselves, and particularly their added values with respect to purely topological metrics. To make comparisons fair, the networks produced should have, at least approximately, similar properties. We split our comparison into two groups: deterministic and random DAG models. We match deterministic models on the number of cycles and the random models on the density of edges post transitive reduction, which is a natural choice when trying to understand the effect of the generating mechanisms on the properties and organisation of the MCBs.

The Lattice and Russian doll models have a lot of similarities but are also clearly differentiated by key metrics. While some difference are size independent and fixed features of the models, e.g. $\lambda^C_{\textrm{max}}$ and balance, other are size dependent and the difference will increase with the respective size of the DAGs, e.g. the stretch and height. We note that some of these differences depend on the choice of the root node in the Russian doll model.

\begin{figure}[!ht]

    \centering
    \includegraphics[width = 0.5\linewidth]{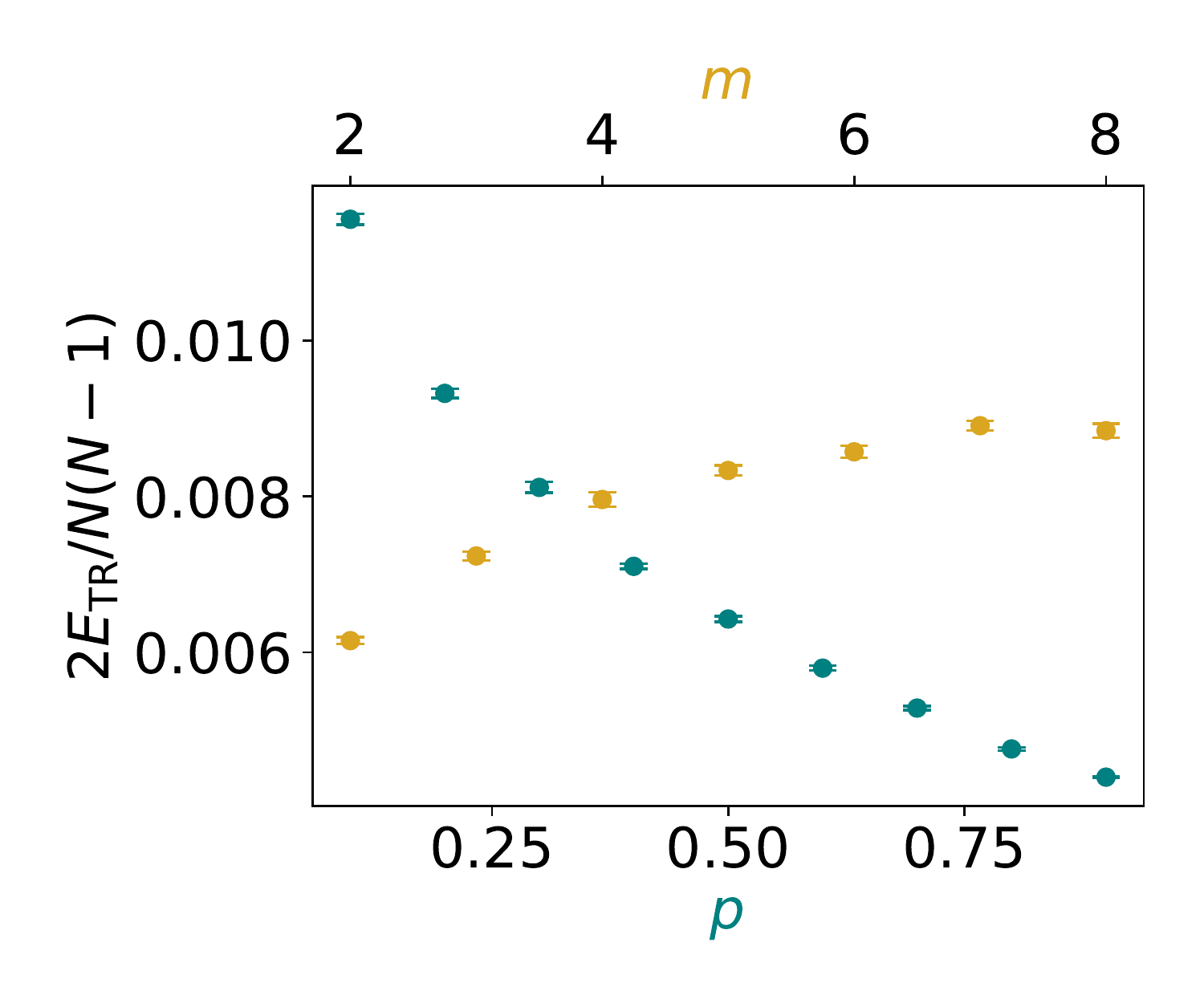}
    \caption{Comparison of the density of reduced DAGs from the Price model (green symbols, top abscissa) and Erd\"os-R\`enyi DAGs (yellow symbols, bottom abscissa), values averaged over $n=10$ realisations for networks of size $N=500$ and $\delta=m(1+m)$ for the Price Model. The densities statistically coincide at approximately $p=0.3$ for Erd\"os-R\`enyi DAGs, and $m=4$ for Price  DAG. We compare MCB statistics for these parameter values in \tabref{t_random_price_same_density}.}
    \label{fig:comparison}
\end{figure}

Let us now turn to the random models. To obtain transitively reduced DAGs with similar densities, we first scan through the parameter values $p$ and $m$ to determine a target density value post Transitive Reduction, see \figref{fig:comparison}. As the density is computed post Transitive Reduction, it is a random variable, hence the error bars and the impossibility of matching exactly the densities of the transitively reduced DAGs. The value of the metrics obtained for 1 realisation of each network models at their target density are presented in \tabref{t_random_price_same_density}. First we remark that the standard errors of the mean are always small comparatively to the value of the average, showing that the MCBs obtained have well-defined characteristics. The MCB of two random models also share similarities with metrics having overlapping standard errors of the mean, such as the maximum cycles size, the stretch. Average cycle size and edge participation are also fairly close. Transitively reduced Price DAGs tend to have a balance mixture of diamonds and mixers, while transitively reduced Erd\"os-R\`enyi DAGs have a clear majority of diamonds. Mixers being more prevalent in transitively reduced Price DAGs can be understood from the generating mechanism, as a more intricate structure is due to the attachment constraint. Another very significant difference due to the preferential attachment of the Price model is the much higher interconnectedness of cycles, reflected by a much higher value for the leading eigenvalue of $\mathbf{M}$.

The MCB thus reflects specificities of DAG models and the metrics we introduced naturally capture them. As expected, while the metrics are correlated, none is enough to capture the complexity of the organisation of MCBs when considered in isolation to the others. We also conjecture that their usefulness generalises to other DAG models, as well as to real world DAGs, thus providing objects capable not only of characterising DAGs but also differentiating them. Our results also show that pure topological metrics, i.e. not using any meta-data information, are not enough to differentiate the different models and that localising the topological features using meta data is necessary to uncover differences, similarly to what is done in \cite{Petri:2014hq} with persistent homology.

\begin{table}[h]

    \caption{Comparison of the MCB characteristics between the Lattice, Russian doll, the Transitively Reduced Erd\"os-R\`enyi DAG and Transitively Reduced Price model; with respective parameters $p=0.3$ and $m=3$ for the random models, see \figref{fig:comparison}. Values for the two random models are averaged over the cycles comprising the MCB. Errors are standard error of the mean over $n$ realisations.}
    \label{t_random_price_same_density}

\setlength{\tabcolsep}{2pt}
    \centering
    \begin{tabular}{lllll}
        \hline\hline
         Measure        & Lattice & Russian doll & Random $p=0.3$& Price $m=4$    \\ \hline
         $\sigma(C)$         &  0  &  0      & $0.7\pm 0.0(1)$   & $0.8\pm0.01$    \\
         $\langle C \rangle$ & 4   &  6      & $4.6\pm  0.0(1)$  & $5.0\pm0.04$    \\
         $C_{\mathrm{max}}$  & 4   &  6      & $8.5\pm 0.2$      & $ 7.6 \pm0.1$   \\
         $\langle s\rangle$  & 2   & $2+d/2$ & $4.5\pm0.04$      & $4.8\pm0.1$     \\
         $\langle b\rangle$  & 0   &  1/3    & $0.086\pm0.0$     & $.13\pm 0.00(3)$\\
         $\langle h\rangle$  & $d$ & $\frac{5d}{6}+\frac{1}{3}-\frac{1}{6d}$
                                             & $97.3\pm1.4$      & $5.6\pm0.2$     \\
         \# Diamonds         & $d$ & $d$     & $385.6\pm 7.2$    & $261.7\pm10.8$  \\
         \# Mixers           & 0   & 0       & $127.9\pm2.8$     & $232.5 \pm 8.4$ \\
         $E_p$               & 2   & 2       & $2.3\pm 0.01 $    & $2.5\pm 0.0(2)$ \\
  $\lambda^C_{\mathrm{max}}$ & 8   & 10      & $18.3\pm0.2$      & $61\pm 3$       \\
$\lambda^C_{\mathrm{max}}/C$ & 2   & 1.66    & $4.0\pm0.1$       & $15\pm 0.8$     \\
        null$(\mathbf{L}^C)$ & 1   & 1       & $1.65\pm0.2$      & 1$\pm0$         \\
        \hline\hline
    \end{tabular}
\end{table}

\section{Conclusion}
In this paper, we defined four general classes of directed cycles by considering the underlying undirected graph and augmenting it with the meta data associated with edge directionality. We then presented a principled way to define, characterise and interpret cycles and Minimal Cycle Bases in Directed Acyclic Graphs (DAGs) by computing an MCB on the underlying undirected graph and augmenting it with the meta data associated with DAGs. We then focused on Transitively Reduced DAGs.

We simplified the representation of a DAG via Transitive Reduction to consider the minimal amount of information needed to retain causal connectivity with the aim to reduce the variance in the composition of the MCB and thus a stable characterisation of network models once enriched with the mate-data. The simplification also gives the potential to only consider specific types of cycle bases based purely on diamonds. Moreover, we have shown numerically that this simplification does not reduce the discriminatory power of the metrics we introduced, as we were able to clearly differentiate and characterise DAG models. Moreover, the comparison between the MCB for the reduced Erd\"os-R\`enyi DAG and the reduced Price model DAG, and also between the lattice and Russian doll models, clearly shows that topological features are not enough to pinpoint differences, particularly the ones linked to the generating mechanisms, between systems and localising the topological features of cycles.

Some of the metrics we defined are not --- reduced --- DAG specific and can be used to characterise any cycle basis, and can thus be applied to any directed network. The measures which are DAG-specific, i.e.\ that require the notion of order, height, antichain, and longest path, may still be the most interesting: they can literally be used as ``coordinates'' of cycle, allowing their geometrical embedding and localisation.

\input{supplementary_material}

\clearpage

\bibliographystyle{unsrt}

\end{document}

%% file: supplementary_material.tex

\begin{center}
\large\textbf{Appendices}
\end{center}
\appendix
\renewcommand{\thesection}{\Alph{section}}
\setcounter{equation}{0}
\renewcommand{\theequation}{\thesection\arabic{equation}}
\renewcommand{\thefigure}{\thesection\arabic{figure}}
\renewcommand{\thetable}{\thesection\arabic{table}}
\setcounter{section}{0}


\section{Description of Horton's algorithm for minimum diamond basis}\label{app:mdb}

Horton's algorithm~\cite{H87} is the first proposed polynomial time algorithm for finding a fundamental MCB. The algorithm first produces $O(EN)$ of fundamental cycles $\Hcal$, called \vdef{Horton cycles}. 
 Horton showed that $\Hcal$ contains at least one fundamental cycle basis, and the goal of the algorithm is to sift the Horton set to find it.

To find the shortest paths in each tree, Dijkstra's shortest path search algorithm can be used. 
Then, Gaussian elimination is performed to find a Minimum Cycle Basis from $\Hcal$. A variation of Horton's algorithm was proposed in~\cite{MD05}, where authors suggested to replace Gaussian elimination by an iterative construction of a cycle basis. To find an MCB in this way, one uses support vectors in order to retrieve MCB from $\Hcal$. These support vectors are described in description of De Pina's algorithm.

It is possible to adapt Horton's algorithm to DAGs to obtain a cycle basis composed solely of diamonds. To proceed, suppose that we have a directed acyclic graph with a single global source node, a root. Such DAG can be reduced to an arborescence, a directed tree which has one source node that has paths to every other node~\cite{C18}. If we consider the directionality of edges while building $\Hcal$, all directed paths must originate at a node with a smaller height and end at a node with larger height. In other words, for each $v$ and edge $(u,v)$ a Horton cycle with nodes $SP(v,u),SP(v,w),\{u,v\}$ can be constructed. Such a cycle is always a diamond. There is also no need for building $N$ arborescences (composed of $v$ as a source node as well as its descendants), as each such arborescence is contained within the arborescence grown from the global root. Since we build a single tree, we reduce the size of $\Hcal$ from $O(NE)$ to $O(E)$, minimising both temporal and memory demands of the algorithm. Of course, such cycle basis must not be minimum, as it is not able to detect mixers. We will also achieve a heuristic speedup at node level, as the paths are searched for only in its descendant list which is (heuristically) smaller than $N$. Let us call this algorithm for minimum diamond basis (MDB) detection.

\section{Lattice Model Spectral Properties}\label{app:Lspectral}

Here we give more details on the spectral properties of the Lattice model of \secref{sub_sec:lattice}. Consider a network which is equivalent to a square lattice of $L$ by $L$ nodes. The MCB is unique and can be visualised as an $(L-1) \times (L-1)$ square `cycle' lattice (in fact the dual of the original lattice). In this lattice each cycle is a lattice point and each cycle has one edge in common with the cycle represented by the nearest neighbour nodes in this cycle lattice. The dimension of the basis is $(L-1)^2$. We find this we find the $\Mmatr$ for $L=5$ is of the form
\begin{equation}
\begin{array}{cccccccccccccccc}
4  & 1  & 0  & 0  & 1  & 0  & 0  & 0  & 0  & 0  & 0  & 0  & 0  & 0  & 0  & 0  \\
1  & 4  & 1  & 0  & 0  & 1  & 0  & 0  & 0  & 0  & 0  & 0  & 0  & 0  & 0  & 0  \\
0  & 1  & 4  & 1  & 0  & 0  & 1  & 0  & 0  & 0  & 0  & 0  & 0  & 0  & 0  & 0  \\
0  & 0  & 1  & 4  & 0  & 0  & 0  & 1  & 0  & 0  & 0  & 0  & 0  & 0  & 0  & 0  \\
1  & 0  & 0  & 0  & 4  & 1  & 0  & 0  & 1  & 0  & 0  & 0  & 0  & 0  & 0  & 0  \\
0  & 1  & 0  & 0  & 1  & 4  & 1  & 0  & 0  & 1  & 0  & 0  & 0  & 0  & 0  & 0  \\
0  & 0  & 1  & 0  & 0  & 1  & 4  & 1  & 0  & 0  & 1  & 0  & 0  & 0  & 0  & 0  \\
0  & 0  & 0  & 1  & 0  & 0  & 1  & 4  & 0  & 0  & 0  & 1  & 0  & 0  & 0  & 0  \\
0  & 0  & 0  & 0  & 1  & 0  & 0  & 0  & 4  & 1  & 0  & 0  & 1  & 0  & 0  & 0  \\
0  & 0  & 0  & 0  & 0  & 1  & 0  & 0  & 1  & 4  & 1  & 0  & 0  & 1  & 0  & 0  \\
0  & 0  & 0  & 0  & 0  & 0  & 1  & 0  & 0  & 1  & 4  & 1  & 0  & 0  & 1  & 0  \\
0  & 0  & 0  & 0  & 0  & 0  & 0  & 1  & 0  & 0  & 1  & 4  & 0  & 0  & 0  & 1  \\
0  & 0  & 0  & 0  & 0  & 0  & 0  & 0  & 1  & 0  & 0  & 0  & 4  & 1  & 0  & 0  \\
0  & 0  & 0  & 0  & 0  & 0  & 0  & 0  & 0  & 1  & 0  & 0  & 1  & 4  & 1  & 0  \\
0  & 0  & 0  & 0  & 0  & 0  & 0  & 0  & 0  & 0  & 1  & 0  & 0  & 1  & 4  & 1  \\
0  & 0  & 0  & 0  & 0  & 0  & 0  & 0  & 0  & 0  & 0  & 1  & 0  & 0  & 1  & 4  \\
\end{array}
\label{e:Mlattice4}
\end{equation}
If we consider the entry $M_{\alpha\beta}$ with indices $\alpha,\beta \in \{ 0,1,\ldots,(L-2)^2-1\}$ then we can think of label $\alpha$ as representing the cycle placed at $(x,y)$ in the cycle lattice where $x,y \in \{ 1,2,\ldots,(L-1)\}$ and $\alpha=(x-1)+(y-1)*(L-1)$ so $x,y \in \{1,\ldots,(L-1)\}$. The nodes of the original lattice are then at $(X,Y)$ where $X,Y \in \{ 1,2,\ldots,L\}$ with the cycle $(x,y)$ running between nodes of the original network at $(X,Y)=\{ (x,y), (x+1,y), (x,y+1), (x+1,y+1) \}$.
If we write the eigenvectors as $v_\alpha \equiv v_{x,y}$ then the eigenvalue equation is
\beq
 \lambda v_{x,y} = 4 v_{x,y} + v_{x+1,y}  + v_{x-1,y}  + v_{x,y+1}  + v_{x,y-1}  \,,
 \quad \mbox{with} \;\;\;
 v_{0,y}=v_{x,0}=v_{L,y}=v_{x,L}=0 \, .
\eeq
This can be solved by assuming the eigenvectors take the form $v_{x,y} = \sin(\pi m x/L)\sin(\pi n y/L)$
which gives the eigenvalues as
\begin{eqnarray}
     \lambda^{(\omega)}
     &=&
     4
     + 2\cos \left( \frac{ \pi m}{L} \right)
     + 2\cos \left( \frac{ \pi n}{L} \right)  \, ,
     \nonumber
     \\
     && \qquad 
     \quad \omega = (m-1)+(n-1)\,(L-1), \quad m,n \in \{ 1, \ldots, (L-1) \} \, .
     \label{a:sqlatev}
\end{eqnarray}
Note that the solution shows that the constant $4$ comes from the number of edges each cycles shared with itself, i.e.\ the size of each cycle, while each of the four $\cos$ factors comes from one edge shared with each neighbouring cycle.
Once again we notice in the large system limit, $L\to \infty$, the largest eigenvalue is $8$ (the Peron-Frobenius theorem tells us the largest eigenvalue is between five and eight inclusive) with $4$ coming from the size of the cycle, and the remaining contribution of four is the number of edges shared with other cycles.

\begin{landscape}
For example, for the square lattice with $L=5$, we have the following numerical results to three decimal places for the eigenvalues and eigenvectors for cycle overlap matrix $\Mmatr$.
\begin{center}\small
\begin{tabular}{c||c|c|c|c|c|c|c|c|c|c|c|c|c|c|c|c}
 $\phi$ & 0 & 5 & 4 & 2 & 6 & 9 & 12 & 13 & 15 & 14 & 10 & 11 & 3 & 8 & 7 & 1   \\ \hline \hline
$\lambda^{(\phi)}$ & $7.236 $ & $6.236 $ & $6.236 $ & $5.236 $ & $5 $ & $5 $ & $4 $ & $4 $ & $4 $ & $4 $ & $3 $ & $3 $ & $2.764 $ & $1.764 $ & $1.764 $ & $0.764 $   \\ \hline
$\eigenvector{\phi}_{0}$ & $0.138$ & $0.002$ & $0.316$ & $0.362$ & $-0.316$ & $-0.033$ & $-0.548$ & $0.008$ & $-0.055$ & $-0.008$ & $-0.316$ & $-0.010$ & $0.362$ & $-0.005$ & $0.316$ & $0.138$   \\
$\eigenvector{\phi}_{1}$ & $0.224$ & $-0.155$ & $0.354$ & $0.224$ & $-0.158$ & $-0.368$ & $-0$ & $-0.119$ & $0.297$ & $0.356$ & $0.158$ & $0.358$ & $-0.224$ & $0.164$ & $-0.354$ & $-0.224$   \\
$\eigenvector{\phi}_{2}$ & $0.224$ & $-0.352$ & $0.158$ & $-0.224$ & $-0.158$ & $-0.368$ & $0.183$ & $0.079$ & $-0.282$ & $0.244$ & $-0.158$ & $-0.358$ & $-0.224$ & $-0.356$ & $0.158$ & $0.224$   \\
$\eigenvector{\phi}_{3}$ & $0.138$ & $-0.316$ & $0$ & $-0.362$ & $-0.316$ & $-0.033$ & $0$ & $-0.412$ & $-0.195$ & $-0.240$ & $0.316$ & $0.010$ & $0.362$ & $0.316$ & $0$ & $-0.138$   \\
$\eigenvector{\phi}_{4}$ & $0.224$ & $0.161$ & $0.354$ & $0.224$ & $-0.158$ & $0.335$ & $0$ & $0.119$ & $-0.297$ & $-0.356$ & $0.158$ & $-0.348$ & $-0.224$ & $-0.152$ & $-0.354$ & $-0.224$   \\
$\eigenvector{\phi}_{5}$ & $0.362$ & $0.002$ & $0.316$ & $0.138$ & $0.316$ & $0.033$ & $0.365$ & $-0.087$ & $0.337$ & $-0.236$ & $0.316$ & $0.010$ & $0.138$ & $-0.005$ & $0.316$ & $0.362$   \\
$\eigenvector{\phi}_{6}$ & $0.362$ & $-0.316$ & $0$ & $-0.138$ & $0.316$ & $0.033$ & $0$ & $0.531$ & $-0.102$ & $-0.117$ & $-0.316$ & $-0.010$ & $0.138$ & $0.316$ & $0$ & $-0.362$   \\
$\eigenvector{\phi}_{7}$ & $0.224$ & $-0.355$ & $-0.158$ & $-0.224$ & $-0.158$ & $0.335$ & $-0.183$ & $-0.079$ & $0.282$ & $-0.244$ & $-0.158$ & $0.348$ & $-0.224$ & $-0.351$ & $-0.158$ & $0.224$   \\
$\eigenvector{\phi}_{8}$ & $0.224$ & $0.355$ & $0.158$ & $-0.224$ & $-0.158$ & $0.335$ & $0.183$ & $0.079$ & $-0.282$ & $0.244$ & $-0.158$ & $0.348$ & $-0.224$ & $0.351$ & $0.158$ & $0.224$   \\
$\eigenvector{\phi}_{9}$ & $0.362$ & $0.316$ & $0$ & $-0.138$ & $0.316$ & $0.033$ & $-0$ & $-0.531$ & $0.102$ & $0.117$ & $-0.316$ & $-0.010$ & $0.138$ & $-0.316$ & $-0$ & $-0.362$   \\
$\eigenvector{\phi}_{10}$ & $0.362$ & $-0.002$ & $-0.316$ & $0.138$ & $0.316$ & $0.033$ & $-0.365$ & $0.087$ & $-0.337$ & $0.236$ & $0.316$ & $0.010$ & $0.138$ & $0.005$ & $-0.316$ & $0.362$   \\
$\eigenvector{\phi}_{11}$ & $0.224$ & $-0.161$ & $-0.354$ & $0.224$ & $-0.158$ & $0.335$ & $-0$ & $-0.119$ & $0.297$ & $0.356$ & $0.158$ & $-0.348$ & $-0.224$ & $0.152$ & $0.354$ & $-0.224$   \\
$\eigenvector{\phi}_{12}$ & $0.138$ & $0.316$ & $0$ & $-0.362$ & $-0.316$ & $-0.033$ & $0$ & $0.412$ & $0.195$ & $0.240$ & $0.316$ & $0.010$ & $0.362$ & $-0.316$ & $-0$ & $-0.138$   \\
$\eigenvector{\phi}_{13}$ & $0.224$ & $0.352$ & $-0.158$ & $-0.224$ & $-0.158$ & $-0.368$ & $-0.183$ & $-0.079$ & $0.282$ & $-0.244$ & $-0.158$ & $-0.358$ & $-0.224$ & $0.356$ & $-0.158$ & $0.224$   \\
$\eigenvector{\phi}_{14}$ & $0.224$ & $0.155$ & $-0.354$ & $0.224$ & $-0.158$ & $-0.368$ & $0$ & $0.119$ & $-0.297$ & $-0.356$ & $0.158$ & $0.358$ & $-0.224$ & $-0.164$ & $0.354$ & $-0.224$   \\
$\eigenvector{\phi}_{15}$ & $0.138$ & $-0.002$ & $-0.316$ & $0.362$ & $-0.316$ & $-0.033$ & $0.548$ & $-0.008$ & $0.055$ & $0.008$ & $-0.316$ & $-0.010$ & $0.362$ & $0.005$ & $-0.316$ & $0.138$   \\
\end{tabular}
\end{center}
Note that the index on the eigenvectors in the table, $\phi$, is assigned in some unspecified order by the numerical routine but is in a one-to-one map with the $\omega$ labels used in \eqref{a:sqlatev}.  The eigenvalues are in the row $\lambda^{(\omega)}$ while  $\eigenvector{\omega}_{\alpha}$ gives the $\alpha$-th entry of the eigenvector $\omega$, that is \beq
 \lambda^{(\omega)} \eigenvector{\omega}_{\alpha} = \sum_\alpha M_{\alpha\beta}\eigenvector{\omega}_{\beta} \, .
\eeq 
The symmetry group, $D_4$, ensures that some eigenvalues are repeated in pairs and in that case the eigenvectors given are just two examples from a two-dimensional subspace. Accidental degeneracy is also present, e.g.\ in the eigenvectors 12 to 15.

\end{landscape}

\section{Russian Doll Model Spectral Properties}\label{app:Rdspectral}


Here we give more details on the spectral properties of the Russian Doll model of \secref{sub_sec:rd}. Suppose we label the cycles so that $C_d$ is the largest cycle in the MCB (minimal cycle basis) of $R_d$ described in \secref{sub_sec:rd}.  The matrix $\Mmatr$ of \eqref{e:Mdef} describing cycle overlap is a symmetric tridiagonal matrix and here it has a 6 on the diagonal, 2 on the leading (non-zero) off-diagonals, with the exception that $M_{00}=4$ as the first cycle, $C_1$, the only cycle in $R_1$, is the only one of size 4. 
That is general we have a matrix of the form 
\begin{equation}
     \Mmatr
     =
     b \Imatr
     +
    \left( \begin{array}{cccccccc}
    -c       & a       & 0      & 0       & \cdots & 0      & 0       & 0       \\
    a        & 0       & a      & 0       & \cdots & 0      & 0       & 0       \\
    0        & a       & 0      & a       & \cdots & 0      & 0       & 0       \\
    \vdots   & \vdots  & \vdots & \vdots  & \cdots & \vdots & \vdots  & \vdots  \\
    0        & 0       & 0      & 0       & \cdots & a      & 0       & a       \\
    0        & 0       & 0      & 0       & \cdots & 0      & a       & 0       \\
    \end{array}
\right)
\end{equation}
with $a=2$, $b=6$, $c=2$ in our case. Following \cite{K06b} we find the eigenvalues satisfy 
\begin{equation}
     \lambda^{(\omega)}
     =
     6 + 4\cos \left( \frac{2 \pi \omega}{(2d+1)} \right) \, ,
     \quad 
     \omega \in \{ 1, \ldots, d \} \, .
\end{equation}
where the $6$ comes from the size of the majority of cycles. 

For example, $d=4$ we have
\begin{equation}
    \Mmatr
    =
    \left( \begin{array}{cccc}
    4  & 2  & 0  & 0  \\
    2  & 6  & 2  & 0  \\
    0  & 2  & 6  & 2  \\
    0  & 0  & 2  & 6  \\
    \end{array}
    \right)
    \label{e:MR4}
\end{equation}
and we find approximately that the eigenvalues and eigenvectors are
\begin{center}
\begin{tabular}{c || c | c | c | c}
 $\omega$                  & 1       & 2        & 3        & 4   \\ \hline \hline
$\lambda$                  & $9.064$ &  $6.695$ &  $4.000$ & $2.241 $   \\ \hline
$\eigenvector{\omega}_{0}$ & $0.228$ & $-0.429$ &  $0.577$ & $0.657$   \\
$\eigenvector{\omega}_{1}$ & $0.577$ & $-0.577$ &  $0.000$ & $-0.577$   \\
$\eigenvector{\omega}_{2}$ & $0.657$ &  $0.228$ & $-0.577$ & $0.429$   \\
$\eigenvector{\omega}_{3}$ & $0.429$ &  $0.657$ &  $0.577$ & $-0.228$   \\
\end{tabular}
\end{center}
where $9.064\approx 6+ 4\cos(2\pi/9)$, $6.695\approx 6+ 4\cos(4\pi/9)$, $4 = 6+ 4\cos(6\pi/9)$, and $6.695 \approx 6+ 4\cos(8\pi/9)$.

\section{Price model for $m=7$}

We can vary the amount of random attachment in the Price model by considering a fixed value of $m$ and varying $\delta$. In this section we report the results obtained by fixing $m=7$ and varying $\delta=1-c/m$ where $c\in[1,6]$.

\begin{figure}[!ht]
\centering
        \textbf{Reduced Price model}\par\medskip
         \includegraphics[width=4cm]{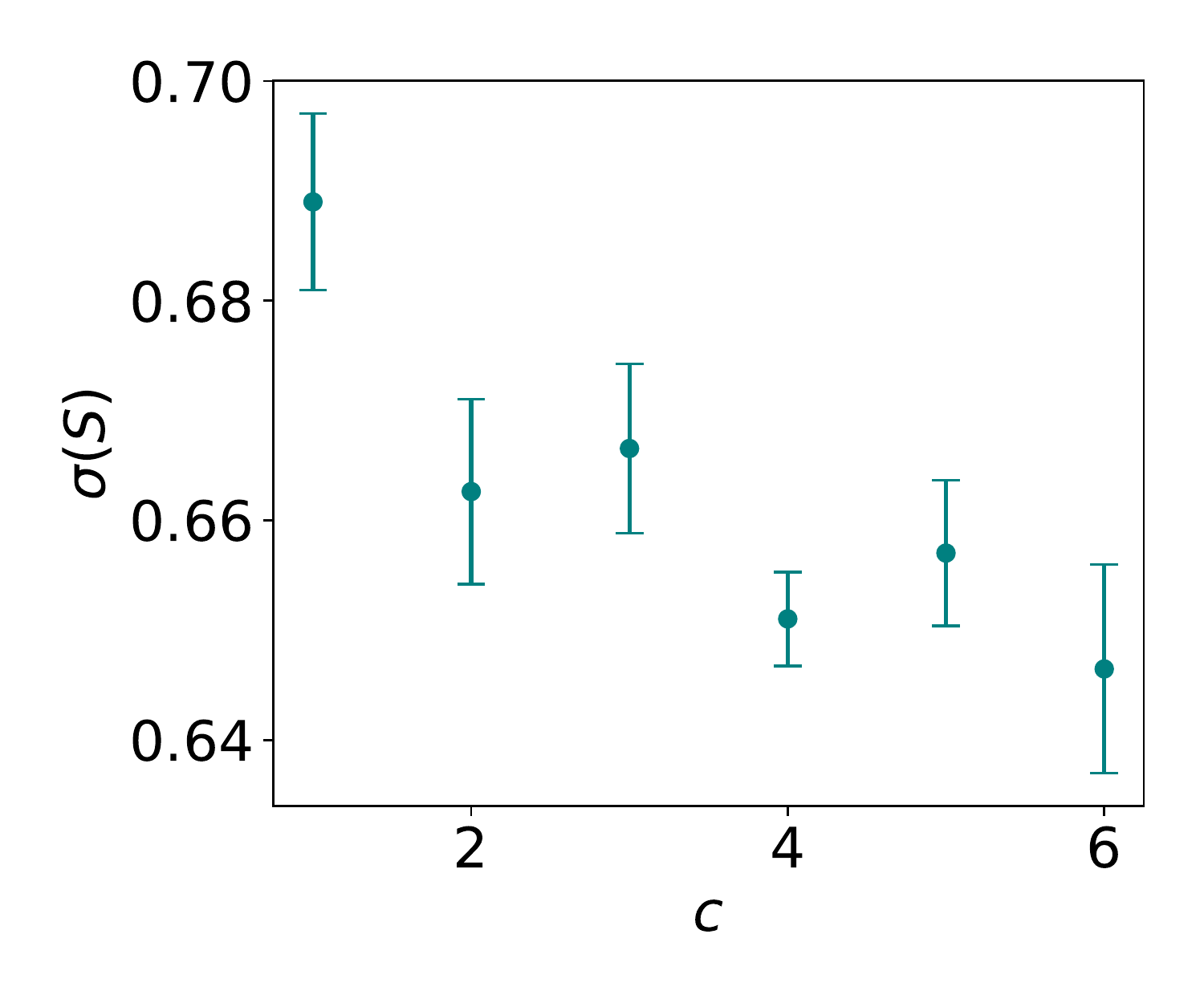}
        \includegraphics[width=4cm]{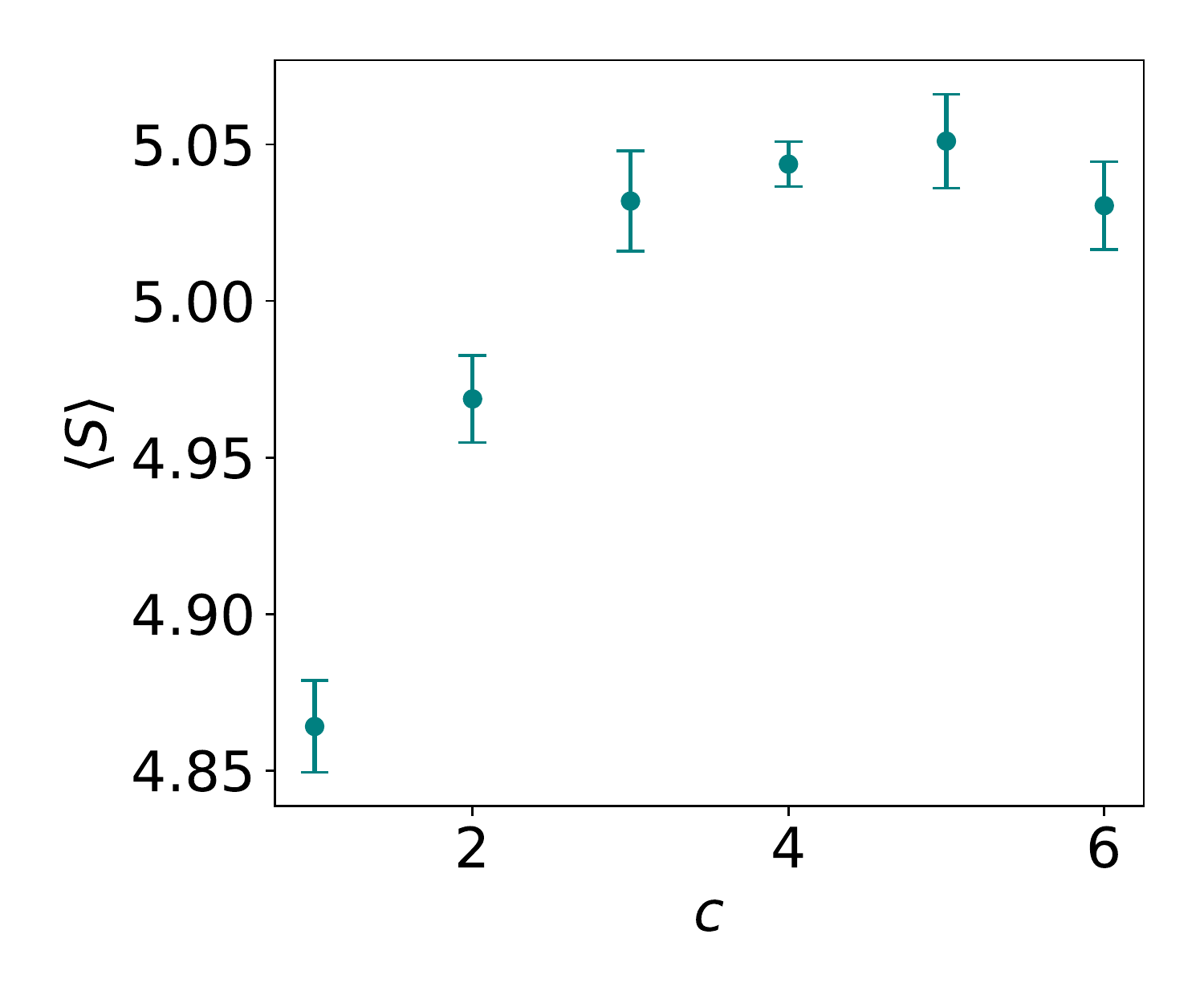}
        \includegraphics[width=4cm]{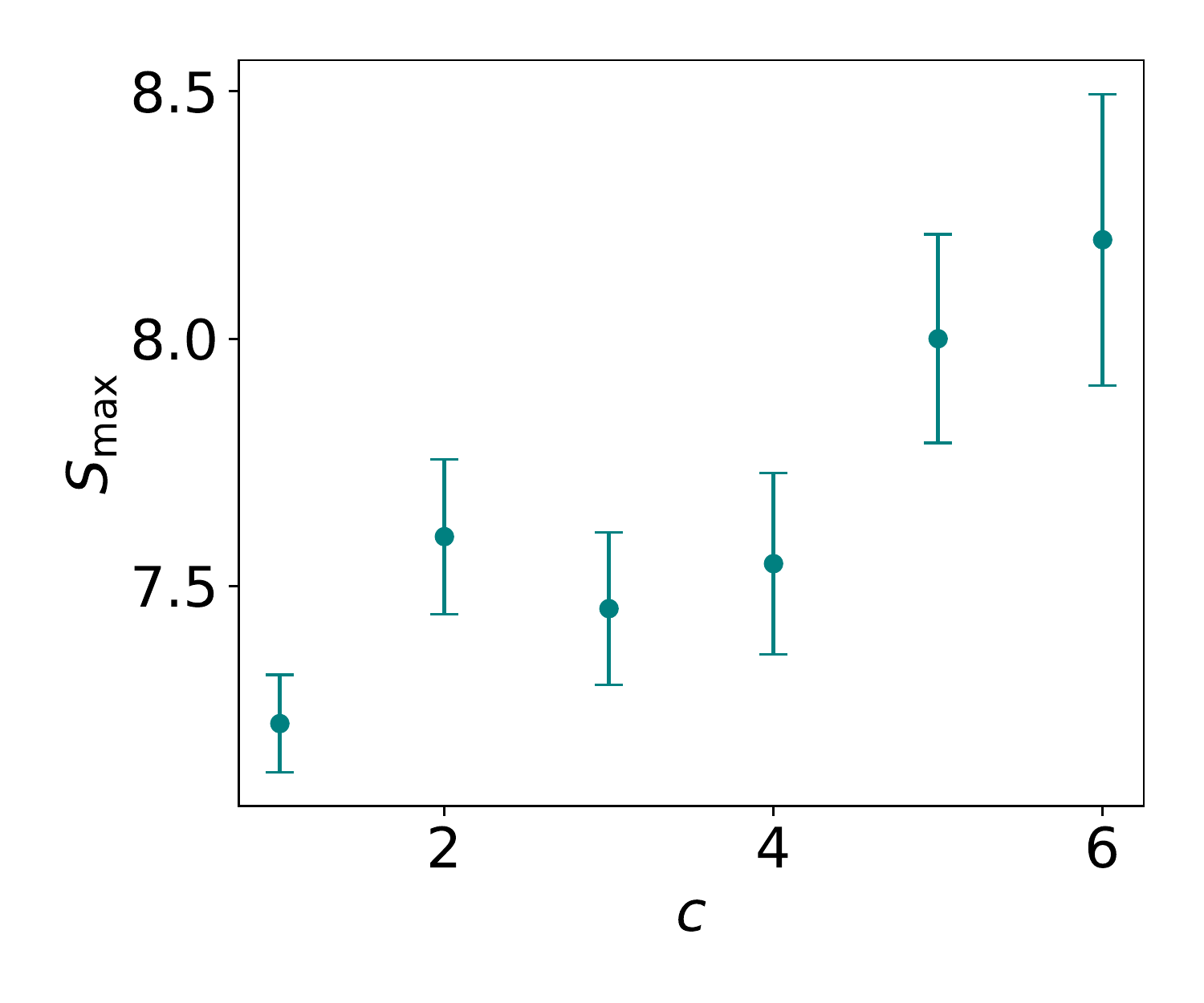}\\
        \includegraphics[width=4cm]{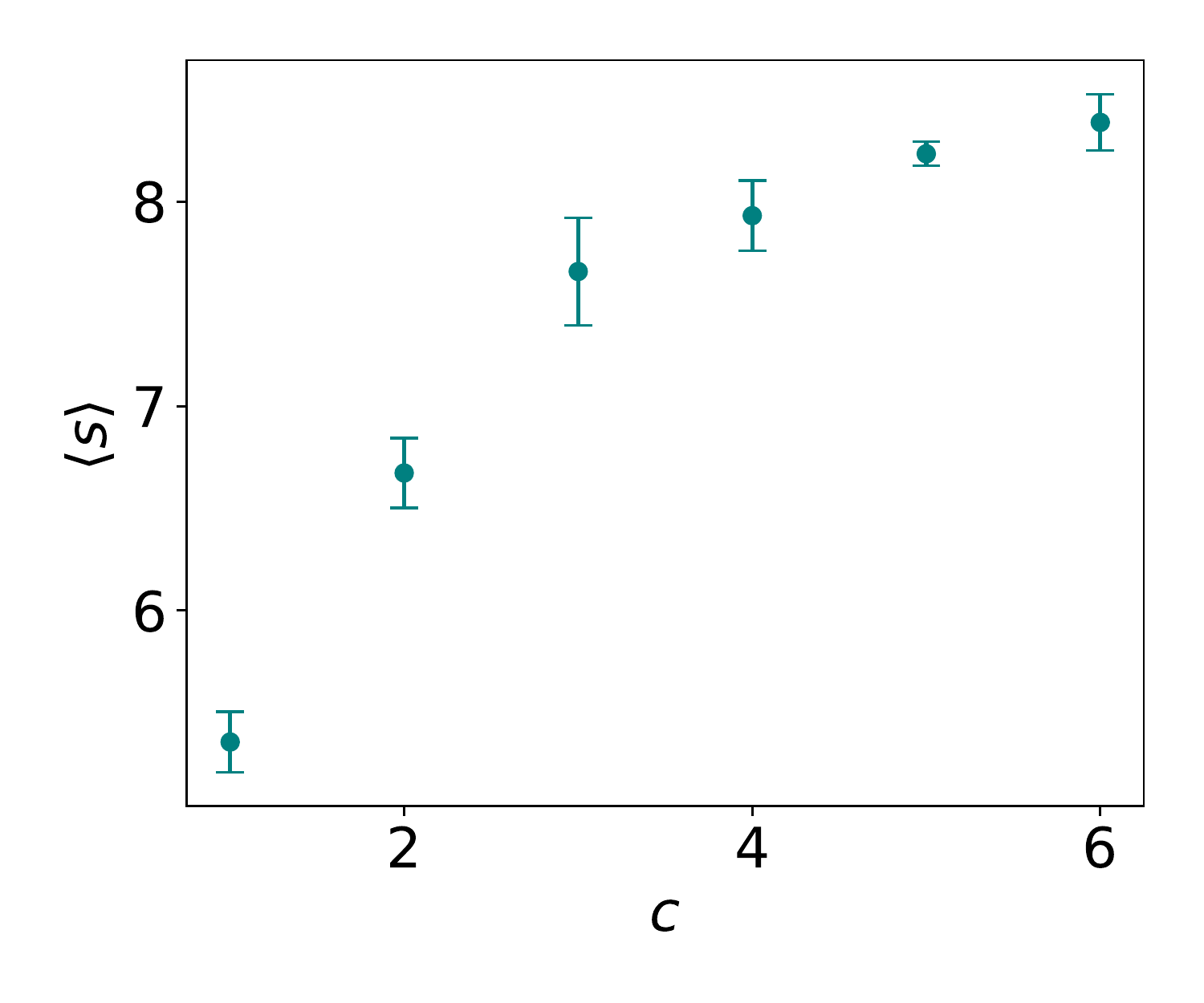}
        \includegraphics[width=4cm]{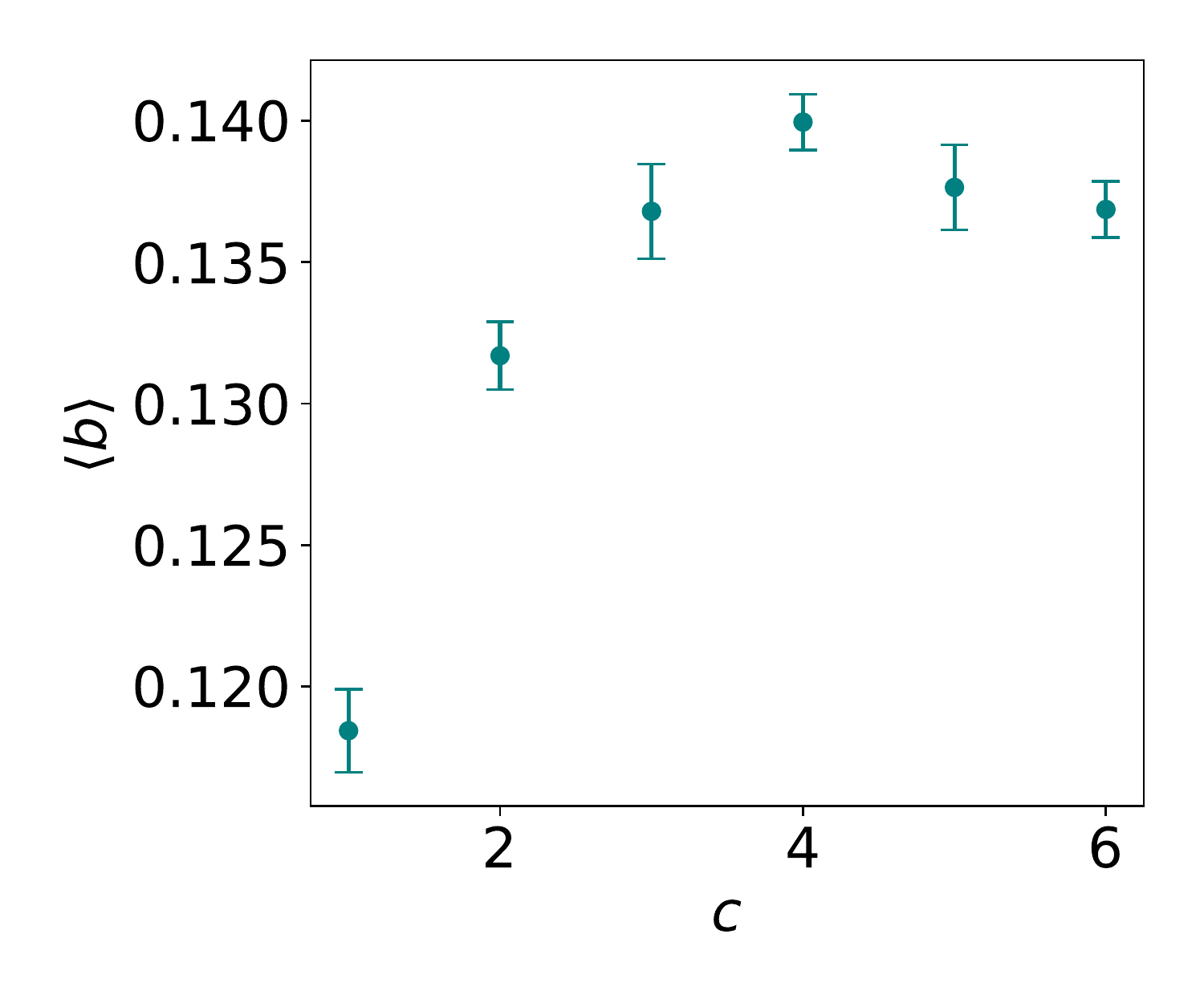}
        \includegraphics[width=4cm]{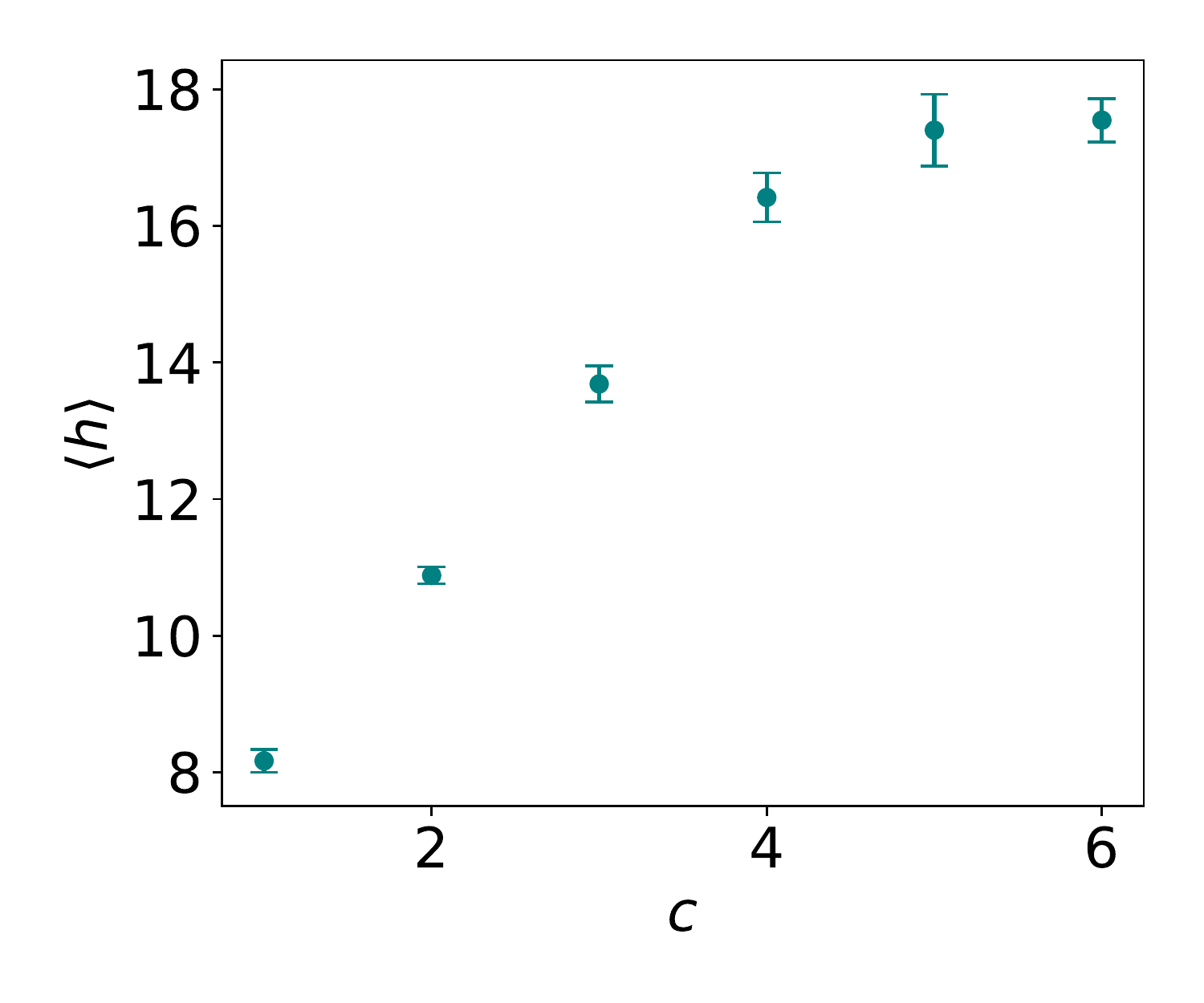}\\
        \includegraphics[width=4cm]{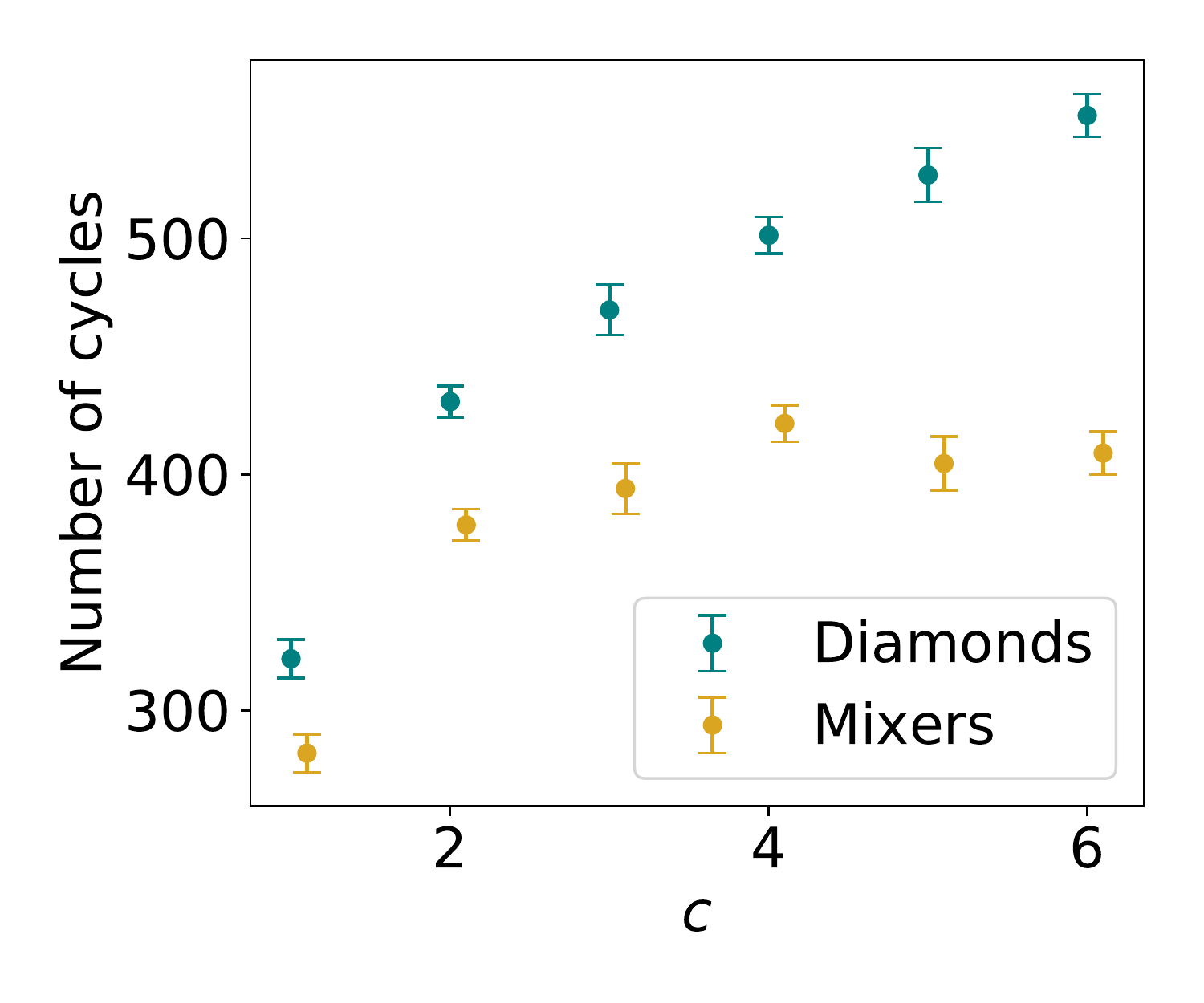}
        \includegraphics[width=4cm]{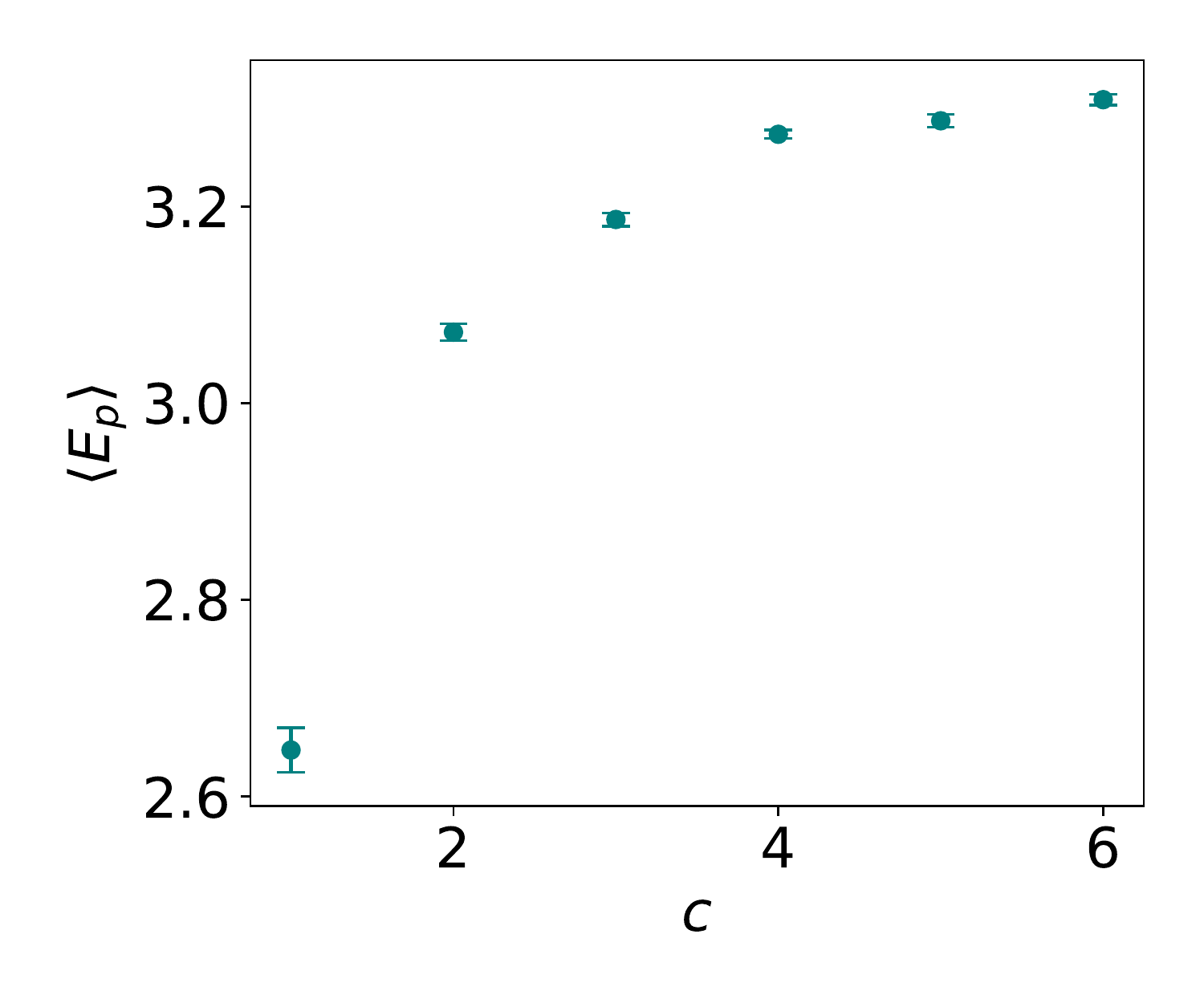}
    \caption{Collected cycle basis statistics - $\sigma(S)$, $\langle S\rangle$, $S_{max}$, $\langle s\rangle$, $\langle b\rangle$ and $\langle h\rangle$, see \secref{sec:cycle_metrics} for the definitions -- for Transitively Reduced Price model for a fixed $m=7$ and varied values of $c$. We considered networks with $N=500$ nodes, and for stochastic network models, we generated $n=10$ realisations for each parameter value and computed one MCB for each realisation.}\label{fig:varyc_price}
\end{figure}